\documentclass{aa}

\usepackage[varg]{txfonts}
\usepackage[utf8]{inputenc}

\usepackage{hyperref}
\hypersetup{
        colorlinks = true,
        citecolor=blue,
        urlcolor=purple,
        linkcolor=magenta
}

\usepackage[dvipsnames]{xcolor}
\usepackage{graphicx}
\usepackage{amsmath}
\usepackage{amssymb}
\usepackage{epstopdf}
\usepackage{afterpage} 

\usepackage{booktabs}
\usepackage[flushleft]{threeparttable}

\usepackage{natbib}
\bibpunct{(}{)}{;}{a}{}{,} 

\newcommand{\rhk}{$\log R'_\text{HK}$}
\newcommand{\ca}{\ion{Ca}{ii} H\&K~}

\begin{document}

\title{Stellar chromospheric activity of 1 674 FGK stars from the AMBRE-HARPS sample I.}
\subtitle{A catalogue of homogeneous chromospheric activity\thanks{Table \ref{tab:full} is only available in electronic form at the CDS via anonymous ftp to \url{cdsarc.u-strasbg.fr} (130.79.128.5) or via \url{http://cdsweb.u-strasbg.fr/cgi-bin/qcat?J/A+A/}.}}

\author{
    J.~Gomes~da~Silva\inst{\ref{1}}\thanks{Electronic address: \texttt{Joao.Silva@astro.up.pt}; Corresponding author} \and 
    N.C.~Santos\inst{\ref{1},\ref{2}} \and
    V.~Adibekyan\inst{\ref{1}, \ref{2}} \and
    S.G.~Sousa\inst{\ref{1}} \and
    Tiago~L.~Campante\inst{\ref{1}, \ref{2}} \and
    P.~Figueira\inst{\ref{5},\ref{1}} \and
    D.~Bossini\inst{\ref{1}} \and
    E.~Delgado-Mena\inst{\ref{1}} \and
    M\'ario~J.P.F.G.~Monteiro\inst{\ref{1}, \ref{2}} \and
    P.~de~Laverny\inst{\ref{3}} \and
    A.~Recio-Blanco\inst{\ref{3}} \and
    C.~Lovis\inst{\ref{4}}
    }

\institute{
    Instituto de Astrof\'isica e Ci\^{e}ncias do Espa\c{c}o, Universidade do Porto, CAUP, Rua das Estrelas, PT4150-762 Porto, Portugal\label{1}
    \and
    Departamento de Fi\'isica e Astronomia, Faculdade de Ci\^encias, Universidade do Porto, 4169-007 Porto, Portugal\label{2}
    \and
    European Southern Observatory, Alonso de Cordova 3107, Vitacura, Santiago, Chile\label{5}
    \and
    Universit\'e C\^ote d'Azur, Observatoire de la C\^ote d'Azur, CNRS, Laboratoire Lagrange, France\label{3}
    \and Observatoire Astronomique de l'Universit\'e de Gen\`eve, 51 Ch. des Maillettes, 1290, Versoix, Switzerland\label{4}
    }


\abstract
{}
{The main objective of this project is to characterise chromospheric activity of FGK stars from the HARPS archive. We start, in this first paper, by presenting a catalogue of homogeneously determined chromospheric emission (CE), stellar atmospheric parameters, and ages for 1 674 FGK main sequence (MS), subgiant, and giant stars. The analysis of CE level and variability is also performed.}
{We measured CE in the \ca lines using more than 180  000 high-resolution spectra from the HARPS spectrograph, as compiled in the AMBRE project, obtained between 2003 and 2019. We converted the fluxes to bolometric and photospheric corrected chromospheric emission ratio, $R'_\text{HK}$. Stellar atmospheric parameters $T_\text{eff}$, $\log g$, and [Fe/H] were retrieved from the literature or determined using a homogeneous method. $M_\star$, $R_\star$, and ages were determined from isochrone fitting.}
{We show that our sample has a distribution of CE for MS stars that is consistent with an unbiased sample of solar-neighbour MS stars. We analysed the CE distribution for the different luminosity classes and spectral types and confirmed the existence of the very inactive (VI) star and very active (VA) star populations at \rhk$< -5.1$ and $> -4.2$ dex, respectively. We found indications that the VI population is composed mainly of subgiant and giant stars and that \rhk$= -5.1$ dex marks a transition in stellar evolution.
Overall, CE variability decreases with decreasing CE level but its distribution is complex.
There appears to be at least three regimes of variability for inactive, active, and very active stars, with the inactive and active regimes separated by a diagonal, extended Vaughan-Preston (VP) gap.
We show that stars with low activity levels do not necessarily have low variability.
In the case of K dwarfs, which show high CE variability, inactive and active stars have similar levels of activity variability.
This means that activity levels alone are not enough to infer the activity variability of a star.
We also explain the shape of the VP gap observed in the distribution of CE using the CE variability-level diagram.
In the CE variability-level diagram, the Sun is located in the high-variability region of the inactive MS stars zone.
A method to extract the probability density function of the CE variability for a given \rhk level is discussed, and a python code to retrieve it is provided.}
{}{}


\maketitle

\section{Introduction}
Characterisation of stellar magnetic activity is important for our understanding of stellar physics and evolution in general, but also for identification of the radial velocity (RV) signals used to detect and characterise planetary systems.
Magnetic activity in low-mass stars is generated by the coupling between the convective envelope \citep{kippenhahn2012} and stellar differential rotation, which produces a dynamo mechanism that generates a magnetic field \citep[see, e.g. the review by][]{brun2017}.
Classical models of rotation suggest that, as a star evolves, it loses angular momentum via magnetized stellar winds \citep[e.g.][]{schatzman1962, weber1967, skumanich1972} which results in loss of rotation rate.
This loss of rotation rate is accompanied by a decrease in stellar activity \citep[e.g.][]{skumanich1972}.
But recently, \citet{vanSanders2016} used the \textit{Kepler} mission to study a subset of 21 stars with high-precision asteroseismology measurements that are near or beyond the age of the Sun and found shorter rotation periods than those predicted by gyrochronology.
The authors speculate that angular momentum loss from magnetic braking depends sensitively on the global magnetic field configuration, suggesting a possible change in the dynamo mechanism as stars evolve while in the main sequence (MS).
Therefore, activity, rotation, and age are interconnected properties of a star and manifestations of stellar evolution.
Using the relationships between these parameters, stellar activity can be used to determine the rotation periods of stars \citep[e.g.][]{noyes1984, mamajek2008} and estimate their ages \citep{barnes2003, mamajek2008, barnes2010, angus2015, lorenzooliveira2016, lorenzooliveira2018}.

Stellar activity is known to induce signals in the observed RV of stars \citep[e.g.][]{saar1997, santos2000}.
These signals (also known as "jitter") can overlap with the centre of mass Keplerian variations produced by orbiting companions, affecting the detection of low-mass or distant planets, but can also mimic these planet signals, giving rise to false-positive detections \citep[e.g.][]{queloz2001, figueira2010, santos2014, faria2020}.
With the advent of sub-millisecond precision RV instruments such as ESPRESSO \citep{pepe2020} and EXPRES \citep{jurgenson2016}, the detection of RV signals with semi-amplitudes as small as a few tenths of a centimetre per second (cm/s) has become achievable.
However, as the semi-amplitudes of RV signals induced by stellar activity can be higher than 1 m/s \footnote{For example, activity cycles of FGK stars can induce RV semi-amplitudes of up to $\sim$25 m/s \citep{lovis2011}, and up to $\sim$5 m/s for early M dwarfs \citep{gomesdasilva2012}}, activity has become one of main obstacles (if not the main obstacle) to the detection of Earth-like planets \citep[e.g.][]{meunier2010, haywood2016, milbourne2019, colliercameron2019}.
Stellar activity also interferes with planet detection and characterisation using other methods.
In photometry, star spots can lead to incorrect estimation of planet parameters \citep[e.g.][]{pont2007, pont2008, oshagh2013}, and even lead to false-positive detections of non-transiting planets when using the transit-timing variation method \citep[e.g.][]{alonso2009, oshagh2012}.
As another example, microlensing events from planetary bodies can be mimicked by the sudden brightness caused by stellar flares \citep[e.g.][]{bennett2012}.
These have led most of the planet-hunting surveys to include the characterisation of activity as an important component of planet detection.

In the Sun, the emission in the \ion{Ca}{ii} H \& K lines is known to be a proxy of the amount of non-thermal chromospheric heating, which is associated with surface magnetic fields \citep{leighton1959, skumanich1975}.
By extrapolating this to other stars, \citet{vaughan1978} defined a chromospheric emission $S$-index, which is  proportional to the ratio of the measured flux in the \ion{Ca}{ii} H \& K lines to that in two continuum regions adjacent to the red and blue wings of the H and K lines.
However, by this construction the $S$-index is also sensitive to the amount of photospheric radiation passed on through the H and K bandpasses.
Furthermore, the index is affected by the change in continuum flux level observed for stars of different spectral types.
To compare stellar activity between stars with different temperatures (or colours) and to decipher the true activity level of a star it is therefore essential to correct for these effects.
\citet{middelkoop1982} derived a conversion factor dependent on stellar colour to convert the $S$-index to the quantity $R_\text{HK}$, correcting the amount of photospheric radiation in the continuum regions used to calculate $S$.
The photospheric contribution to the H and K lines can be corrected by subtracting the flux in the line wings, which is mainly photospheric in origin \citep[][]{white1981}, leaving a purely chromospheric component, $R'_\text{HK}$, closely related to the surface magnetic flux \citep{noyes1984}.
The chromospheric emission ratio, $R'_\text{HK}$, is nowadays the most used activity proxy of FGK stars .

The most comprehensive study of stellar chromospheric emission is the Mt. Wilson (MW) programme \citep{wilson1968, duncan1991, baliunas1995a}.
This programme followed more than 2 300 stars with multiple observations operating from 1966 through 1995.
Other large-scale surveys of \ion{Ca}{ii} H\&K based on chromospheric activity of FGK stars have been carried out since the MW programme.
These include those of \citet[][825 stars]{henry1996}, \citet[][1 058 stars]{strassmeier2000}, \citet[][$>$400 stars]{santos2000}, \citet[][1 231 stars]{wright2004}, \citet[][664 stars]{gray2003}, \citet[][1 676 stars]{gray2006}, \citet[][225 stars]{jenkins2006}, \citet[][143 stars]{hall2007}, \citet[][353 stars]{jenkins2008}, \citet[][2 620 stars]{isaacson2010}, \citet[][850 stars]{jenkins2011}, \citet[][673 stars]{arriagada2011}, \citet[][311 stars]{lovis2011}, \citet[][13 000 stars]{zhao2013}, \citet[][4 454 stars]{borosaikia2018}, and more recently \citet[][617 stars]{luhn2020}.
The majority of these programmes have just a few observations per star, which is not enough to sample a single magnetic cycle (or even a rotational period), which in turn is important  to be able to deliver an accurate mean level of activity.
In the case of \citet{borosaikia2018}, the authors compiled data from previous catalogues and calculated the $S$-index for 304 stars of the HARPS archive (most of them are included here).

Because of quasi-periodic variations in activity produced by rotational modulation and magnetic cycles, the time-span of some of these catalogues is very short, thus affecting the accuracy of calculations of the average activity levels for the observed stars.
Furthermore, to study activity cycles, timescales comparable to or longer than the activity cycles are needed.
The surveys with the longer time-spans are those of the MW programme ($\sim$40 years), \citet[][$\sim$6 years]{wright2004}, \citet[][$\sim$10 years]{hall2007}, \citet[][$\sim$6-8 years]{isaacson2010} \citet[][$\sim$7 years]{arriagada2011}, \citet[][$\sim$7 years]{lovis2011}, and \citet[][$\sim$20 years]{luhn2020}.

In this work, we present the first paper in a series where we study the chromospheric activity of solar-type stars.
Here we present a catalogue of 1 674 FGK MS, subgiant, and giant stars with more than 180 000 observations from the HARPS archive taken between 2003 and 2019.
The vast majority of the stars are in the solar neighbourhood at less than 100 pc away.
The catalogue has a maximum time-span of $\sim$15.5 years.
We have 428 stars with an observation time-span longer than 10 years, 178 of them with more than 50 individual nights of observation.
These stars with long time-spans (and therefore high-accuracy median activity levels) constitute a prime sample with which to study stellar rotation, cycles, and evolution.
These time series will be reported in the following papers of this series.

Additionally, we also report homogeneous stellar parameters such as spectroscopic effective temperatures, surface gravities, and metallicities, and isochronal masses, radius, and ages.
The purpose of this catalogue is to provide precise and accurate activity and stellar parameters for FGK MS and evolved stars to the community to help advance research in subjects related to stellar evolution and planetary systems.

This paper is organised as follows. In \S\ref{sec:obs} we present the sample and data-cleaning procedure. The retrieval and estimation of observable and stellar parameters is explained in \S\ref{sec:stellar_params}. In \S\ref{sec:chrom_act} we go through the determination and calibration of the chromospheric emission index, in \S\ref{sec:rhk_dist} we analyse its distribution, and in \S\ref{sec:rhk_disp} we explore the activity variability distribution and the relation between variability and activity level. Our concluding remarks follow in \S\ref{sec:conclusions}. Further information about flux and index determination, how to infer activity variability from a star activity level, and how to obtain synthetic populations of stars with activity levels and variability are given in Appendices \S\ref{app:actin} and \S\ref{app:2d_kde}.

\section{Data} \label{sec:obs}
The AMBRE project is a collaboration between the European Southern Observatory (ESO) and the Observatoire de la C\^ote d'Azur (OCA)  established to determine the stellar atmospheric parameters of the archived spectra of three ESO spectrographs including HARPS \citep[for more information see][]{delaverny2013}.
HARPS \citep{mayor2003} is a high-resolution, high-stability, fibre-fed, cross-dispersed echelle spectrograph with a resolution of $\lambda / \Delta \lambda = 115,000$ and a spectral range from 380 nm to 690 nm, mounted at the ESO 3.6-m telescope in La Silla, Chile. For a detailed description of the instrument we refer to \citet{pepe2002}.

We identified 1 977 late-F to early-K stars from the AMBRE-HARPS sample \citep{depascale2014} and downloaded 183 176 reduced 1D spectra and corresponding cross-correlation function (CCF) files from the ESO-HARPS archive\footnote{\url{http://archive.eso.org/wdb/wdb/adp/phase3_spectral/form}}, corresponding to observations that span the period between October 2003 and May 2019.
The 1D spectra are produced by the HARPS Data Reduction Software (DRS) after the interpolation of the 2D echelle spectra (one spectrum per order) over a grid with a constant step in wavelength of 0.01 \AA.
The CCF files provide the RV for each spectrum determined by the DRS.
The spectra are also corrected for the Earth barycentric velocity by the DRS.
Eighteen stars had no corresponding CCF files in the ESO archive and were discarded.
For all remaining spectra, the RV of the star was used to calibrate the wavelength to the stellar rest frame.
The median RV precision for the sample is $\sim$0.7 m/s, which corresponds to a precision in wavelength calibration of $\sim$$10^{-5}$ \AA~at 500 nm, three orders of magnitude smaller than the spectral resolution.

After carefully analysing the spectra, a simple quality selection was made by choosing spectra with less than 1\%  negative flux in the $\ion{Ca}{ii}$ H\&K bandpasses.
Unphysical negative flux may happen in the blue end of the spectrum when the signal-to-noise ratio (S/N) is low and the background light correction may lead to erroneous flux levels.
Only spectra with S/N at spectral order 6 (the order of the \ca lines) higher than 3 were used.
Very often, several spectra were obtained for the same star during one single night.
As we are always interested in detecting periods longer than one day, to increase S/N, mitigate high-frequency noise, and reduce data size we binned the data per night using weighted means with quadratically added errors.
Following the binning, outliers were excluded via a 3-$\sigma$ cut on the $S_\text{CaII}$ index (derivation of the index is explained in \S\ref{sec:caii}).
This cleaning and binning process resulted in a sample of 1 674 stars with a total of 179 291 spectra, corresponding to 44 353 observations binned per night.

\begin{figure}
        \resizebox{\hsize}{!}{\includegraphics{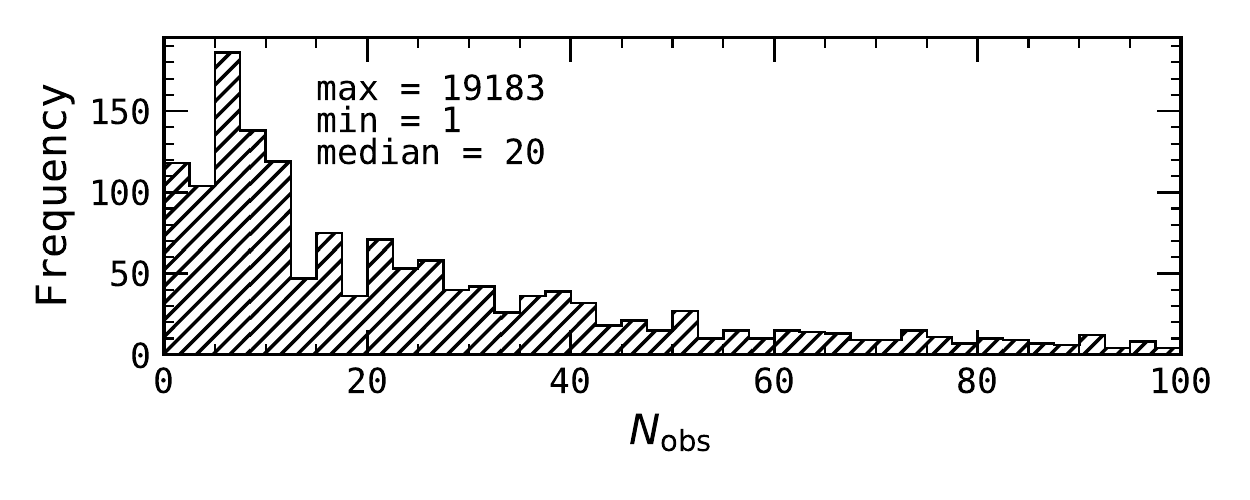}}
        \resizebox{\hsize}{!}{\includegraphics{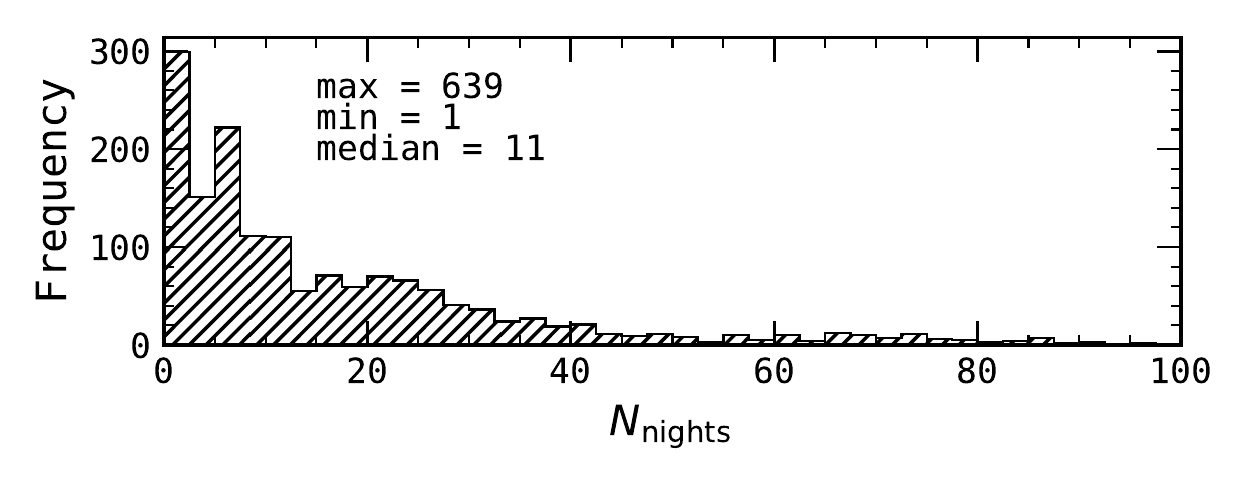}}
    \resizebox{\hsize}{!}{\includegraphics{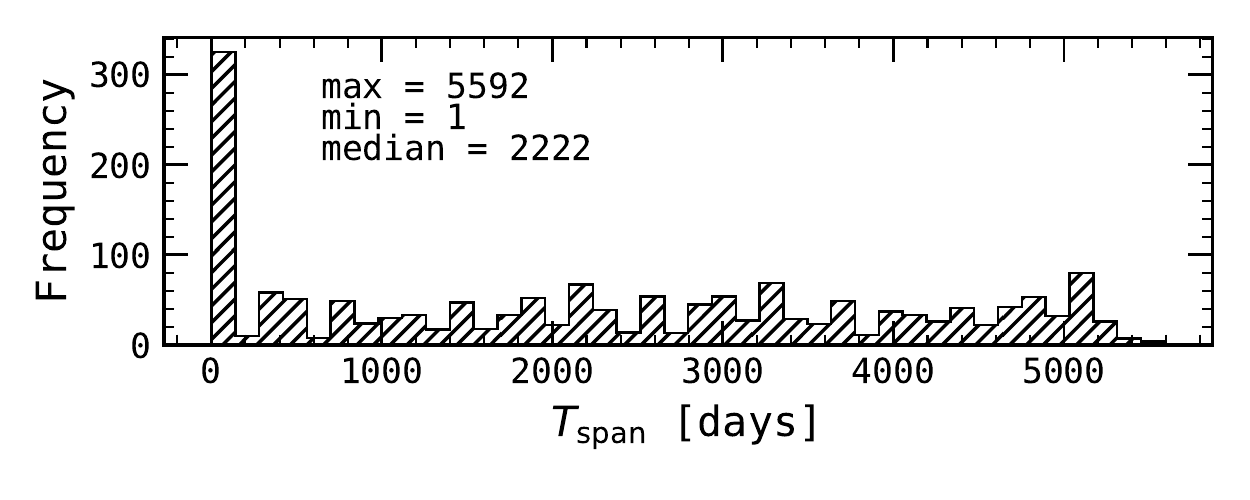}}

    \caption{\textit{Upper panel:} Number of (unbinned) observations per star of our cleaned sample of 1 674 stars. \textit{Middle panel:} Number of nights per star. Both upper and middle plots are truncated at 100 for visualisation purposes.
    \textit{Lower panel:} Time-span of observations for each star.}
        \label{fig:nbins}
\end{figure}

In Fig. \ref{fig:nbins} (upper and middle panels) we present the distributions of the number of observations (unbinned) and number of nights per star.
The minimum number of observations per star is one (44 stars) and the maximum is 19 183 ($\alpha$ Cen B).
Although some stars have a very high number of observations, such as for example $\alpha$ Cen B or $\tau$ Ceti, the vast majority have a lower number, resulting in a median number of observations of 20 per star.
In the middle panel of Fig. \ref{fig:nbins} we can see the effect of binning the data per night and cleaning of outliers.
The maximum number of nights was reduced to 639 with a median of 11 nights per star.
After cleaning and binning there are 172 stars with one night of observation.
Some stars were observed with very high cadence for asteroseismic studies, but we ended up with a low number of nights for these stars because of the binning and low time-span of this type of observations.
The time-span of cleaned, nightly observations (lower panel) varies between 1 day and 5 592 days ($\sim$15 years), with a median time-span of 2 222 days per star ($\sim$6 years).

\section{Stellar parameters} \label{sec:stellar_params}
Because of the high precision of the \textit{Gaia} instruments, we first looked for parallaxes in the \textit{Gaia} catalogues.
Around 98\% of the parallaxes were obtained from \textit{Gaia} DR1 and DR2 \citep{GaiaDR12016, GaiaDR22018}.
For the stars brighter than the brightness magnitude limit of \textit{Gaia}, the parallaxes were obtained from the \textit{Hipparcos} catalogue \citep{hipparcos1997, hipparcos2007} or \verb+Simbad+.
Apparent $V$ magnitudes and $B-V$ colours were obtained from \textit{Hipparcos} and if not available from \verb+Simbad+.
We calculated trigonometric distances from the retrieved parallaxes with a median relative error of 14.3\%.
Apparent $V$ magnitudes and $B-V$ colours were dereddened using \verb+MWDUST+\footnote{\url{https://github.com/jobovy/mwdust}} which uses dust maps of the galaxy to estimate extinction values for different filters.
We used the \verb+combined15+ option, which uses a combination of the \citet{marshall2006}, \citet{green2015}, and \citet{drimmel2003} maps \citep[see][]{bovy2016}.

\begin{figure}
        \resizebox{\hsize}{!}{\includegraphics{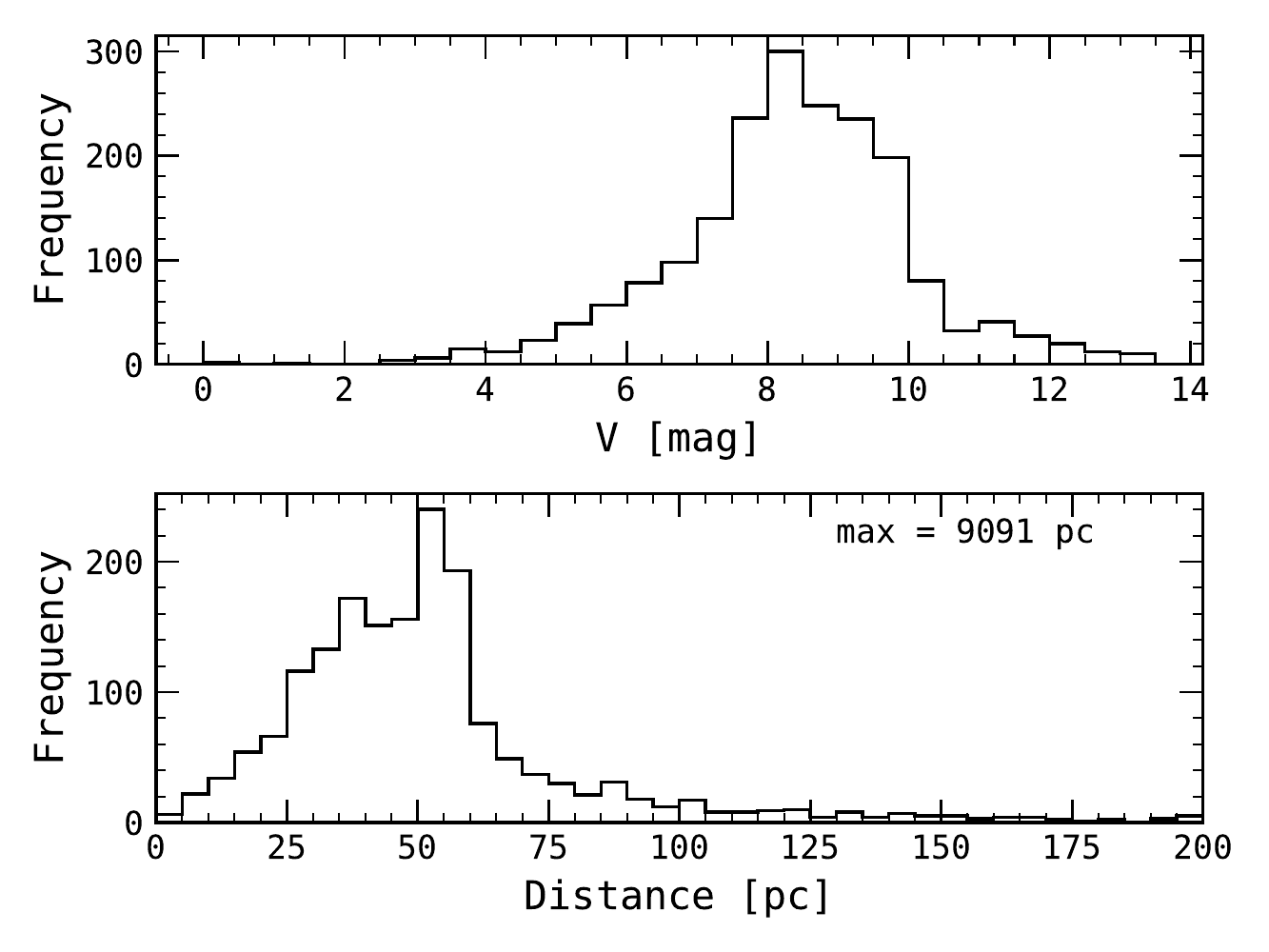}}

        \caption{\textit{Upper panel:} Apparent V magnitude distribution. \textit{Lower panel:} Trigonometric distance distribution, truncated at 200 pc for visualisation purpose.}
        \label{fig:vmag_dist}
\end{figure}

The distributions of apparent $V$ magnitudes and distances are given in Fig. \ref{fig:vmag_dist}.
The majority of the stars in our sample have apparent magnitudes between $\sim$4 and $\sim$12 mag with the peak of the distribution close to $\sim$8 mag (upper panel).
The trigonometric stellar distances have a long range, between 1.26 pc ($\alpha$ Cen AB) and 2041 pc (CD-30\,6015), however the vast majority (91\%) of the stars are at a distance of closer than 100 pc (lower panel).

The stellar parameters of the majority of the stars in this catalogue were determined by our group.
We obtained stellar parameters ($T_{\text{eff}}$, $\log g$, and [Fe/H]) from \cite{sousa2008}, \cite{sousa2011}, and \cite{delgadomena2017} and the \verb+SWEETCat+ catalogue\footnote{\url{https://www.astro.up.pt/resources/sweet-cat/}} \citep{santos2013, andreasen2017, sousa2018} for the stars we have in common.
For the other 564 stars, we co-added up to 100 best spectra (based on high S/N values) for each star and used \verb+ARES++\verb+MOOG+ to derive the spectroscopic parameters \citep[for more details, see][]{sousa2014}.
This analysis is based on the excitation and ionisation balance of iron abundance.
The equivalent widths of the lines were consistently measured
with the \verb+ARES+ code \citep{sousa2007, sousa2015} and the abundances were derived in local thermodynamic equilibrium (LTE) with
the \verb+MOOG+ code \citep{sneden1973}.
For this step we used Kurucz ATLAS9 model atmospheres \citep{kurucz1993} and two line lists were used in this analysis.
For stars with temperatures higher than 5 200 K we use the one presented in \citet{sousa2008}.
For the cooler stars we used the line list presented in \citet{tsantaki2013}.

\begin{figure}
        \resizebox{\hsize}{!}{\includegraphics{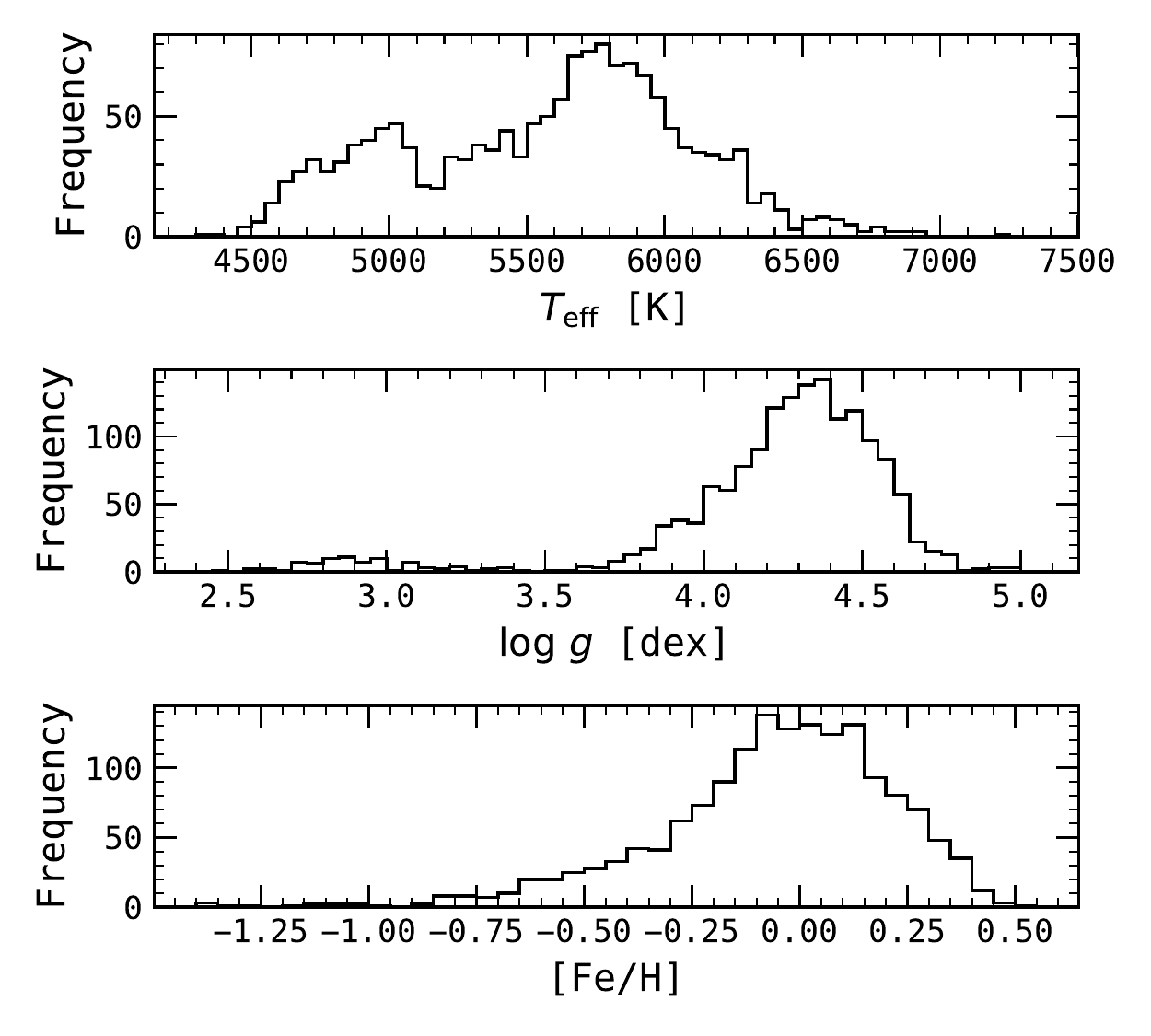}}
        \caption{Distribution of spectroscopic stellar parameters $T_\text{eff}$, $\log g$, and [Fe/H].}
        \label{fig:params_hist}
\end{figure}

The spectroscopic $\log g$ is the least constrained parameter derived with \verb!ARES+MOOG!.
The reason for this is related with the fewer \ion{Fe}{ii} lines available in the optical spectrum that are required to find the ionisation equilibrium in order to constrain this parameter.
Although $\log g$ is not very well constrained, the positive side of the \verb!ARES+MOOG! methodology is that the temperature and metallicity derived remain very precise and are basically independent of the derived surface gravity.
In order to try to improve and get more accurate values of the surface gravity, \citet{mortier2014}  derived empirical corrections using the comparison of spectroscopic $\log g$ derived with these very same tools with values derived either by asteroseismology or transits.
Both corrections are very similar and can be applied to the spectroscopic $\log g$.
Our spectroscopic surface gravity values were corrected using \cite[][eq. 4]{mortier2014}:
\begin{eqnarray}
\log g_\text{corr} = -3.89 \cdot 10^{-4} T_\text{eff} + \log g + 2.10,
\end{eqnarray}
and the $\log g$ errors were determined via error propagation.
Figure \ref{fig:params_hist} shows the distribution of $T_\text{eff}$, $\log g$, and [Fe/H].
Our stars lie in the ranges $4,348 \leq T_\text{eff} \leq 7,212$ K, $2.40 \leq \log g \leq 5.06$ dex, and $ -1.39 \leq \text{[Fe/H]} \leq 0.55$ dex, with median errors of 32 K for $T_\text{eff}$, 0.06 dex for $\log g,$ and 0.02 dex for [Fe/H].
We estimated spectral types for our stars using effective temperatures via \citet{pecaut2013}.
There are 325 F, 779 G, and 485 K stars in the sample.

\begin{figure}
        \resizebox{\hsize}{!}{\includegraphics{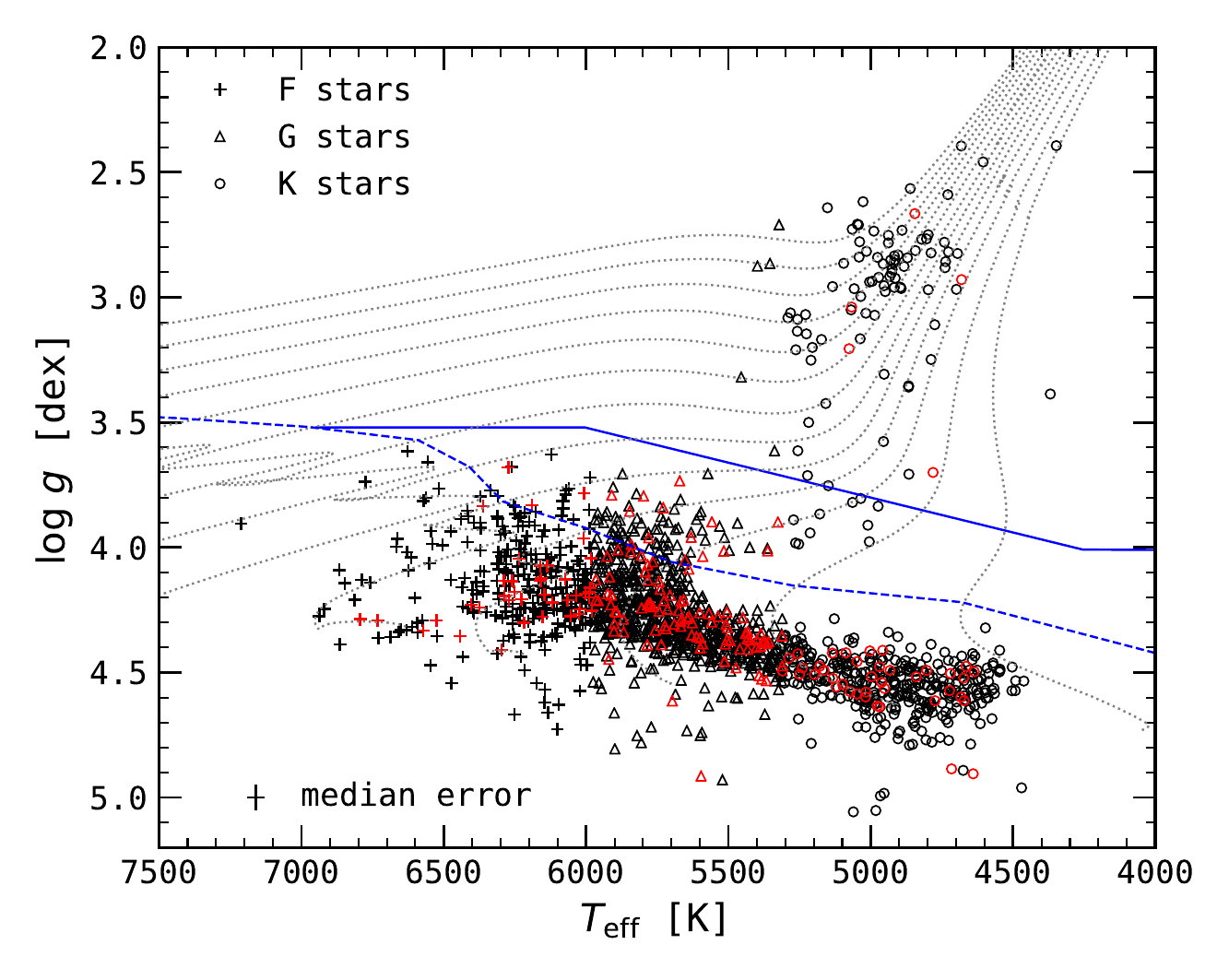}}
        \caption{HR-diagram of stars with spectroscopic surface gravity and effective temperature. The dotted grey lines are the evolutionary tracks obtained with MIST showing the evolution of stellar masses between 0.4 and 3.2 M$_\odot$ in steps of 0.2 M$_\odot$ and with [Fe/H] = 0.0. The blue dashed line marks the TAMS location, while the blue solid line is the separation between subgiant and giant stars (see text). F, G, and K stars are marked with plus symbols, triangles, and circles, respectively. The 194 stars with published planetary companions are marked in red.}
        \label{fig:sample_sel}
\end{figure}

To estimate luminosity classes (required in the following section to calculate the $\log R'_\text{HK}$ activity index for MS and evolved stars) we interpolated the terminal age main sequence (TAMS) locations of the evolutionary tracks obtained with \verb+MIST+ \citep{dotter2016, choi2016, Paxton2011, Paxton2013, Paxton2015} to separate MS stars from subgiants. To separate subgiants from giants we applied the empirical selection cut in $\log g$ as used by \cite{ciardi2011} on \textit{Kepler} stars.
Figure \ref{fig:sample_sel} shows the HR-diagram with luminosity class selections. The \verb+MIST+ evolutionary tracks are shown as grey dotted lines with the TAMS cut marked as a dashed blue line.
The blue solid line shows the \cite{ciardi2011} separation between giants and subgiants.
In red are the 194 planet host stars obtained by cross-correlation with the NASA Exoplanet Archive\footnote{\url{http://exoplanetarchive.ipac.caltech.edu}} \citep{akeson2013} as on December 12, 2019.
The simple selection presented above will probably misclassify some stars in the boundaries between luminosity classes because evolutionary tracks and MS turn-off locations are dependent on metallicity values.
However, for the purposes of this work, a rough luminosity classification is likely not critical.
This selection resulted in the classification of 1 389 MS, 109 subgiant, and 91 giant stars.

\begin{figure}
        \resizebox{\hsize}{!}{\includegraphics{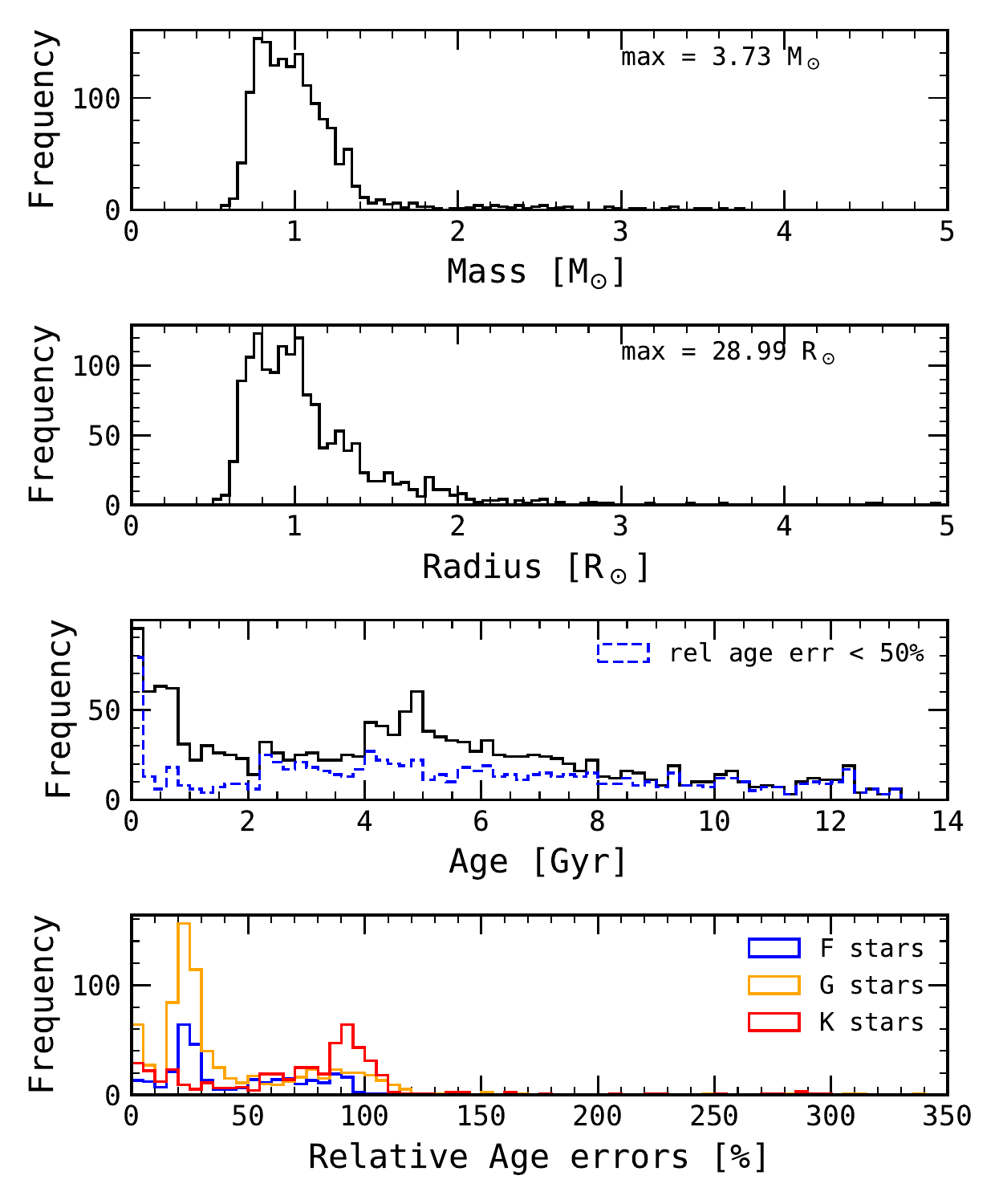}}

        \caption{From top to bottom: Distribution of isochrone masses, radii, and ages. The radius histogram is truncated at $R_* = 5$ R$_{\odot}$ for visualisation purposes. In the age histogram, the solid line represents the full data and the dashed line represents data selected with relative errors below 50\%. The lower panel shows the relative age error distribution for stars segregated into F (blue), G (orange), and K (red) spectral types.}
        \label{fig:PARAM}
\end{figure}

We used apparent V magnitudes, parallaxes (mainly from \textit{Gaia}), $T_\text{eff}$ , and [Fe/H] to calculate stellar masses, radii, and ages using the theoretical \verb+PARSEC+ isochrones \citep{bressan2012} and a Bayesian estimation method described in \cite{dasilva2006} via the 1.3 version of the \verb+PARAM+ web interface\footnote{\url{http://stev.oapd.inaf.it/cgi-bin/param_1.3}}.
We obtained data from \verb+PARAM+ for 1 574 stars, with masses in the range $0.57 \leq M_*/M_\odot \leq 3.7$ with median errors of 0.02 $M_\odot$, radii in the range $0.5 \leq R_*/R_\odot \leq 29.0$ with median errors of 0.02 $R_\odot$, and ages in the range $0.01 \leq \text{Age} \leq 13.2$ Gyr with median errors of 1.4 Gyr.
The median relative age uncertainty is 35\%.
Figure \ref{fig:PARAM} (upper three panels) shows the distribution of stellar masses, radii, and ages, respectively.
The highest peak in the age distribution around 5 Gyr is due to the uncertainty on the age determination, as it disappears when we plot the distribution with a selection of relative age errors below 50\% (blue dashed histogram).
When selecting ages with relative errors below 50\%, the distribution becomes more uniform with a peak for very young stars and a small increase for stars older than 2 Gyr.
The lower panel illustrates the distribution of the relative errors on age, segregated by spectral type.
This histogram shows that the peak in the age distribution around 5 Gyr is mainly composed of K stars, which typically have high uncertainties close to 100\%.

We should note that we expect the stars in the HARPS archive to not be representative of the distribution underlying the galactic stellar population because the majority of the observations are based on planet search programmes that tend to discard active stars, and the majority of these stars are in the solar neighbourhood.
Furthermore, here we only consider F, G, and early K spectral types, with the majority of the stars lying in the MS phase.
Planet search programmes tend to bias data in the following ways:
\begin{itemize}
        \item Normally there is a preference for stars with low activity and/or low rotation rates. These types of stars are  generally considered to have lower activity variability interfering with their RVs. Very often, a $\log R'_\text{HK} < -4.75$ cut is used and/or a cut in $v\sin i$ is done. These cuts will bias towards inactive and slow-rotating stars \citep[e.g.][]{mayor2011}.
        \item Even when active stars are selected, after some observations and constatation of RV and activity index variability, they are generally discarded, meaning that active stars in this sample will probably have a lower number of points and/or time-span when compared to inactive stars.
        \item Most planet search programmes also tend to focus on MS solar-like stars, and therefore the majority of the stars in the sample are of spectral type G and luminosity class V.
        \item There are also limits on the apparent magnitude. These cuts are justified as they allow spectra with higher S/N to be obtained, which leads to lower photon noise in RV measurements. This will bias the sample towards bright and/or nearby stars.
\end{itemize}
Therefore, this sample contains mostly nearby, MS, inactive, and (supposedly) slow-rotating stars, and can be regarded as a typical FGK planet search programme sample but with some other scientific progamme stars mixed in.
These biases have to be taken into account for any statistical analysis of the sample.

\section{The chromospheric activity catalogue} \label{sec:chrom_act}
As a proxy of stellar activity we use the chromospheric emission ratio $R'_\text{HK}$ index \citep{noyes1984}; this index is a calibration of the $S_\text{MW}$ \citep{duncan1991} (\S \ref{sec:rhk}) to correct the bolometric and photospheric effects to enable the comparison of activity between stars with different temperature.
The $S_\text{MW}$ index is obtained by calibrating the $S_\text{CaII}$ (known as the $S$-index) to the Mt. Wilson scale (\S \ref{sec:smw_cal}), which is the scale historically used when comparing $S$-indices between different instruments and methodologies.
In this section we explain how we calculated and calibrated the $S_\text{CaII}$, $S_\text{MW}$, and $R'_\text{HK}$ indices.

\subsection{Chromospheric emission level measured in the \ion{Ca}{ii} H\&K lines}
\label{sec:caii}
The $S_\text{CaII}$ index was calculated using the open-source package \verb+ACTIN+\footnote{\url{https://github.com/gomesdasilva/ACTIN}} \citep{gomesdasilva2018}.
A description of how the code measures flux and determines activity indices is presented in Appendix \ref{app:actin}.

\begin{figure}
        \resizebox{\hsize}{!}{\includegraphics{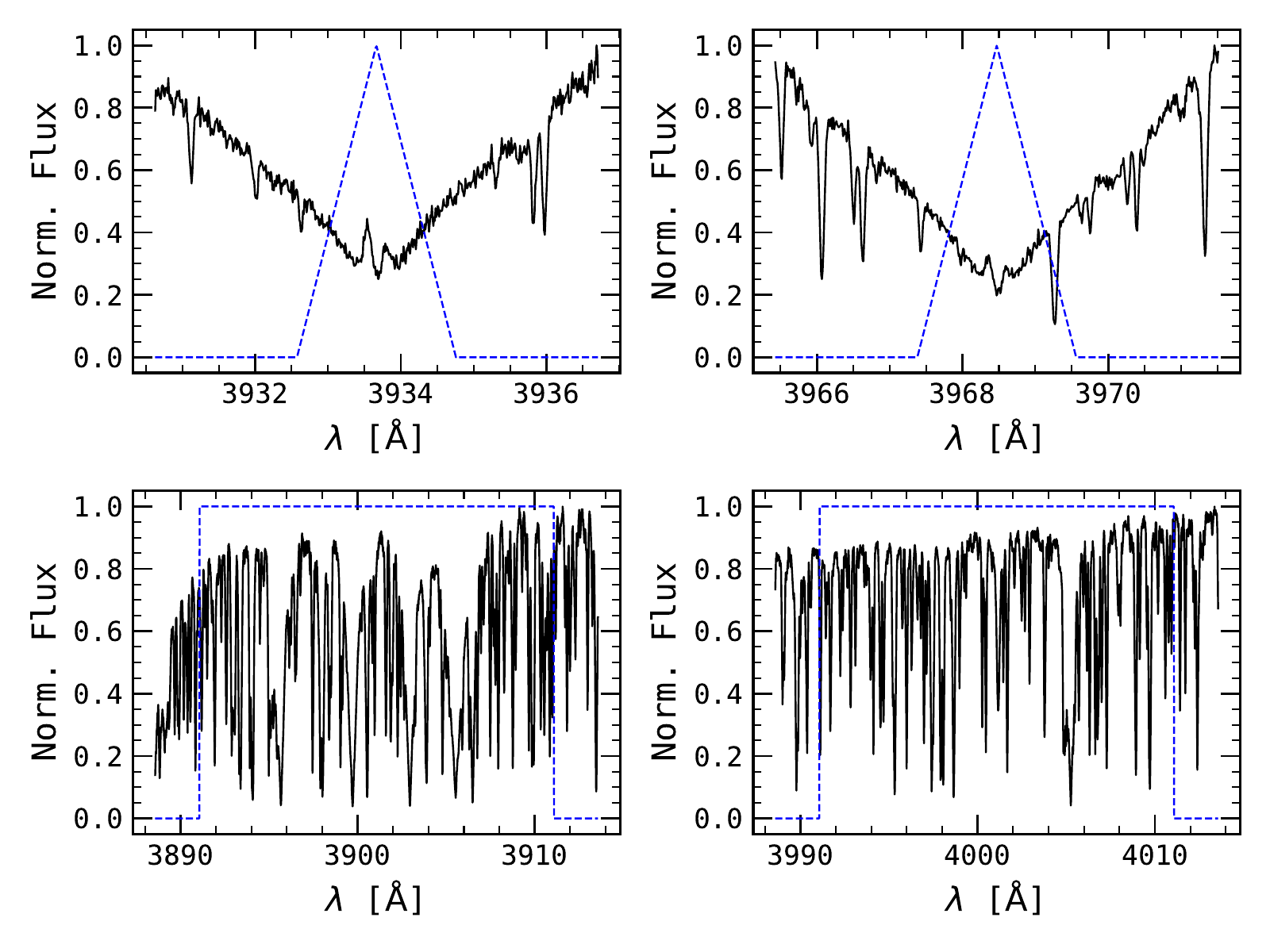}}

        \caption{Example of the bandpasses used to calculate $S_\text{CaII}$ using a spectrum of $\tau$ Ceti. \textit{Upper panels:} \ion{Ca}{ii} K line (left) and \ion{Ca}{ii} H line (right). \textit{Lower panels:} \ion{Ca}{ii} blue reference band (left) and red band (right). Blue dashed lines are the bandpass functions.}
        \label{fig:line_plot}
\end{figure}

This package comes pre-installed with the $S_\text{CaII}$ index, defined as
\begin{eqnarray}
        S_\text{CaII} = \frac{F_\text{H} + F_\text{K}}{R_\text{B} + R_\text{R}},
\end{eqnarray}
where $F_\text{H}$ and $F_\text{K}$ are the mean fluxes in the H and K lines centred at 3968.470 \AA~and 3933.664 \AA, respectively, and $R_\text{B}$ and $R_\text{R}$ are the blue and red pseudo-continuum reference regions centred at 3901.070 \AA~ and 4001.070 \AA, respectively.
The H and K fluxes were integrated using a triangular bandpass with a full width at half maximum (FWHM) of 1.09 \AA\,\,while the fluxes in the blue and red reference regions were integrated with a square bandpass with a 20 \AA\,\,width, similar to the prescription in \citet{duncan1991}.
Figure \ref{fig:line_plot} shows an example of the spectral regions and bandpasses used to calculate $S_\text{CaII}$ using a spectrum of $\tau$ Ceti.

\subsection{Calibration of $S_\text{CaII}$ to the Mt. Wilson scale} \label{sec:smw_cal}
\begin{figure}
        \resizebox{\hsize}{!}{\includegraphics{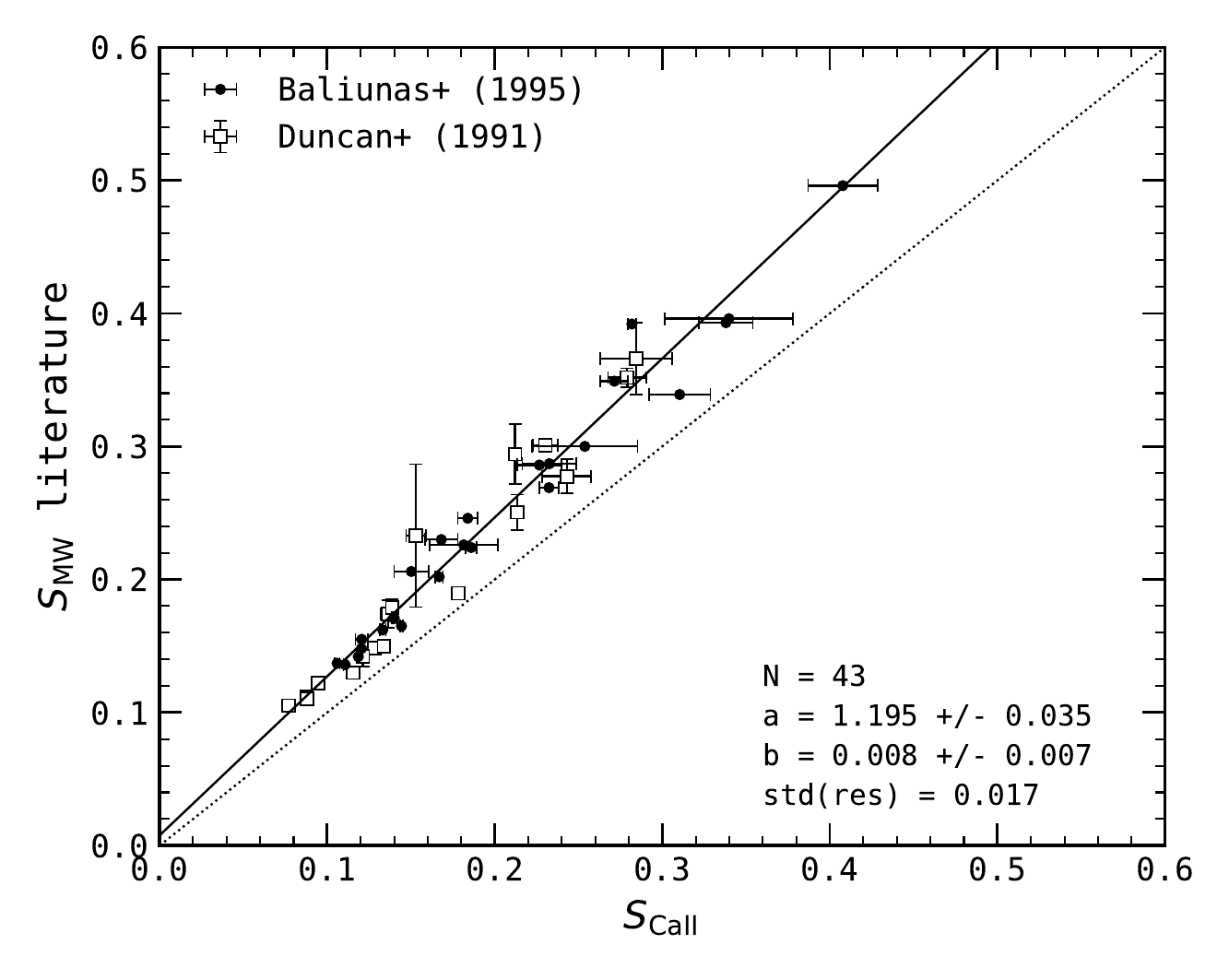}}

        \caption{Calibration of $S_\text{CaII}$ to the Mt. Wilson scale using 43 stars with $S_\text{MW}$ values from \citet{duncan1991} and \citet{baliunas1995a}. The solid black line is the best linear fit with slope ($a$) and intercept ($b$) coefficients shown in the plot with respective errors obtained from the diagonal of the covariance matrix. The residuals standard deviation is 0.017. Errors are the standard deviations of the data. Errors for the indices in \citet{baliunas1995a} were not published. In some cases the errors are smaller than the symbols.}

        \label{fig:smw_cal}
\end{figure}

Before we calculate the usual photospheric-corrected $R'_\text{HK}$ index \citep{noyes1984} we first need to convert our $S_\text{CaII}$ to the Mt. Wilson scale \citep{duncan1991}.
\cite{lovis2011} made a calibration for the HARPS spectrograph using seven stars from \cite{baliunas1995a} in common with their sample.
They used literature stars with $S_\text{MW}$ values between 0.137 and 0.393.
Their calibration is the one currently used by the HARPS DRS pipeline.
However, we have 25 stars in common with \cite{baliunas1995a} and another 18 with \cite{duncan1991}, who reported on the Mt. Wilson $S$-index measurements for the period between 1966 and 1983.
These stars vary in $S_\text{MW}$ between 0.105 and 0.496, and have a lower and upper range that are both slightly higher than those of the values used in the calibration of \cite{lovis2011}.
We therefore decided to derive a new HARPS--Mt. Wilson calibration using the 43 stars in common between the two catalogues.
These stars are presented in Table \ref{tab:calib_smw}.
Figure \ref{fig:smw_cal} shows the $S_\text{MW}$ calibration using main-sequence, subgiant and giant stars.
Black circles are $S_\text{MW}$ values of stars from \cite{baliunas1995a} and white squares are those from \cite{duncan1991}.
The solid black line is the best linear fit and is given by
\begin{eqnarray}\label{eq:smw_cal}
S_\text{MW} = 1.195 \, (\pm 0.035) \cdot S_\text{CaII} + 0.008 \, (\pm 0.007).
\end{eqnarray}
We used the \verb+curve_fit+ function from the \verb+scipy+ package, which uses  a non-linear least squares fit with the Levenberg-Marquardt algorithm.
The dispersion of the residuals around the fit is 0.017.
This dispersion can be a result of the fact that we are measuring activity around two decades after the observations from the literature were made, and some stars have long-term trends which will change the measured median values. It could also be the result of the intrinsic variability of the stars, which, when measured with different cadence or using different time-spans, can result in different median values.
We can observe from Fig. \ref{fig:smw_cal} that the dispersion around the fit increases with increasing $S$-values.
This is due to the internal dispersion of the data:  stars of higher activity tend to have higher levels of variability (see \S \ref{sec:rhk_disp}).

\begin{table}
    \tiny
    \caption{List of stars used to calibrate $S_\text{CaII}$ to the Mt. Wilson scale.
    $N$ is the number of nights of observations, and $T_\text{span}$ the time-span of our sample, $S_\text{CaII}$ our uncalibrated median $S$-index, and $S_\text{MW}$ the literature median $S$-index in the Mt. Wilson scale. The errors are standard deviations. Errors for the indices in \citet{baliunas1995a} were not published.}

\begin{center}
\begin{tabular}{lrrrrrrrrr}
\toprule\toprule
Star   &      $N$ & $T_{\text{span}}$& $S_{\text{CaII}}$ & $\sigma_{S_\text{CaII}}$ & $S_{\text{MW}}$& $\sigma_{S_\text{MW}}$ & Ref. \\
        & [days] & [days]            &      &     &    &   & \\
\midrule
12 Oph &      11 &   2530 &       0.310 &       0.018 &  0.339 &             &   1 \\
14 Cet &      13 &   4360 &       0.186 &       0.003 &  0.224 &             &   1 \\
18 Sco &     193 &   5058 &       0.136 &       0.004 &  0.174 &      0.010 &   2 \\
37 Cet &      16 &   3304 &       0.128 &       0.002 &  0.148 &      0.002 &   2 \\
61 Vir &     227 &   4946 &       0.133 &       0.002 &  0.162 &             &   1 \\
70 Oph A &       6 &      4 &       0.282 &       0.003 &  0.392 &           &   1 \\
89 Leo &      15 &   4286 &       0.167 &       0.002 &  0.202 &             &   1 \\
94 Aqr &      25 &   1718 &       0.121 &       0.004 &  0.155 &             &   1 \\
 9 Cet &      10 &    322 &       0.271 &       0.008 &  0.349 &             &   1 \\
 $\alpha$ CMi &      11 &    309 &       0.140 &       0.000 &  0.171 &      &   1 \\
 $\beta$ Aql &      21 &   1017 &       0.111 &       0.001 &  0.136 &       &   1 \\
 $\delta$ Eri &     152 &   4877 &       0.106 &       0.002 &  0.137 &      &   1 \\
 $\epsilon$ Eri  &      31 &   4304 &       0.408 &       0.021 &  0.496 &   &   1 \\
 $\eta$ Ser &       6 &      4 &       0.094 &       0.001 &  0.122 &      0.003 &  2 \\
 $\iota$ Cap &       8 &   1507 &       0.230 &       0.008 &  0.301 &      0.004 & 2 \\
 $\iota$ Psc &      28 &   2533 &       0.134 &       0.002 &  0.150 &      0.001 & 2 \\
 $\xi^2$ Sgr &       2 &      1 &       0.088 &       0.000 &  0.112 &      0.002 & 2 \\
 $o^2$ Eri &     100 &   4530 &       0.150 &       0.010 &  0.206 &         &   1 \\
 $o$ Aql &      11 &    432 &       0.120 &       0.001 &  0.148 &           &   1 \\
 $\phi^4$ Cet &       9 &   3116 &       0.088 &       0.003 &  0.110 &  0.003 & 2 \\
 $\psi^1$ Aqr &       1 &      0 &       0.077 &       0.000 &  0.105 &  0.002 &   2 \\
 $\sigma$ Boo &       7 &    771 &       0.178 &       0.002 &  0.190 &  0.002 &   2 \\
 $\sigma$ Peg &      28 &   1782 &       0.119 &       0.001 &  0.142 &       &   1 \\
 $\tau$ Cet &     595 &   5146 &       0.140 &       0.001 &  0.171 &        &   1 \\
HD 126053 &       2 &      1 &       0.144 &       0.001 &  0.165 &         &   1 \\
HD 152391 &      36 &   1674 &       0.338 &       0.016 &  0.393 &         &   1 \\
HD 160346 &      31 &   1674 &       0.254 &       0.031 &  0.300 &         &   1 \\
HD 16160 &      44 &   1776 &       0.182 &       0.020 &  0.226 &          &   1 \\
HD 170231 &      17 &   3018 &       0.153 &       0.006 &  0.233 &      0.054 &   2 \\
HD 176983 &       1 &      0 &       0.212 &       0.000 &  0.294 &      0.022 &   2 \\
HD 195564 &      37 &   4363 &       0.116 &       0.001 &  0.130 &      0.004 &   2 \\
HD 210277 &      28 &   4148 &       0.121 &       0.002 &  0.142 &      0.007 &   2 \\
HD 25825 &       5 &   3622 &       0.243 &       0.015 &  0.278 &      0.013 &   2 \\
HD 26913 &       5 &    168 &       0.340 &       0.038 &  0.396 &          &   1 \\
HD 26923 &      35 &   4009 &       0.233 &       0.016 &  0.287 &          &   1 \\
HD 27282 &       3 &   3617 &       0.284 &       0.022 &  0.366 &      0.027 &   2 \\
HD 28635 &       4 &   3428 &       0.213 &       0.002 &  0.251 &      0.013 &   2 \\
HD 32147 &      35 &   4700 &       0.227 &       0.013 &  0.286 &          &   1 \\
HD 42807 &      13 &   4686 &       0.279 &       0.011 &  0.352 &      0.007 &   2 \\
HD 4628 &      41 &   4477 &       0.168 &       0.010 &  0.230 &           &   1 \\
HD 73667 &       1 &      0 &       0.139 &       0.000 &  0.179 &      0.007 &   2 \\
HD 76151 &       7 &   3309 &       0.184 &       0.006 &  0.246 &          &   1 \\
V2213 Oph &      10 &    611 &       0.232 &       0.006 &  0.269 &         &   1 \\
\bottomrule
\end{tabular}
\end{center}
\label{tab:calib_smw}
\begin{tablenotes}
    \small
    \item References: (1) \cite{baliunas1995a}, (2) \cite{duncan1991}.
\end{tablenotes}
\end{table}

The $S_\text{MW}$ errors were calculated via propagation of the $S_\text{CaII}$ uncertainties added in quadrature with the minimum detected $S_\text{MW}$ standard deviation, as described in the following section.

\subsection{$S_\text{MW}$ dispersion and errors} \label{sec:smw_disp}
\begin{figure}
\resizebox{\hsize}{!}{\includegraphics{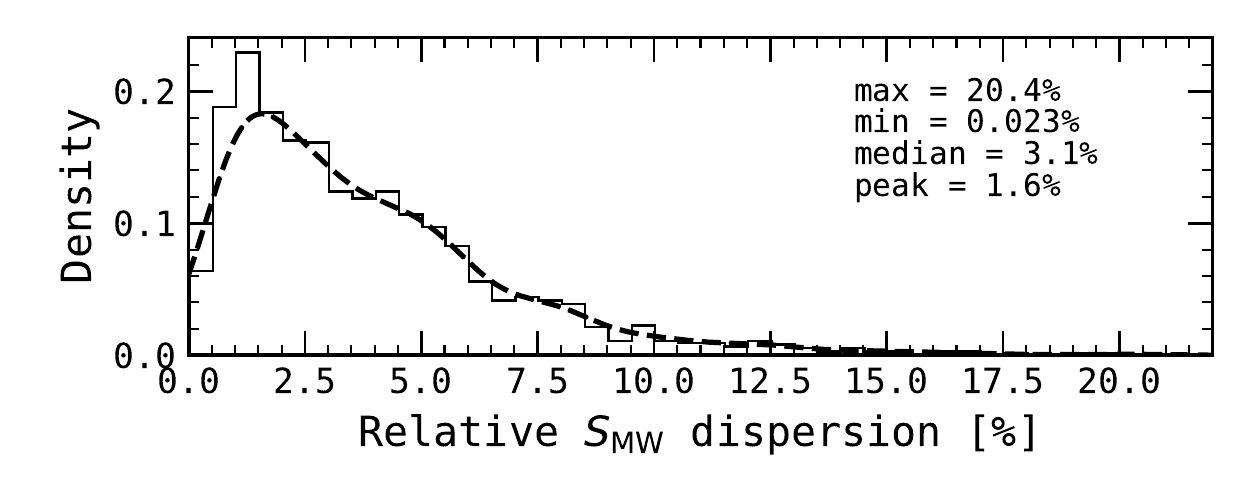}}
\caption{Relative $S_{\text{MW}}$ dispersion of the 1 502 stars in our sample with more than one night of observations. The dashed line is the Gaussian KDE of the distribution, with the peak at 1.6\%.}
\label{fig:smw_disp}
\end{figure}

The precision of the  $S_\text{MW}$ index can be determined by evaluating the smallest $S_\text{MW}$ dispersion detected in the time series.
The dispersion of activity can be due to various phenomena at different timescales, from active regions modulated by stellar rotation with timescales of days to months, to the long-term activity cycles, with timescales of years to decades.
Figure \ref{fig:smw_disp} shows the relative $S_\text{MW}$ dispersion for the 1 502 stars in our sample with more than one night of observations.
The black dashed line is the Gaussian kernel density estimation (KDE) of the distribution. 
The peak of the KDE is at 1.6\%, and the maximum dispersion is $\sim$20\%.
This means that when measuring the activity level of a star, if the timescales of activity variation are not fully covered, the final value can have an uncertainty of up to 20\%.
It is therefore very important to have a long time-span of measurements to reduce this uncertainty to ensure precise activity level estimates are obtained.

\begin{figure}
\resizebox{\hsize}{!}{\includegraphics{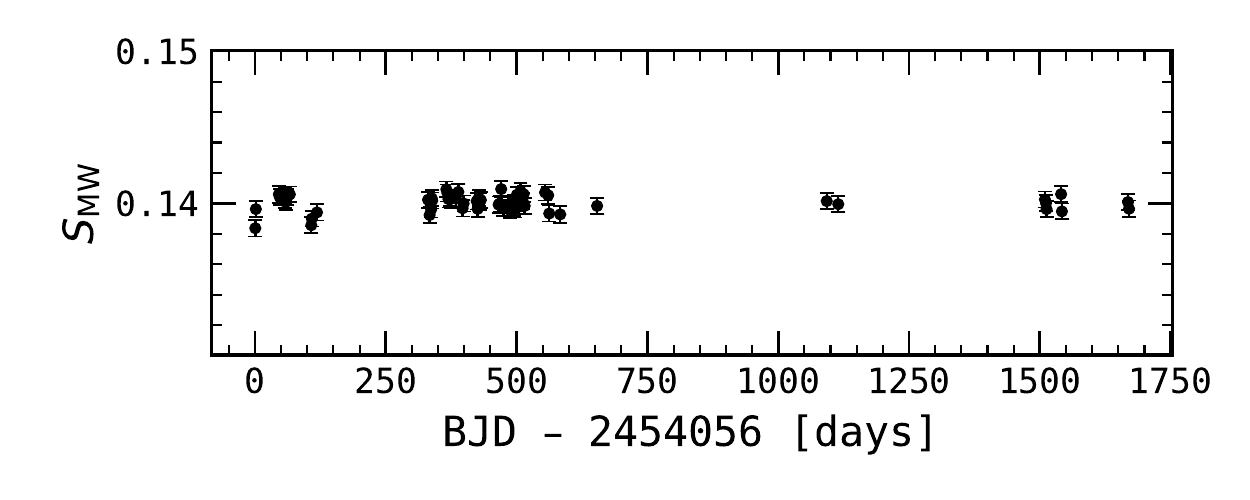}}
\resizebox{\hsize}{!}{\includegraphics{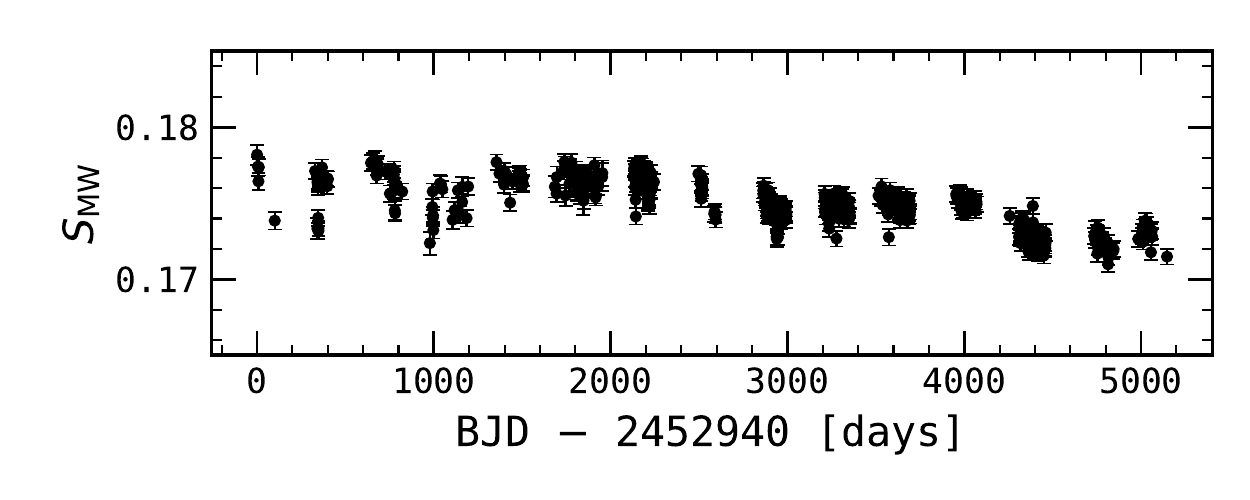}}
\caption{\textit{Upper panel:} $S_{\text{MW}}$ time-series of the quiet F7IV star HD\,60532. \textit{Lower panel:} Time-series of $\tau$ Ceti (K0V). Both plots are set to the same $y$-scale. The relative dispersion for HD\,60532 is 0.36\% and that for $\tau$ Ceti is 0.83\%.}
\label{fig:smw_least_var}
\end{figure}

To estimate the lowest dispersion level, we selected stars with more than 50 nights of observation,  and with time-spans longer than 1 000 days to detect long-term-stable quiet stars; we identified the star with the lowest relative dispersion as measured by the standard deviation in $S_\text{MW}$.
The star with the lowest relative dispersion is the F7IV planet-host star HD\,60532 with a relative dispersion of $\sigma_{S_\text{MW}}/S_\text{MW} = 0.36\%$ and an absolute dispersion of $\sigma_{S_\text{MW}} = 0.0005$.
Figure \ref{fig:smw_least_var} (upper panel) shows the $S_\text{MW}$ time-series of HD\,60532.
This star was observed on 66 nights over more than 4.6 years, and the time-series appears flat during that timescale.

Normally, the standard star used to determine the precision of activity indices (and also RV) is the K0V candidate planet host star $\tau$ Ceti (HD\,10700).
For example, \cite{lovis2011} and \cite{isaacson2010} used this star to estimate the precision of their $S_\text{MW}$ measurements by calculating the dispersion in the time-series.
\cite{lovis2011} detected a relative dispersion of 0.35\% over more than 7 years of observations (157 nights), while \cite{isaacson2010} found a relative dispersion of 1\% over 5 years of data.
\cite{baliunas1995a} also published relative dispersion of $\tau$ Ceti with $\sigma_{S_\text{MW}}/S_\text{MW} = 1.2$\%.
However, the most quiet stars in their sample are HD\,142373 and HD\,216385 ($\sigma$ Peg), both with $\sigma_{S_\text{MW}}/S_\text{MW} = 0.8$\% (we found 0.86\% for $\sigma$ Peg).
We calculated the relative dispersion for $\tau$ Ceti and found a value of 0.83\% ($S_\text{MW}$ = 0.175, $\sigma_{S_\text{MW}}$ = 0.0015) for our time-span of 14 years (595 observing nights).
We note that the data available to \cite{lovis2011} appear constant in time, however our full time-span shows clear variability including a decreasing trend.
In general, our $S_\text{MW}$ values show stability well below the 1\% level.
As a comparison, the relative peak-to-peak variation of the $S_\text{MW}$ of the Sun during its 11 yr activity cycle is $\sim$25\% \citep{baliunas1995b}.

As noted in the beginning of this section, the precision level of $S_\text{MW}$ can be used as an empirical assessment of the measurement errors \citep[e.g.][]{isaacson2010}.
The activity index uncertainties from \verb+ACTIN+ only include photon errors but other sources of noise can be present in the data \citep[e.g.][]{lovis2011}.
To consider unaccounted-for sources of error, we added in quadrature the absolute long-term dispersion of 0.0005 observed for the quiet star HD\,60532 to the photon noise errors obtained from \verb+ACTIN+.

\subsection{The $R'_{HK}$ chromospheric emission ratio} \label{sec:rhk}
Now that we have $S_\text{MW}$ values, we can calculate the photospheric and bolometric corrected $R'_\text{HK}$ chromospheric emission ratio in order to compare the activity of stars with different spectral types.
The $R'_\text{HK}$ index can be calculated from the $S_\text{MW}$ via the \cite{noyes1984} expression:
\begin{eqnarray}
        R'_\text{HK} = R_\text{HK} - R_\text{phot}
,\end{eqnarray}
where
\begin{eqnarray}
        R_\text{HK} = 1.34 \cdot 10^{-4} \, C_\text{cf} \, S_\text{MW}
\end{eqnarray}
is the index corrected for bolometric flux \citep{middelkoop1982}, $C_\text{cf}$ is the bolometric correction, and $R_\text{phot}$ is the photospheric contribution.
The photospheric contribution is a function of B-V colour and has been discussed by \cite{hartmann1984} and derived by \cite{noyes1984} as
\begin{align}
        \log R_\text{phot} = -4.898 + 1.918 \, (\text{B}-\text{V})^2- 2.893 \, (\text{B}-\text{V})^3.
\end{align}
The bolometric correction factor, $C_{cf}$ , was originally derived by \cite{middelkoop1982} for the colour range $0.4\leq(\text{B}-\text{V})\leq~1.2$.
\cite{rutten1984} extended the original correction to include subgiant and giant stars for the colour range $0.3 \leq (\text{B}-\text{V}) \leq 1.7$.
More recently, \cite{suarezmascareno2015, suarezmascareno2016} extended the higher limit of the colour range of the original correction factor to $(\text{B}-\text{V}) = 1.9$, to include more M dwarf stars.
As our sample is composed of FGK dwarfs and evolved stars we decided to use the \cite{rutten1984} bolometric corrections:
\begin{align}
        \log C_{cf} = &\, 0.25 (\text{B}-\text{V})^3 - 1.33 (\text{B}-\text{V})^2 + 0.43 (\text{B}-\text{V}) \,+ \nonumber \\ &+ 0.24
\end{align}
for main-sequence stars with $0.3 \leq (\text{B}-\text{V}) \leq 1.6$, and
\begin{align}
        \log C_{cf} = &-0.066 (\text{B}-\text{V})^3 - 0.25 (\text{B}-\text{V})^2 - 0.49 (\text{B}-\text{V}) \,+ \nonumber \\ & + 0.45
\end{align}
for subgiant and giant stars with $0.3 \leq (\text{B}-\text{V}) \leq 1.7$.

Hereafter, we use the quantity $R_5 = R'_\text{HK} \times 10^5$ to compute variations in chromospheric activity to be able to compare variations for different activity levels on a linear scale.
When analysing median values of activity we use the standard $\log R'_\text{HK}$ because of its widespread use in the literature.

\begin{figure}
    \resizebox{\hsize}{!}{\includegraphics{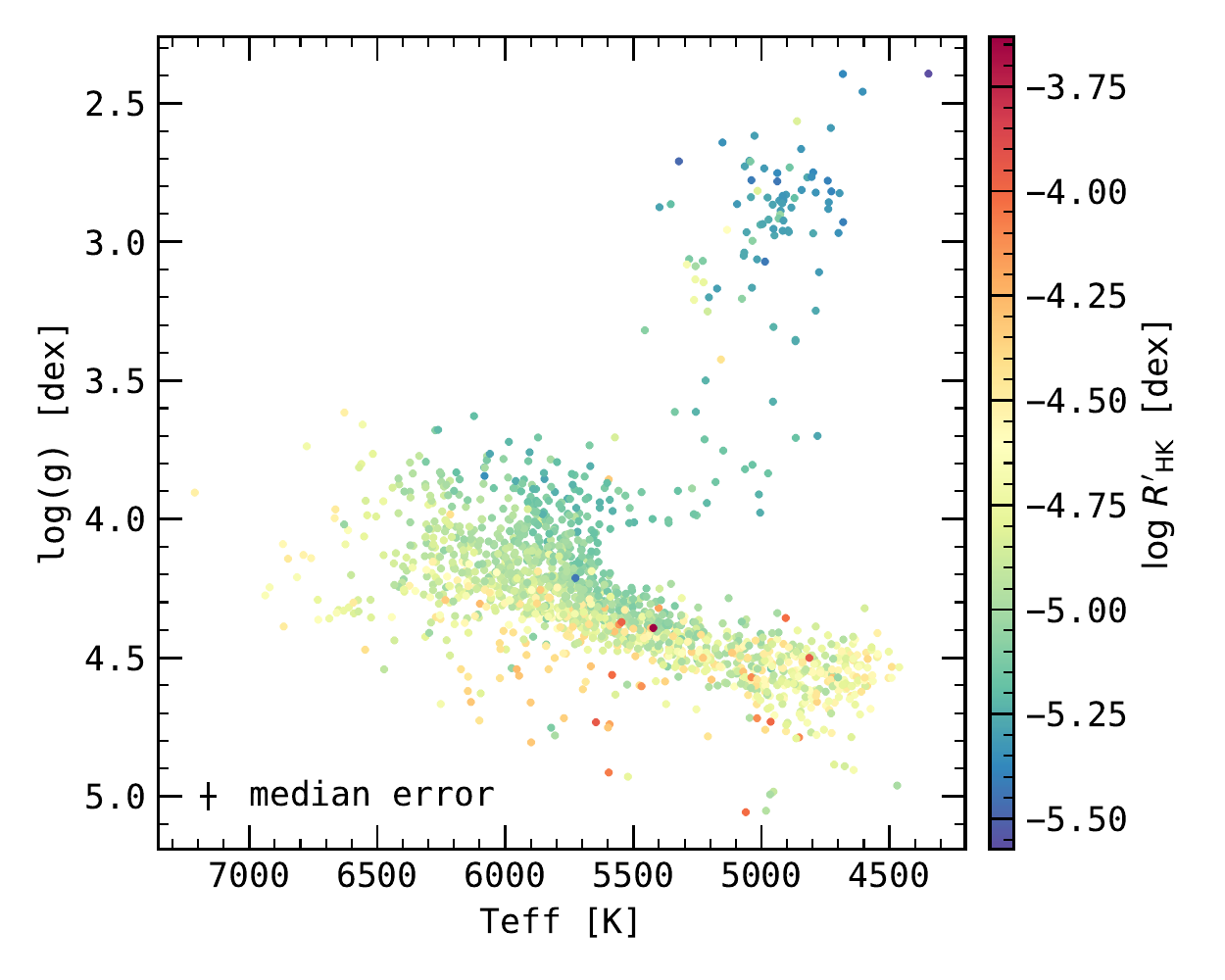}}
    \caption{HR-diagram coloured after \rhk.}
    \label{fig:HR_rhk}
\end{figure}

The CE in the HR-diagram is illustrated in Fig. \ref{fig:HR_rhk}.
Stellar evolution is clearly followed by CE, as MS stars are more active and as stars start to evolve, their CE levels decrease continuously until the giant phase.

\subsection{Chromospheric emission ratio precision}

\begin{figure}
        \resizebox{\hsize}{!}{\includegraphics{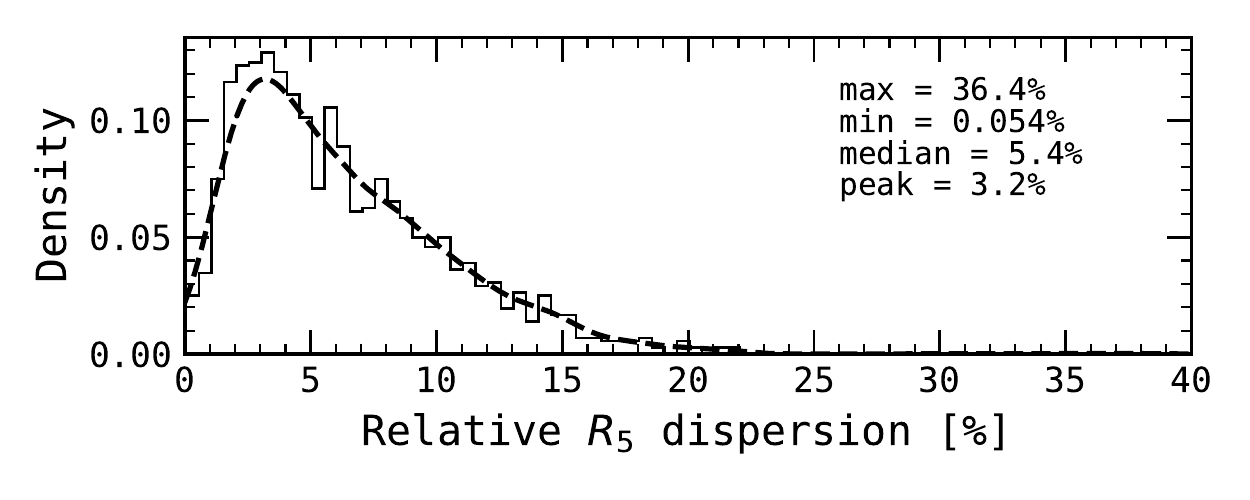}}
        \caption{Relative dispersion of $R_5$ for stars with more than one night of observations. The dashed line is the KDE with a peak at 3.2\%. The histogram is truncated at 25\% for visualisation purposes.}
        \label{fig:rhk_disp}
\end{figure}

\begin{figure}
        \resizebox{\hsize}{!}{\includegraphics{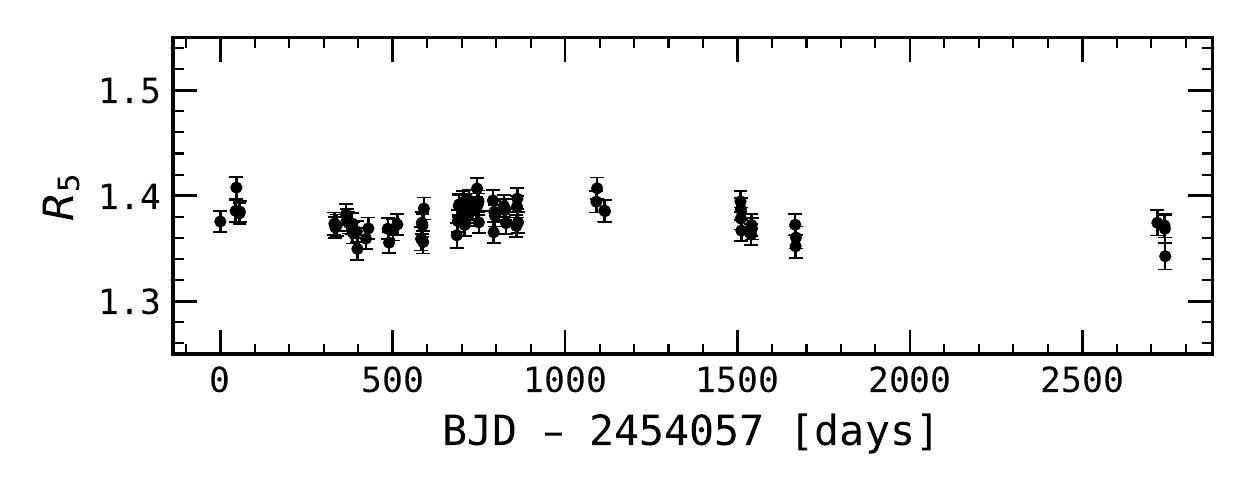}}
        \caption{$R_5$ time-series for 18\,Pup (F5\,V).}
        \label{fig:rhk_least_var}
\end{figure}

Similarly to the case of $S_\text{MW}$ we also checked the precision of our $R'_\text{HK}$ values, which can be estimated by measuring the standard deviation of an inactive quiet star.
Figure \ref{fig:rhk_disp} shows the relative dispersion for stars with $R'_\text{HK}$ values and more than one night of observations.
The dashed line is the KDE with a peak at 3.2\%.
Although the maximum relative dispersion value is 36.4\%, that star is an outlier and the tail of the distribution ends at $\sim$21.6\%.
This means that, when measuring the chromospheric emission ratio level of a star using a very short time-span, activity variations can deviate from the measured mean value by as much as $\sim$22\%.
Accurate activity level measurements therefore require a long time-span of at least the length of the activity cycle if the star has one.
We searched for the least variable star, similarly to in Sect. \ref{sec:smw_disp}.
We selected stars with more than 50 nights of observations and time-spans longer than 1,000 days to find the star with the lowest variability.
We found that 18\,Pup, with 73 data points and a time-span of 7.5 years is the least variable star according to our conditions.
Figure \ref{fig:rhk_least_var} shows the $R_5$ time-series of this star.
The star has a relative dispersion of 0.98\% with a standard deviation of $\sigma_{R_5} = 0.009$, comparable to the values obtained by \cite{lovis2011} for $\tau$ Ceti (0.93\%, $\sigma_{R_5} = 0.0089$).
We also measured the $R_5$ relative dispersion for $\tau$ Ceti and obtained a value of 1.6\% for the relative dispersion and 0.017 for the standard deviation (1.06\%, and $\sigma_{R_5}$ = 0.011 when using \cite{lovis2011} time-span).

Although 18\,Pup has very low variability, we can still observe a long-term low-amplitude variation, probably due to an activity cycle.
The Generalized Lomb-Scargle (GLS) periodogram \citep{zechmeister2009_gls} has its strongest peak above the 0.1\% false-alarm probability (FAP) level at a period of $\sim$1 209 days ($\sim$3.3 years).
This shows that even at very low dispersions below 1\%, we are still able to detect periodic variability in the $R'_\text{HK}$ index.

A description of the catalogue of CE and stellar parameters is provided in Table \ref{tab:full}.
The catalogue is only available in digital format.

\begin{table*}
    \caption{Description of the columns of the catalogue available at CDS.}

\begin{center}
\begin{tabular}{llll}
\toprule\toprule
Keyword   & Data type & Unit      & Description \\
\midrule
\verb+star+        & string    & $-$       & Stellar main identification in \verb+SIMBAD+ \\
\verb+n_obs+      & integer   & $-$       & Number of observations \\
\verb+n_nights+   & integer   & $-$       & Number of nights (binned data) \\
\verb+t_span+     & float     & days      & Time-span of observations \\
\verb+snr_med+    & float     & $-$       & Median signal-to-noise ratio \\
\verb+date_start+ & float     & days      & Barycentric Julian date of start of observations \\
\verb+date_end+   & float     & days      & Barycentric Julian date of end of observations \\
\verb+vmag+           & float     & magnitude & Apparent V-band magnitude (dereddened) \\
\verb+vmag_err+      & float     & magnitude & Apparent V-band magnitude uncertainty \\
\verb+vmag_ref+      & string    & $-$       & Apparent V-band magnitude reference \\
\verb+bv+        & float     & magnitude & B$-$V colour (dereddened) \\
\verb+bv_err+        & float     & magnitude & B$-$V colour uncertainty \\
\verb+bv_ref+      & string    & $-$       & B$-$V reference \\
\verb+plx+         & float     & arcsec    & Parallax \\
\verb+plx_err+    & float     & arcsec    & Parallax uncertainty \\
\verb+plx_ref+    & string    & $-$       & Parallax reference \\
\verb+d+           & float     & parsec    & Geometric distance \\
\verb+n_plt+      & integer   & $-$       & Number of planets in system (as of 2019-12-12) \\
\verb+sptype+      & string    & $-$       & Spectral type \\
\verb+teff+        & float     & K         & Effective temperature \\
\verb+teff_err+    & float     & K         & Effective temperature uncertainty \\
\verb+logg+        & float     & dex       & Logarithm of surface gravity (cgs) \\
\verb+logg_err+    & float     & dex       & Logarithm of surface gravity uncertainty (cgs) \\
\verb+feh+         & float     & dex       & Metallicity \\
\verb+feh_err+     & float     & dex       & Metallicity uncertainty \\
\verb+par_ref+      & string   & $-$       & Reference for spectroscopic stellar parameters $T_\text{eff}$, $\log g$, and [Fe\/H] \\
\verb+mass+        & float     & M$_\odot$ & Isochrone mass \\
\verb+mass_err+   & float     & M$_\odot$ & Isochrone mass uncertainty \\
\verb+radius+      & float     & R$_\odot$ & Isochrone radius \\
\verb+radius_err+ & float     & R$_\odot$ & Isochrone radius uncertainty \\
\verb+age+         & float     & Gyr       & Isochrone age \\
\verb+age_err+     & float     & Gyr       & Isochrone age uncertainty \\
\verb+smw_med+         & float     & $-$       & Median S-index in Mt. Wilson scale \\
\verb+smw_wmean+         & float     & $-$       & Weighted mean S-index in Mt. Wilson scale \\
\verb+smw_err+     & float     & $-$       & Uncertainty of S-index in Mt. Wilson scale  \\
\verb+smw_wstd+     & float     & $-$       & Weighted standard deviation of S-index in Mt. Wilson scale \\
\verb+log_rhk_med+     & float     & dex       & Median $\log R'_\text{HK}$ index \\
\verb+log_rhk_wmean+ & float    & dex          & Weighted mean $\log R'_\text{HK}$ index\\
\verb+log_rhk_err+ & float    & dex          & Uncertainty on $\log R'_\text{HK}$ index\\
\verb+r5_med+    & float     & $-$       & Median $R_5$ index \\
\verb+r5_wmean+    & float     & $-$       & Weighted mean $R_5$ index \\
\verb+r5_err+    & float     & $-$       & $R_5$ uncertainty \\
\verb+r5_wstd+  & float     & $-$       & $R_5$ weighted standard deviation \\
\bottomrule
\end{tabular}
\end{center}
\label{tab:full}
\end{table*}

\subsection{Comparison with the literature}
\begin{figure*}
        \resizebox{\hsize}{!}{\includegraphics{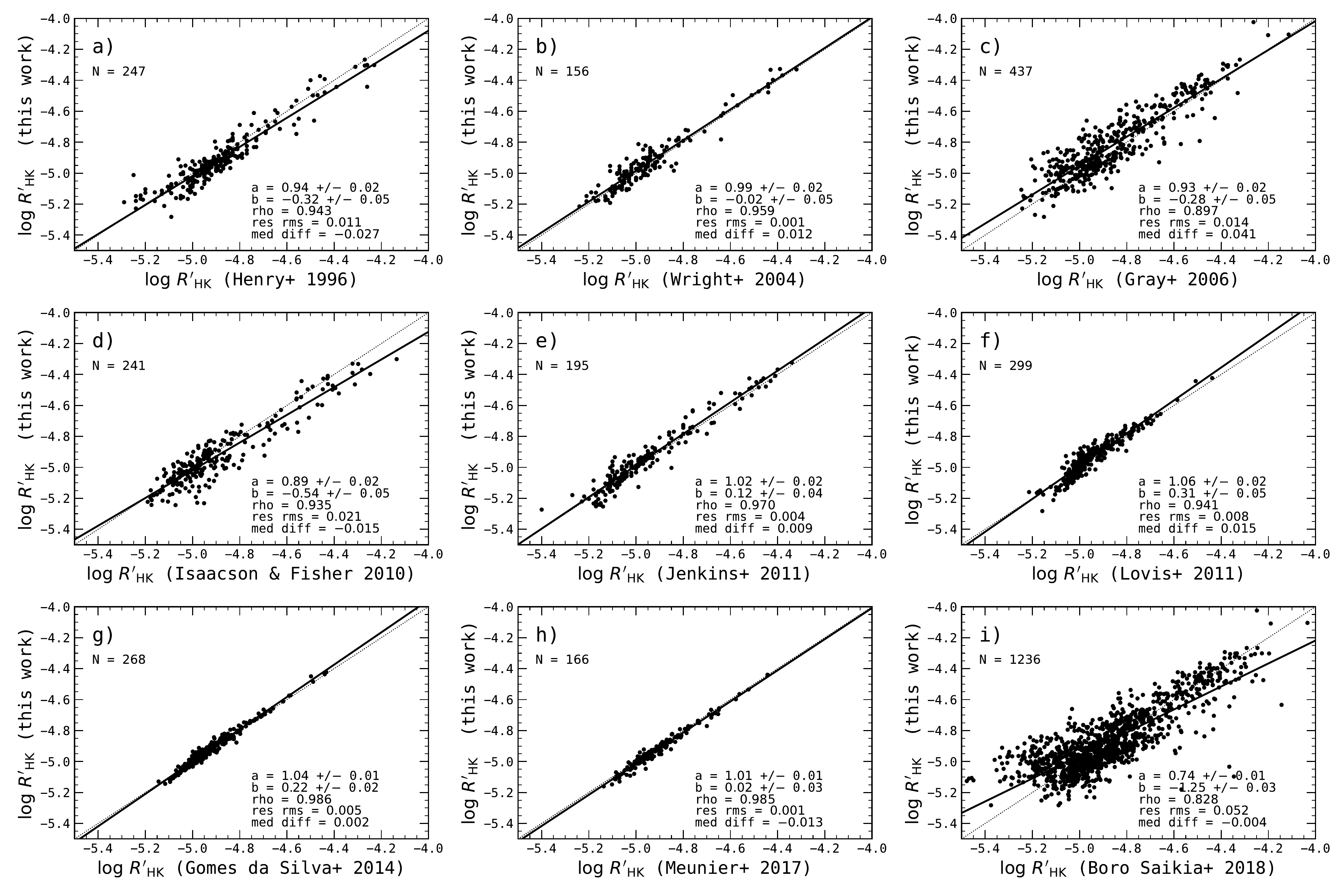}}
        \caption{Comparison of our $\log R'_\text{HK}$ values with those of (a) \citet{henry1996}, (b) \citet{wright2004}, (c) \citet{gray2006}, (d) \citet{isaacson2010}, (e) \citet{jenkins2011}, (f) \citet{lovis2011}, (g) \citet{gomesdasilva2014}, (h) \citet{meunier2017a}, and (i) \citet{borosaikia2018}. The black dotted line is the 1:1 identity, the black solid line the linear best fit, "N" the number of stars in common, "a" and "b" the slope and intercept of the fit, "rho"  the correlation coefficient, "res rms" the rms of the residuals of the fit, and "med diff" is the median difference between the two data sets.}
        \label{fig:rhk_vs_lit}
\end{figure*}

Stellar chromospheric activity surveys based on the \rhk~index have been ongoing for a long time, some of them thanks to the increasingly abundant planet search programmes.
To test whether our values are consistent with the literature we compared our \rhk values with those of \citet{henry1996}, \citet{wright2004}, \citet{gray2006}, \citet{isaacson2010}, \citet{jenkins2011}, \citet{lovis2011}, \citet{gomesdasilva2014}, \citet{meunier2017a}, and \citet{borosaikia2018} (see Fig. \ref{fig:rhk_vs_lit}; the black line is the least squares best linear fit).

Our \rhk values were calculated using the \citet{rutten1984} calibration because of the presence of subgiant and giant stars in our sample.
However, all the comparison data sets shown in Fig. \ref{fig:rhk_vs_lit} use the \citet{middelkoop1982} calibration for all stars, including subdwarfs.
Including subdwarfs in the Middelkoop calibration and comparing it with the Rutten calibration produces a trend towards higher values of activity levels for these types of stars.
This trend can be observed in plots (f), (e), (d), and (c) for \rhk$< -5$ dex.

Furthermore, as the $S$-index to $S_\text{MW}$ calibration generally uses a low number of stars (normally stars in common between the uncalibrated sample and the Mt. Wilson survey), selecting different calibration stars results in different calibration coefficients.

\citet{jenkins2011} studied the effect of varying spectrograph resolution on the derived \rhk values and found that low-resolution spectrographs ($R < 2500$) increase the scatter of the data.
This is the case for the Cassegrain and Boller \& Chivens spectrographs used in Gray et al. ($R \sim 2000$).
The same authors also studied the effect of increasing the bandpass in the H and K lines, and showed that it modified the slope of the correlation between data sets.
Both Henry et al. and Gray et al. use bandpasses of 3.28 and 4.0 \AA~on the H and K lines.

Another difference is that we corrected our $B-V$ colours for reddening (see \S\ref{sec:stellar_params}).
The use of different $B-V$ colours will also affect the \rhk values, as these are a function of $B-V$.
The effect is very small or negligent for the majority of stars, but can contribute to small-scale dispersion when compared to non-derredened \rhk values.
We compared our own \rhk with the dereddened \rhk for the full data set and found an offset of 0.002 dex and a residuals dispersion around the fit of 0.0001 dex.
When we compare the MS stars only, we have an offset of 0.002 dex and the dispersion around the best fit is 0.0003 dex.
Using subgiants only, the offset is 0.0005 dex and the residuals dispersion is 0.002 dex, while using giants only gives an offset of 0.0006 dex and residuals dispersion of 0.005 dex.
Thus, dereddening appears to increase the dispersion for evolved stars and create a small offset for stars in the MS.
Although these effects are similar to the offset and residuals dispersion values we obtain in the comparison plots (b), (e), (f), (g), and (h), they are at least around one order of magnitude lower than the median \rhk errors of the sample, $\sigma$(\rhk)$~= 0.013$ dex and around two orders of magnitude lower than the intrinsic variability of dwarf stars which can be at the $\pm 0.15$ dex level \citep{jenkins2006}.
This shows that although it has a measurable effect on \rhk values, dereddening $B-V$ for our sample does not produce a significant difference.

These effects, together with the effect of sampling activity levels at different phases of rotation modulation and activity cycles when using poor cadence and short time-spans, will increase scatter and affect the slopes and offsets when comparing different data sets.

Our data are, in general, well correlated with the literature, with the correlation coefficient varying between $\rho = 0.83$ (Boro Saikia et al.) and $\rho = 0.99$ (Gomes da Silva et al. and Meunier et al.).
The slope of the best linear fit is also close to one for the majority of the literature, with the exception of Isaacson \& Fisher with a slope of 0.89 and Boro Saikia et al. with 0.74.
The dispersion around the best-fit line varies between 0.001 (Gomes da Silva et al. and Wright et al.) and 0.052 (Boro Saikia et al.).
There are three data sets that use some of the same spectra we use here: Lovis et al., Gomes da Silva et al., and Meunier et al..
These sets are among those showing the best linear correlation, but Jenkins et al. and Wright at al. are also very well correlated with our values, showing low residual dispersion and slopes close to one.
The data sets with the highest scatter are those of Gray et al. and Boro Saikia et al;
the former calculate \rhk using close to one measurement per star, which is not enough to sample rotational and activity cycle variations, whereas
the latter use a combination of compiled data from various catalogues, some of them included here.
As is apparent from the plots in the figure, this will introduce scatter from a variety of sources, some of them explained above.

\section{The distribution of CE in solar-type stars} \label{sec:rhk_dist}
The distribution of chromospheric emission for cool stars has long been known to follow a bimodal distribution with a peak for inactive stars at around $\log R'_\text{HK} = -5$ dex, a peak for active stars at around $-4.5$ dex, and an intermediate region with a dearth of stars at around $-4.75$ dex called the Vaughan-Preston (VP) gap \citep[][]{vaughan1980, duncan1991, henry1996, gray2006, jenkins2006}.
This gap, confirmed by \citet{henry1996} to be real, suggests that either there was a non-uniform star formation rate which decreased at the age corresponding to the intermediate activity levels of the gap, or, as CE decreases with stellar age \citep{skumanich1972}, that the evolution of activity with age passes through different phases, and the phase corresponding to the intermediate activity levels is faster than the active and inactive phases.
\citet{henry1996} also suggested two new additional CE groups: the very active (VA) stars with $\log R'_\text{HK} > -4.2$ dex and the very inactive (VI) stars with $\log R'_\text{HK} < -5.1$ dex.
These latter authors argued that VI stars could be solar-type stars temporarily in the Maunder Minimum phase, while VA stars might be a group of young stars, some of them close binaries.
In their sample of unbiased CE of mostly G dwarfs, these latter authors found active stars to represent ~30\% of the stars while VA stars comprised 2.6\% and VI stars 7.9\%, with the remaining being inactive stars.

In the following sections, we analyse the distribution of CE for our sample, segregating stars by the luminosity classes and spectral types determined previously.

\subsection{Chromospheric emission distribution of main sequence, subgiant, and giant stars}
\begin{figure}
        \resizebox{\hsize}{!}{\includegraphics{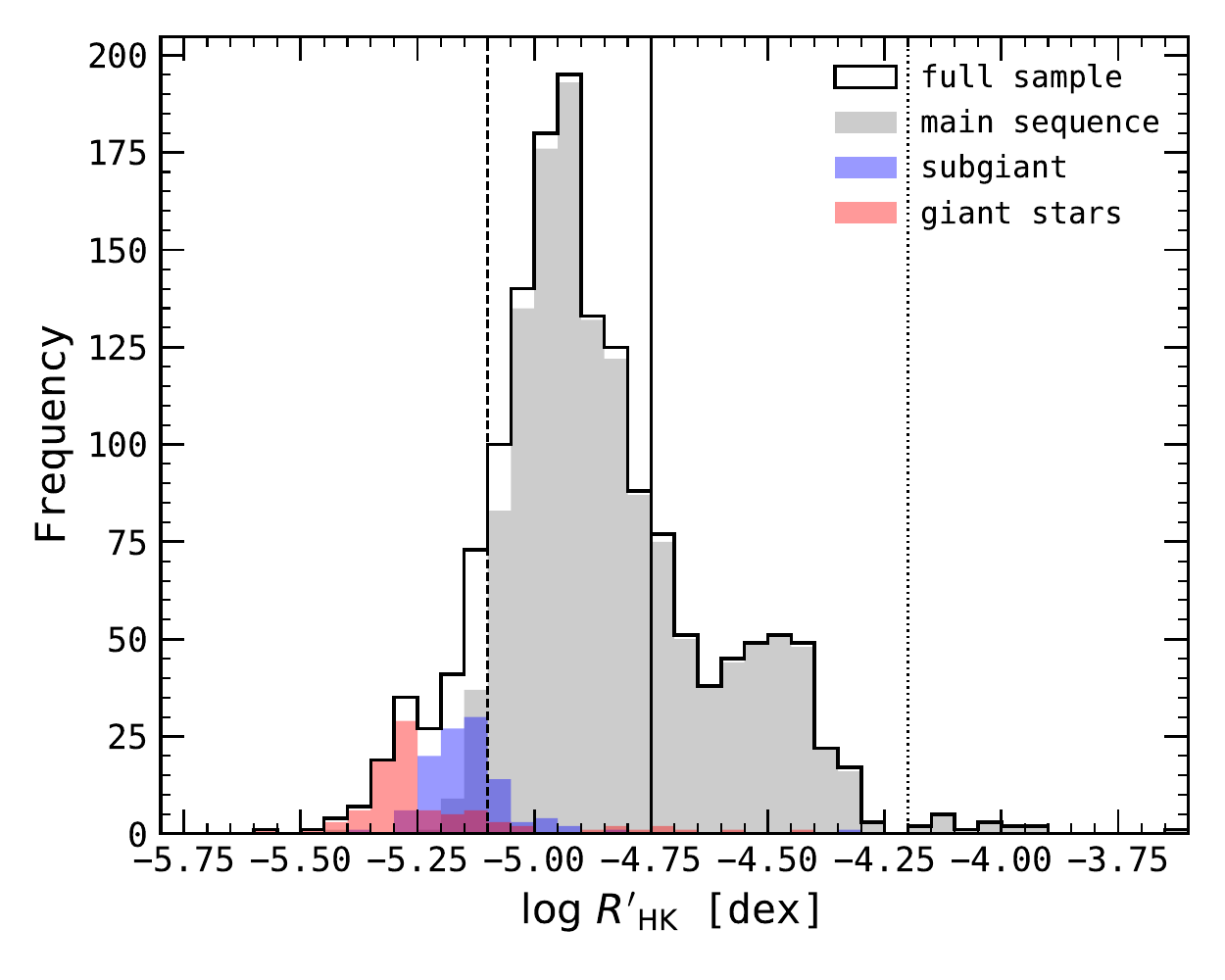}}

        \caption{
                Distribution of median values of $\log R'_\text{HK}$ segregated into MS (grey), subgiant (blue) and giant (red) luminosity classes. The black solid histogram is the full sample. The vertical lines separate different CE groups according to \citet{henry1996}: the dashed line at $-5.1$ dex separates VI from inactive stars, the solid line at $-4.75$ dex separates inactive from active stars, and the dotted line at $-4.2$ dex separates active from VA stars.
        }
        \label{fig:rhk_hist_lum_class}
\end{figure}

Figure \ref{fig:rhk_hist_lum_class} illustrates the distribution of median values of CE for each star as measured by $\log R'_\text{HK}$.
The distribution is segregated into MS stars (grey), subgiants (blue), and giants (red), and extends from very inactive stars with \rhk of around $-5.5$ dex to very active stars with approximately $-3.6$ dex.
At first glance, the distribution appears to be bimodal, with a peak for inactive stars close to $-5$ dex and another for active stars near $-4.5$ dex, as expected.
We mark \citet{henry1996} CE classification in the figure using vertical lines to separate the four groups: a dashed line to separate VI from inactive stars, a solid line to separate inactive from active stars, and a dotted line to separate active from VA stars.
The group of VI stars is clearly dominated by giants and subgiants, while the VA stars group is represented by a small tail of MS stars with $> -4.25$ dex.

In Fig. \ref{fig:rhk_hist_lum_class}, the minima between the active and inactive groups, usually considered to be the separation between inactive and active stars, is located closer to $-4.6$ dex instead of the normally considered location near $-4.75$ dex (the VP gap).
Also, the separation between active and VA stars appears to be at around 
$-4.25$ dex, when the number of stars suffers a more drastic decrease in numbers.

\citet{henry1996} reported that in their unbiased sample of southern nearby MS stars, the proportions of stars in the different CE classes are as follows: VA stars are 2.6\%, active stars are 27.1\%, inactive stars are 62.5\%, and VI stars are 7.9\%.
If considering only active and inactive stars separated at \rhk$ = -4.75$ dex, these authors find 29.7\% active stars.
Our proportions in the MS sample are: VA stars are 1.2\%, active stars are 28.5\%, inactive stars are 66.9\%, and VI stars are 3.5\%.
If we consider only active and inactive stars, we also have 29.7\% active stars, exactly the same proportion as \citet{henry1996}.
We reiterate that our sample includes numerous stars from planet search programmes which normally show a bias for inactive stars.
It is therefore curious that we have very similar proportions of activity levels as an unbiased sample.
This shows that, even with the bias towards more inactive stars, our sample of MS stars closely follows the distribution of an unbiased sample.

\subsubsection{Multi-Gaussian fitting of CE distribution}
Similarly to what was done by \cite{henry1996} and \cite{jenkins2008, jenkins2011} for example, in order to better characterise the activity distribution, we fitted multi-Gaussian models to the $\log R'_\text{HK}$ density distribution using non-linear least-squares with the Trust Region Reflective method, as implemented by \verb!scipy! module \verb!least_squares!.
We used four different models with increasing Gaussian numbers from two up to a total of five (G2, G3, ..., G5).
Although the first two models were easily fitted to the data, we found the higher order models to be very sensitive to the initial conditions, and so we carefully adjusted the initial values to reduce the $\chi^2$-statistic by as much as possible.
We used the Bayesian information criterion \citep[BIC,][]{schwarz1978} statistic for model selection because, while its value decreases for models closer to the data, it also penalises overfitting by penalising models with a higher number of parameters by increasing its value.
By construction, lower BIC values indicate better models.
Table \ref{tab:fit_full} presents the results from the fitting process.
The model with the lowest BIC is the four-Gaussian model (G4) with a value of BIC = 71.7.
The closest model to G4 is G3 with a value of BIC = 79.5.
The difference between the two is $\Delta$BIC $= 7.8$, which according to \citet{kass1995} is considered as `strong' evidence against the higher BIC model.
Thus, the preferred model is G4, whose best-fit parameters are shown in Table \ref{tab:fit_res_full}.

\begin{figure}
        \resizebox{\hsize}{!}{\includegraphics{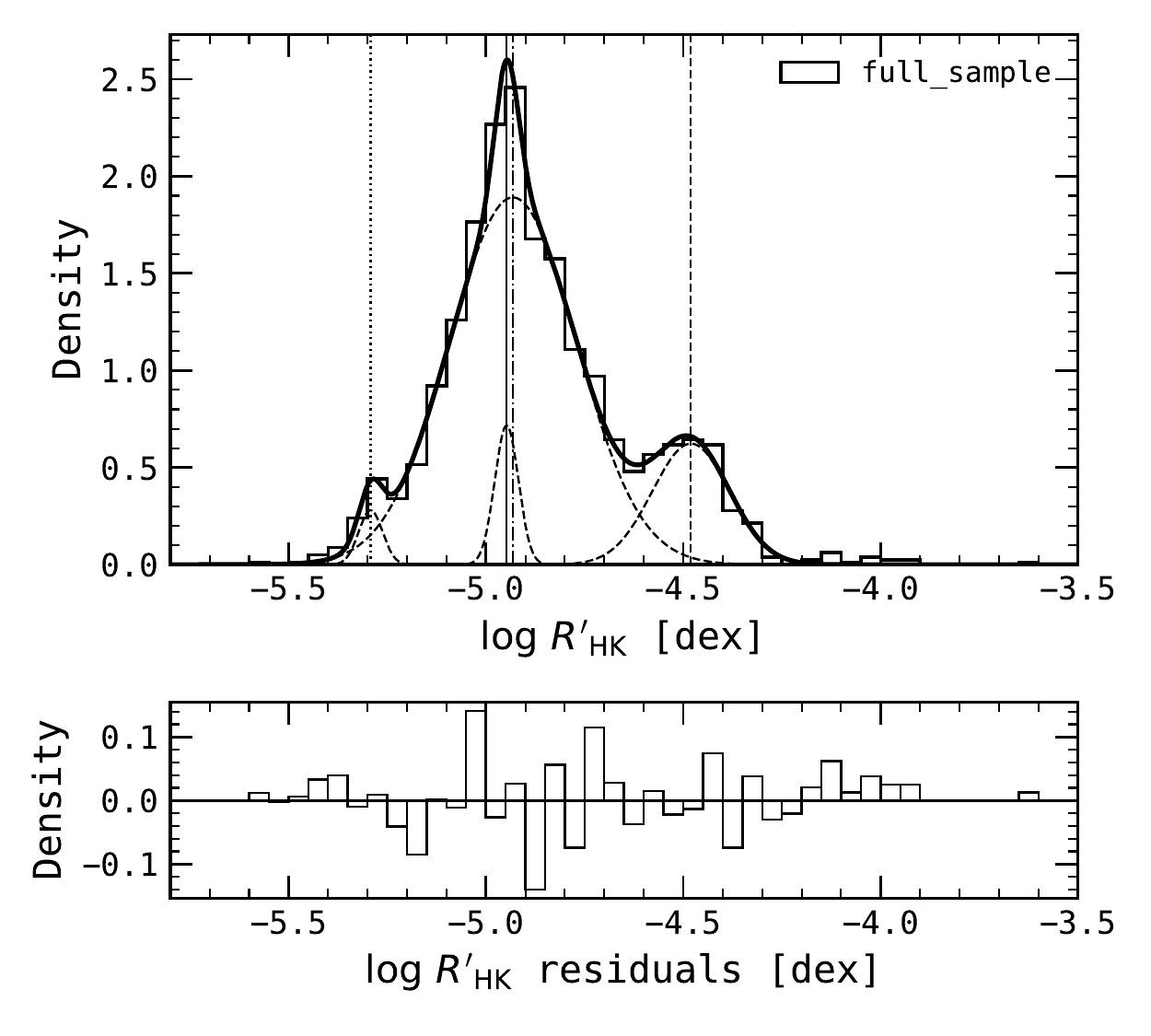}}

        \caption{
                \textit{Upper panel:} Distribution of median values of $\log R'_\text{HK}$ for the full sample including main-sequence, subgiant, and giant stars. The black solid curve is the four-Gaussian fit to the distribution. The dashed black curves are the individual Gaussians. The means of the Gaussians are marked as vertical lines.
                \textit{Lower panel:} Residuals of the fit.
        }
        \label{fig:rhk_hist_fit}
\end{figure}

Figure \ref{fig:rhk_hist_fit} provides the full sample $\log R'_\text{HK}$ distribution with the best-fit model superimposed.
The upper panel is the density distribution fitted by the G4 model curve.
The lower panel shows the residuals of the fit.
The distribution has three peaks, representing three populations at $-5.29$ dex (VI stars, evolved), $-4.95$ dex (inactive stars, MS), and $-4.49$ dex (active, MS).
The highest peak comprises two Gaussians at very close mean values, with the broader Gaussian adjusting the base and the thinner Gaussian adjusting the top of the distribution.
Both peaks for MS stars are in agreement with the values found in the literature \citep[e.g.][]{henry1996, jenkins2006, jenkins2008, jenkins2011}.
The population of  very active stars is not represented (was not fitted) because of the low number of representatives in our sample (16 stars).
There are two local minima in the distribution at $-5.24$ and $-4.62$ dex but no gaps empty of stars separating the different populations of CE levels.
We need to caution that the fitting of the distribution using histograms is highly dependent on the binning used (we used a binning of 0.05 dex).

\begin{table}
        \caption{Parameters for the fits to the $\log R'_\text{HK}$ full sample distribution using different numbers of Gaussians. G2$-$5 refer to the models using 2$-$5 Gaussians. $N_\text{free}$ is the number of free parameters, $\chi^2$ and $\chi_\nu^2$ the chi-squared and reduced chi-squared statistics, respectively, and BIC is the BIC value calculated using BIC $ = \chi^2 + k\log(N)$, where $k$ is the number of fitted parameters and $N$ the total number of bins in the histogram.}
        \begin{center}
                \label{tab:fit_full}
                \begin{tabular}{lrrrrrrrrrrrrrrrrr}
                        \toprule\toprule
                        Parameter       & & & Value \\
                        \midrule        
                        Model                           & G2    & G3    & G4      & G5   \\
                        $N_\text{free}$         & 29    & 26    & 23    & 20   \\
                        $\chi^2$                        & 70.6  & 48.0  & 29.8    & 35.8 \\
                        $\chi_\nu^2$            & 2.61  & 2.00  & 1.42  & 1.99 \\
                        BIC                             & 91.6  & 79.5  & 71.7    & 88.2 \\
                        \bottomrule
                \end{tabular}
        \end{center}
\end{table}

\begin{table}
        \caption{Best-fit four-Gaussian model parameters for the full sample $\log R'_\text{HK}$ distribution. The parameters of the four Gaussians are: $\mu$, the centre values, $A,$ the heights, and $\sigma,$ the standard deviations. Local maxima and local minima are included in the end.}
        \begin{center}
                \label{tab:fit_res_full}
                \begin{tabular}{lrrrrrrrrrrrrrrrrr}
                        \toprule\toprule
                        Parameter       & Value \\
            \midrule
                        $\mu_1$         & $-5.29$       \\
                        $A_1$           & 0.28          \\
            $\sigma_1$  & 0.030         \\
            \midrule
                        $\mu_2$         & $-4.95$       \\
                        $A_2$           & 0.72          \\
                        $\sigma_2$      & 0.030         \\
            \midrule    
                        $\mu_3$         & $-4.93$       \\
                        $A_3$           & 1.89          \\
                        $\sigma_3$      & 0.161         \\
                        \midrule
                        $\mu_4$         & $-4.48$       \\
                        $A_4$           & 0.62          \\
                        $\sigma_4$      & 0.098         \\
                        \midrule
                        Max. 1          & $-5.29$       \\
                        Max. 2          & $-4.95$       \\
                        Max. 3          & $-4.49$       \\
                        \midrule
                        Min. 1          & $-5.24$       \\
                        Min. 2          & $-4.62$       \\
                        \bottomrule
                \end{tabular}
        \end{center}
\end{table}

\subsubsection{Very inactive star population}
\begin{figure}
        \resizebox{\hsize}{!}{\includegraphics{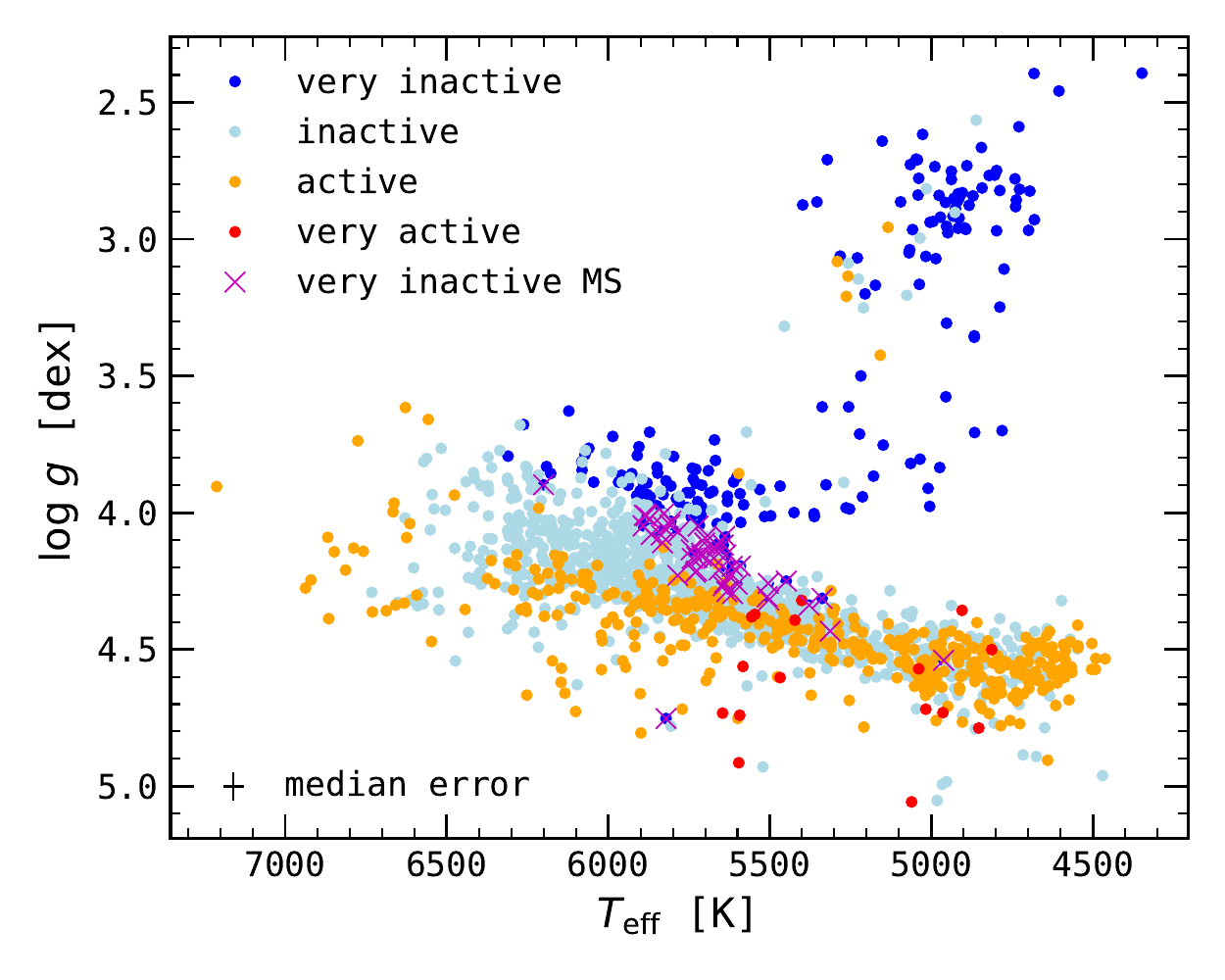}}

        \caption{
        HR-diagram for different CE level classifications following \citet{henry1996}. Dark blue are VI, light blue inactive, orange are active, and red VA stars.
        A selection of MS VI stars is also shown with magenta crosses (see text). The median error bars are illustrated in the lower left corner.
        }
        \label{fig:rhk_hr_diagram}
\end{figure}

As stated above, \citet{henry1996} argued that the VI group could be MS stars in a Maunder Minimum type phase.
We checked our MS stars with activity levels below $-5.1$ dex in the HR-diagram and almost all of them are G stars near the MS/subgiant transition zone (Fig. \ref{fig:rhk_hr_diagram}, marked with `X' in magenta\footnote{The outlier below the MS is a star with large uncertainties in the determination of both $T_\text{eff}$ and $\log g$ ($\sigma_{T_\text{eff}} = 147$ K and $\sigma_{\log g} = 0.24$ dex)}).
We observed their CE time-series and they show signs of variability and trends similar to other MS inactive stars with higher activity levels and therefore these are not CE-quiet stars.
We also compared the average CE dispersion of these MS VI stars with the average values of inactive stars and subgiants.
Main sequence VI stars have average $\sigma_{R_5}$ of 0.041 while inactive stars have 0.067 and the subgiants have 0.041, the same dispersion as the MS VI group.
Following this analysis, we conclude that the MS VI group is likely not made up of MS stars in the Maunder Minimum phase, but stars already starting to evolve beyond the MS \citep[see also][]{wright2004b}.
This type of misclassification between MS and evolved stars is a general issue and not 
specific to our work.
The boundaries between the different luminosity classes are very hard to distinguish, which can lead to misclassifications of stars in these regions.
Similarly to the separation between active and inactive stars, the VI and VA groups might separate other stages of stellar evolution, the post MS evolution and possibly the initial MS evolution.
This is shown in Fig. \ref{fig:rhk_hr_diagram}, where most of the stars evolving beyond the MS and into the giant phase are classified as very inactive (dark blue).
We therefore suggest that \rhk$= -5.1$ dex marks a separation in stellar evolution where stars begin the transition from the MS to the subgiant phase.

\subsubsection{Very active star population}
Although some of the VA stars (Fig. \ref{fig:rhk_hr_diagram}, red) seem to fall below the MS, there are also some of these stars in the MS.
This group includes nine G and seven K dwarfs.
The majority of our K dwarfs have age estimations with relative uncertainties close to 100\%, but in this group even the G dwarfs have very high relative age errors, with insufficient precision to detect small differences between VA and active stars.
The small number of stars in the VA group together with the large uncertainties of our isochronal age estimates for this group prevents us from forming conclusions about the VA evolutionary stage.
Nevertheless, as activity level generally decreases with age, we expect these stars to be younger than the active stars group.

\subsubsection{Chromospherically enhanced giant stars}
The majority of the giant stars, that is, $\sim$72\%, have activity levels below $-5.2$ dex.
These stars are evolved and are expected to have low rotation rates and therefore lower activity levels than their MS or subgiant counterparts.
However, there are some giant stars that show higher activity levels, some of them inside the active stars group with $\log R'_\text{HK} > -4.75$ dex.
\citet{middelkoop1981} suggested that giant stars with enhanced \ca emission are stars with intermediate masses between 1.5 and 4 M$_\odot$ that are rapidly rotating in the MS phase.
However, as they have no significant convective envelope, their activity levels during that phase are insignificant.
The star evolves without losing large quantities of angular momentum until it enters the giant phase and develops a convective envelope.
At this phase, because the stars still have rapid rotation, the activity level becomes very high when compared to lower mass giant stars.
More recently, \citet{luhn2020} also observed that some evolved stars in their sample had higher activity levels than expected for their evolutionary stage.
These latter authors suggested that these are intermediate-mass stars that are still spinning down but evolved quickly beyond the MS before losing their magnetic activity.
We compared the masses of the two groups by selecting giants with $\log(g) < 3.5$ dex and separating very inactive giants (VIGs), that is those with \rhk$< -5.2$ dex, from the chromospheric enhanced giants (CEGs) with \rhk$> -5.2$ dex.
Giant stars with $\log R'_\text{HK} > -5.2$ dex (CEGs) are more massive, with a median mass of 2.53 M$_\odot$, while VIGs have a median mass of 1.88 M$_\odot$.
The Kolmogorov-Smirnov (KS) test for these two groups results in a p-value of $1.4 \times 10^{-4}$ (KS statistic of 0.53), and so we can exclude the null hypothesis that the two samples come from the same continuous parent distribution.
This result confirms that the CEGs are more massive in general than VIGs.
We also calculated the median ages of the two groups.
The CEGs are in general more than two times younger than the VIGs, having median ages of $0.64 \pm 0.04$ Gyr while the VIGs have median ages of $1.6 \pm 0.21$ Gyr.
We also performed a KS test to evaluate whether or not the two groups come from the same parent distribution.
The KS test results in a p-value of $1.6 \times 10^{-4}$ (KS statistic of 0.53) confirming that these groups do not come from the same parent distribution.
These results support the suggestion of \citet{luhn2020}  that, for a given $\log g$, the more active giants are more massive and younger than the VIGs.
A comparison of the rotation periods of these two groups could provide further insight into whether or not they have different spinning rates and whether or not rotation contributes to the difference in activity between these groups.

\subsection{Chromospheric emission distribution of main sequence FGK stars}
\begin{figure}
        \centering
        \resizebox{8.1cm}{!}{\includegraphics{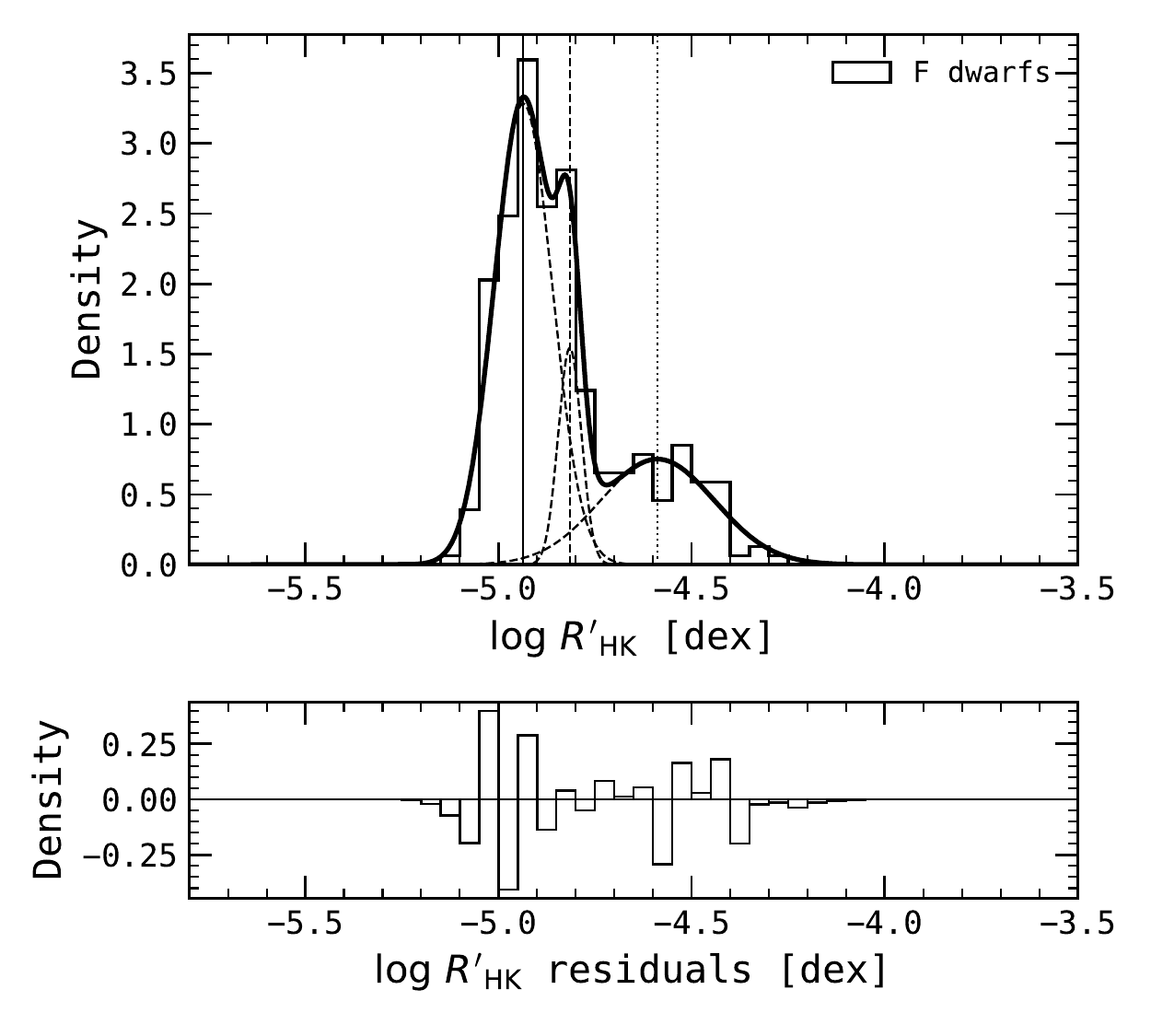}}
        \resizebox{8.1cm}{!}{\includegraphics{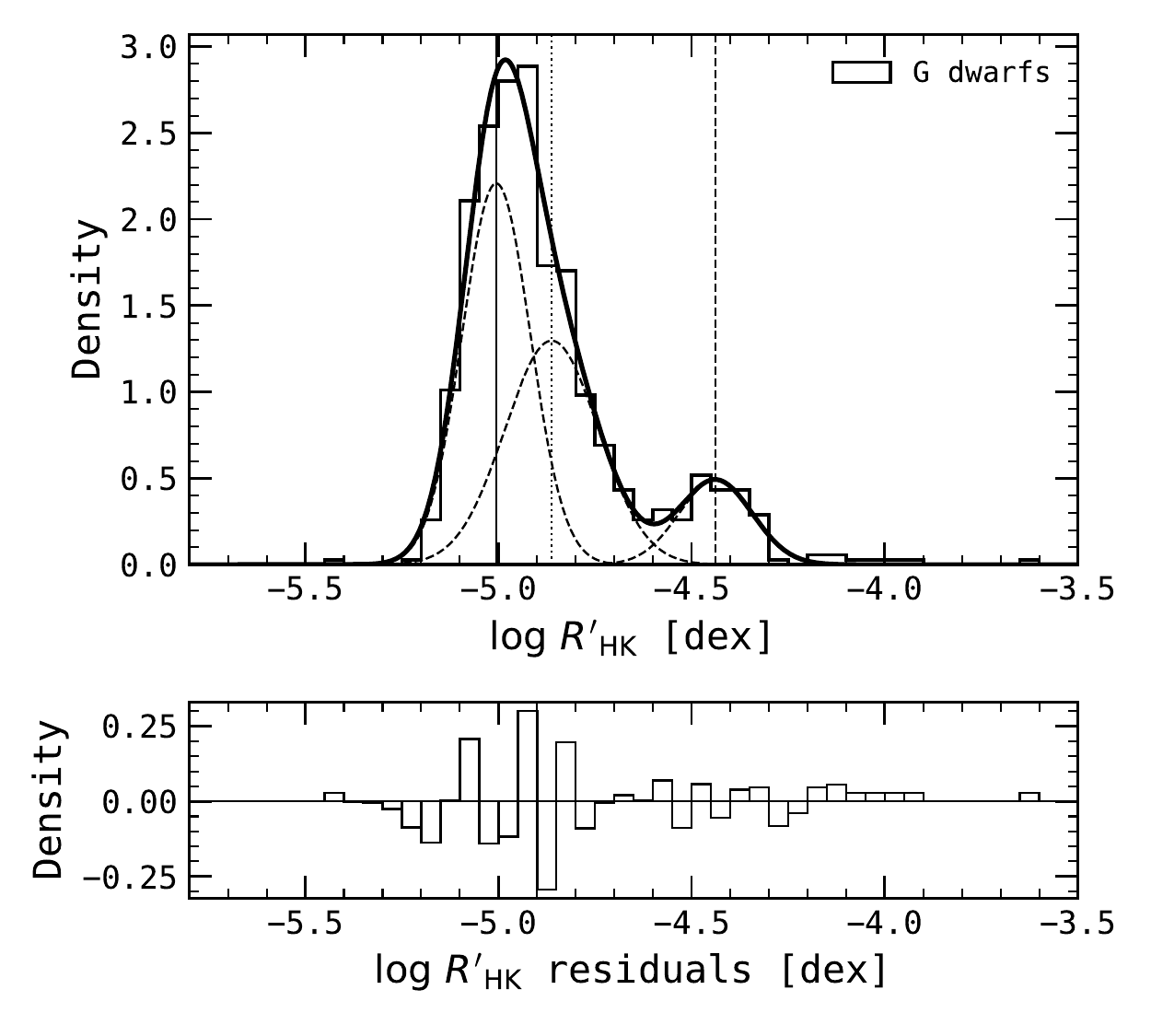}}
        \resizebox{8.1cm}{!}{\includegraphics{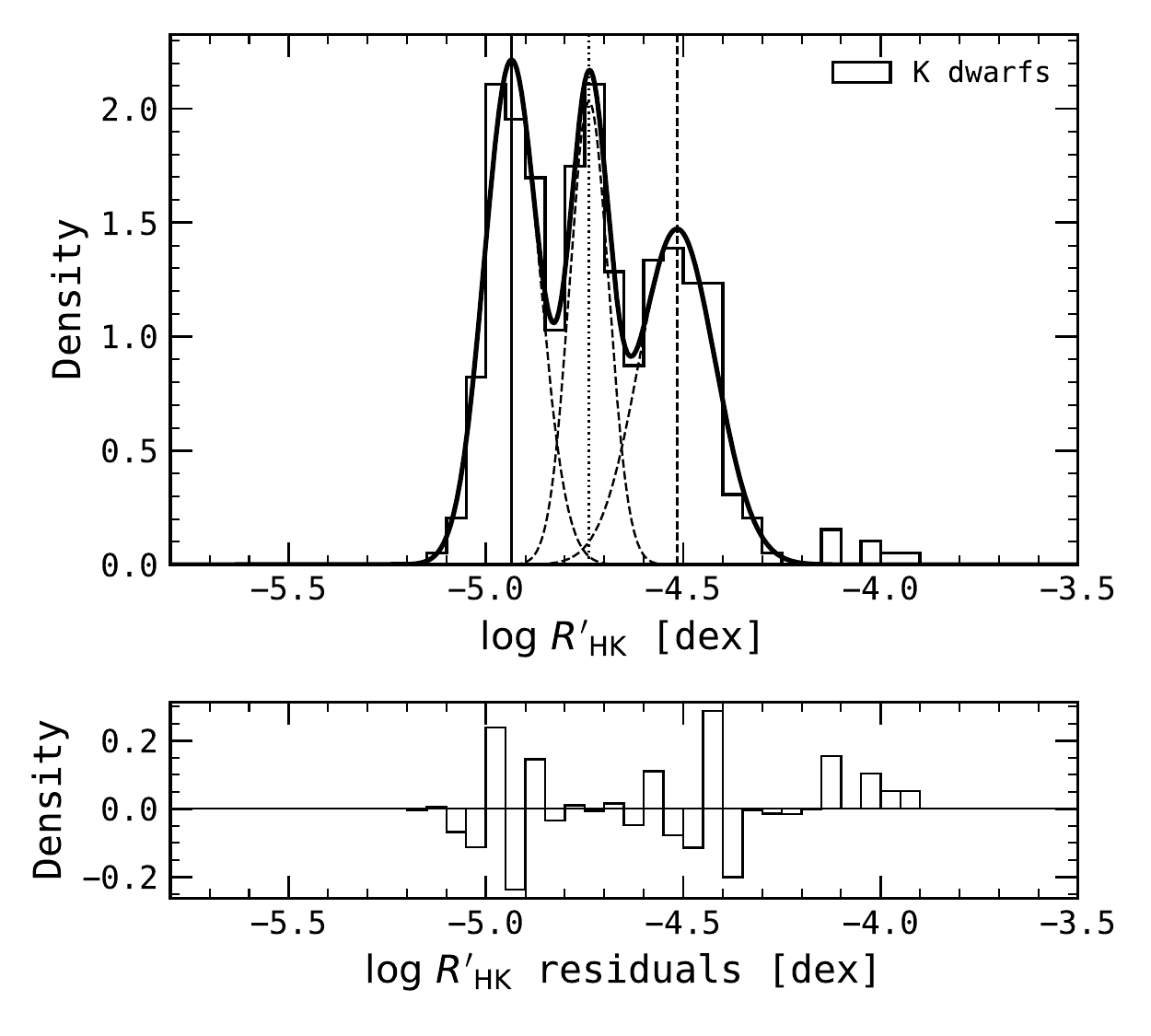}}

        \caption{Distribution of median values of $\log R'_\text{HK}$ for MS stars segregated into F (upper panel), G (middle panel), and K (lower panel) spectral types. The solid curve is the best-fit multi-Gaussian model and the dashed curves represent the individual Gaussians included in the model. The vertical lines mark the centres of the Gaussians. Below each plot are the residuals of the fit.}
        \label{fig:rhk_hist_sptypes}
\end{figure}

The distribution of $\log R'_\text{HK}$ for F, G, and K dwarfs is shown in Fig. \ref{fig:rhk_hist_sptypes}, in the upper, middle, and lower panels, respectively.
Each panel shows the distribution with the best-fit model and the residuals of the fit below.
Similarly to the analysis for the full sample in the previous section, we fitted multi-Gaussian models including from two to five Gaussians.
The fit statistics are shown in Table \ref{tab:fit_FGK}.
We followed the same model-selection procedure as before, choosing the model with the lowest BIC value.
A description of the activity distribution and model selection for the three spectral types is presented below. 

\paragraph{F dwarfs:}
The distribution of CE of F dwarfs (Fig \ref{fig:rhk_hist_sptypes}, upper panel) can be described as double or triple peaked.
The model with the lowest BIC value is G3 with BIC = 45.2.
The $\Delta$BIC value for the next-lowest BIC model is 2.2 for G4.
According to \citet{kass1995}, these values are considered as `positive' evidence against G4, however the evidence is not strong.
Therefore, we prefer the simpler model and consider G3 as the most appropriate.
Although the best-fit model shows three peaks, the middle peak at \rhk$-4.83$ dex could be a binning effect due to low numbers.
The middle Gaussian also contributes to the broadening of the inactive main peak in the direction of higher activity levels.
We should note that we cannot robustly discriminate between the two best models quantitatively due to the close BIC values of the models, and other studies with different number of F dwarfs could give different results.
Model G4 has one more Gaussian separating the active zone into two populations with a minimum at \rhk$= -4.58$ dex with two maxima at each side.
Thus, the three-peak distribution is not very significantly different from a four-peak distribution for F dwarfs.
The proportion of active F dwarfs is 24.2\% if considering the separation between active and inactive at \rhk$= -4.75$ dex.

\paragraph{G dwarfs:}
The distribution of G dwarfs is clearly double peaked, with the inactive peak near $-5$ dex and the active one around $-4.5$ dex (Fig. \ref{fig:rhk_hist_sptypes}; middle panel).
The model with the lowest BIC is G4 with BIC = 64.2, however G3 is very close with BIC = 66.3, resulting in $\Delta$BIC $= 2.1$, which is also considered as "positive evidence against the model with higher BIC" according to \citet{kass1995}.
Both G3 and G4 can then be considered best-fits, but we chose model (G3) as the best-fit model due to its simpler form, despite it having higher BIC.
One of the Gaussians in the fit does not contribute to a local maxima, but instead broadens the inactive peak asymmetrically in the direction of higher activity levels.
The proportion of active G dwarfs is 19.6\%.

\paragraph{K dwarfs:}
Finally, the K dwarfs distribution is clearly triple-peaked, with the inactive peak near $-4.9$ dex, an intermediate activity peak near $-4.75$ dex, and an active peak near $-4.5$ dex.
The best model fit is G3 with BIC $= 41.3$, and the second-best-fit model is G4 with BIC $= 47.7$.
This gives $\Delta$BIC $= 6.4$, meaning that there is `strong' evidence against G4.
The distribution together with the best-fit G3 model is shown in Fig. \ref{fig:rhk_hist_sptypes} (lower panel).
The proportion of active K dwarfs is 51.9\%, more than double the proportion of active stars for F or G dwarfs.\\

\begin{table}
        \caption{Multi-Gaussian fit statistics for the F, G, and K dwarf $\log R'_\text{HK}$ distributions. Parameters are the same as in Table \ref{tab:fit_full}.}
        \begin{center}
                \label{tab:fit_FGK}
                \begin{tabular}{lrrrrrrrrrrrrrrrrr}
                        \toprule\toprule
                        Parameter       & & & Values \\
                        \midrule
                        F dwarfs \\
                        \midrule        
                        Model                           & G2    & G3    & G4      & G5   \\
                        $N_\text{free}$         & 12    & 9     & 6             & 3          \\
                        $\chi^2$                        & 31.9  & 19.2  & 12.8    & 10.2  \\
                        $\chi_\nu^2$            & 2.65  & 2.14  & 2.12  & 3.38 \\
                        BIC                             & 49.2  & 45.2  & 47.4    & 53.5 \\
                        \midrule
                        G dwarfs \\
                        \midrule
                        Model                           & G2    & G3    & G4      & G5   \\
                        $N_\text{free}$         & 22    & 19    & 16    & 13   \\
                        $\chi^2$                        & 92.0  & 36.3  & 24.2    & 43.1 \\
                        $\chi_\nu^2$            & 4.18  & 1.91  & 1.51  & 3.31 \\
                        BIC                             & 112.0 & 66.3  & 64.2    & 93.1 \\
                        \midrule
                        K dwarfs \\
                        \midrule
                        Model                           & G2    & G3    & G4      & G5   \\
                        $N_\text{free}$         & 16    & 13    & 10    & 7    \\
                        $\chi^2$                        & 50.7  & 13.5  & 10.5    & 9.38  \\
                        $\chi_\nu^2$            & 3.17  & 1.04  & 1.05  & 1.34 \\
                        BIC                             & 69.2  & 41.3  & 47.7    & 55.7 \\
                        \bottomrule
                \end{tabular}
        \end{center}
\end{table}

\begin{table}
        \caption{Best-fit three-Gaussian model parameters and local maxima and minima of the best-fit function for the F, G, and K dwarf $\log R'_\text{HK}$ distributions. The parameters are the same as in Table \ref{tab:fit_res_full}. The local maxima and minima values are organised by similarity.}
        \begin{center}
                \label{tab:fit_res_FGK}
                \begin{tabular}{lrrrrrrrrrrrrrrrrr}
                        \toprule\toprule
                        Parameter       & & Values \\
                        \midrule
                        & F dwarfs & G dwarfs & K dwarfs \\
            \midrule
            Best model  & G3        & G3        & G3        \\
            \midrule    
                        $\mu_1$         & $-4.94$       & $-5.01$       & $-4.94$         \\
            $A_1$               & 3.28          & 2.21          & 2.21          \\
            $\sigma_1$  & 0.075         & 0.088         & 0.068         \\
            \midrule
            $\mu_2$             & $-4.82$       & $-4.86$       & $-4.74$       \\
            $A_2$               & 1.54          & 1.30          & 2.03          \\
            $\sigma_2$  & 0.030         & 0.12          & 0.050         \\
            \midrule
            $\mu_3$             & $-4.58$       & $-4.44$   & $-4.52$   \\
            $A_3$               & 0.75          & 0.49          & 1.47          \\
            $\sigma_3$  & 0.148         & 0.095         & 0.096         \\
                        \midrule
                        Max. 1          & $-4.94$       & $-4.98$       & $-4.94$ \\
                        Max. 2          & $-4.83$       &               & $-4.74$ \\
                        Max. 3          & $-4.59$       & $-4.44$       & $-4.52$ \\
                        \midrule
                        Min. 1          & $-4.86$       &               & $-4.83$ \\
                        Min. 2          & $-4.72$       & $-4.60$       & $-4.63$ \\
                        \bottomrule
                \end{tabular}
        \end{center}
\end{table}

For the three spectral types considered, the best-fit model parameters are presented in Table \ref{tab:fit_res_FGK}, together with the local maxima and minima values of the  model.
All three spectral types have their main peak in the inactive zone close to \rhk$= -4.95$ dex.
The secondary peak, also similar for the three groups, is located in the active region near $-4.5$ dex, with the F dwarfs having the peak at a slightly more inactive region at $-4.59$ dex and the G dwarfs in a slightly more active region at $-4.44$ dex.
An intermediate Gaussian, between the inactive and the active peaks, is also present in the three spectral types.
This Gaussian has a similar location for F and G stars, at $-4.82$ and $-4.86$ dex, respectively.
They appear to broaden the inactive zone asymmetrically, but do not contribute to a significant peak.
The same is not true for the K dwarf population.
The Gaussian is located in a slightly more active region, very close to the VP gap at \rhk$-4.74$ dex, and contributes to a peak that is almost the same size as the inactive peak.
It is interesting that this is the first time, to the best of our knowledge, that a triple-peaked activity distribution is observed for K dwarfs.
These peaks are not an effect of binning, as they remain in the same position, and the minima get deeper, if we decrease the binning size from the used 0.05 to 0.025 dex.

The local minima in the most active region of the G and K spectral type distributions are close in activity level, having values of $-4.60$ (G) and $-4.63$ (K) dex which is also close to the minimum observed in the full data distribution near the active stars region ($-4.62$ dex) shown in Fig. \ref{fig:rhk_hist_fit}.
The F spectral type activity distribution also has a bin close to \rhk$= -4.6$ dex with lower density, however that minima is only fitted with the G4 model, which was not chosen.
We can therefore say that there is a clear minimum at around $-4.6$ dex in the activity distribution that is caused by the minima in the G and K (and possibly F) dwarf distributions at that location.
The second local minimum, in the inactive zone, is only observed for F and K dwarfs.
The former have the minimum at $-4.86$ dex while the latter at $-4.83$ dex.
These minima are not observed in the full data distribution because of the larger number of G dwarfs occupying these activity level regions.

These maxima and minima could be interpreted in terms of the time spent by the different spectral type stars in different activity level space regions:
\begin{itemize}
    \item Considering that the F dwarf distribution is double-peaked (the middle peak is not significant), as they decrease in activity levels with time, they have one region of faster evolution at $-4.72$ dex before stabilising in the inactive zone close to $-4.94$ dex.
    \item G dwarfs also only have a zone of faster activity evolution near $-4.6$ dex.
    \item K dwarfs have two phases of rapid activity evolution, the first at a similar level to the other spectral types at $-4.63$ dex, and the second in the inactive stage of evolution at $-4.83$ dex. They do not show the famous VP gap in their distribution.
\end{itemize}
The CE distributions also suggest that K dwarfs spend a more considerable amount of time in the active phase than F and G dwarfs because their proportion of active stars is more than double.
The evolution of CE with time will be discussed further in a following paper where we will compare the median $\log R'_\text{HK}$ evolution with isochronal age.

\section{The variability of CE in solar-type stars} \label{sec:rhk_disp}
The standard deviation of the stellar activity provides information about the activity variability of stars.
Generally, activity variability is proportional to stellar activity levels, and thus is expected to decrease as a star evolves.
In the following sections we analyse CE variability in our sample for the different luminosity classes and spectral types.

\subsection{Distribution of CE variability for main sequence, subgiant, and giant stars}
\begin{figure}
        \centering
        \resizebox{8.2cm}{!}{\includegraphics{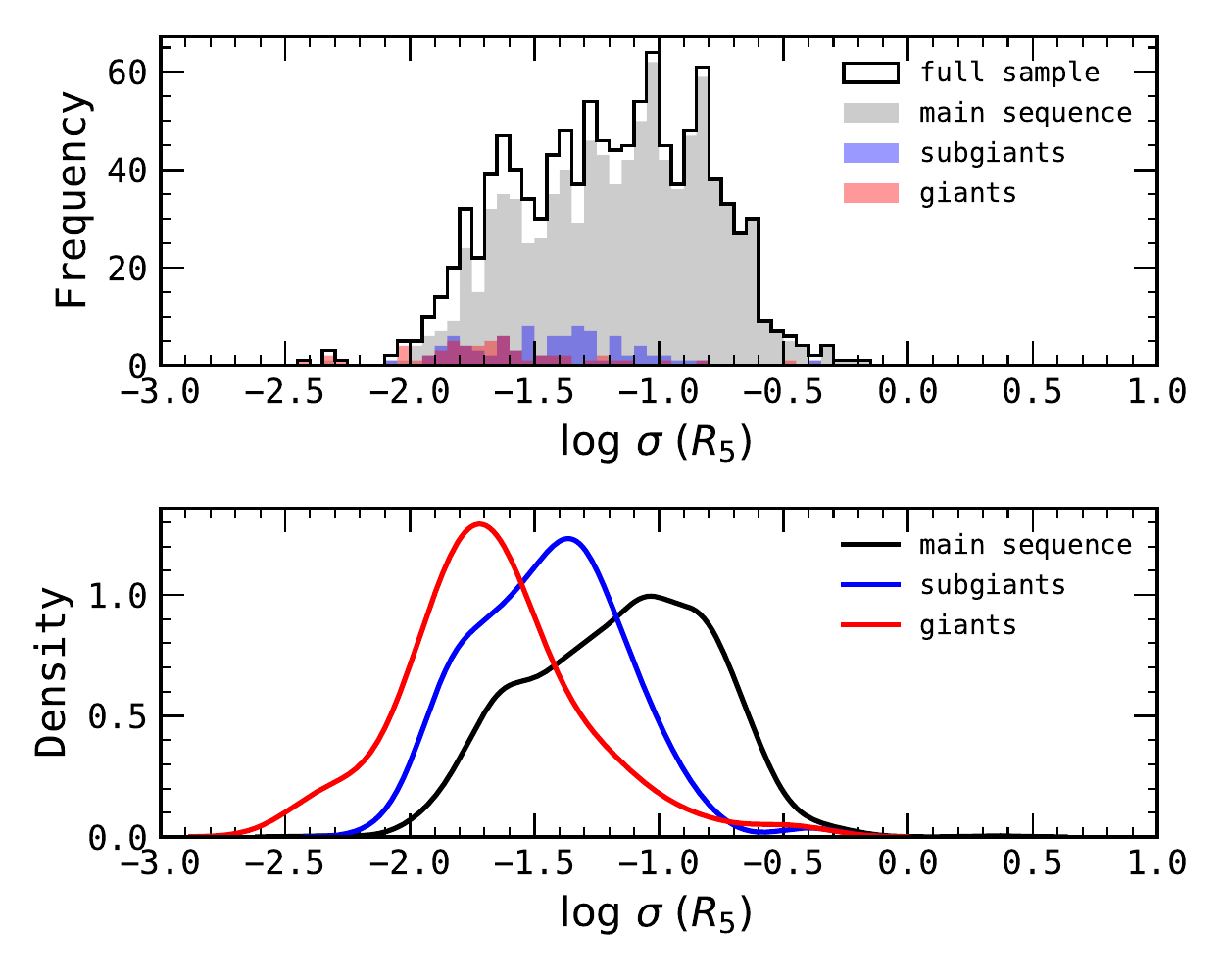}}

        \caption{\textit{Upper panel:} Distribution of the logarithm of the weighted standard deviation of $R_5$ for the full sample (black) segregated into MS (grey), subgiant (blue), and giant (red) luminosity classes.
        \textit{Lower panel:} Gaussian KDE of the distributions for the different luminosity classes.}
        \label{fig:rhk_disp_lum_class}
\end{figure}

The logarithm of the activity dispersion distribution for the full data set and the three luminosity classes present in our sample is presented in Fig. \ref{fig:rhk_disp_lum_class} (upper panel) where the black histogram is the full sample, grey represents the MS stars, blue the subgiants, and red the giants.
The majority of stars are in the range of about $-2.0 \leq \log \sigma(R_5) \leq -0.5$ dex.
The histogram clearly shows the difference between the total numbers in each luminosity class, with subgiants and giants being completely outnumbered by MS stars.
As expected, because they are generally less active and have longer rotational periods\footnote{In this sample, giant stars generally have low number of observations and these observations do not cover a long time-span, meaning that it is more difficult to measure variations due to long rotation periods.}, the evolved stars occupy the least variable region of the distribution.
This can be observed more clearly in the lower panel of Fig. \ref{fig:rhk_disp_lum_class} where we show the Gaussian KDEs of the distributions.
The highest peak of the KDEs moves from higher to lower values as we go from the MS (black) to giants (red) stars.
We can also see that the three KDEs appear to have the highest peak at higher values of variability and a secondary bump at lower values.
In the case of the giant stars, the secondary bump is close to $\log \sigma (R_5) = -2.35$ dex, and as we show in \S \ref{sec:rhk_std_vs_rhk}, the variability of these stars is not significant.
These bumps in the distribution could represent different stages in stellar variability.

\subsection{Distribution of CE variability for FGK dwarfs}
\begin{figure}
        \centering
        \resizebox{8.2cm}{!}{\includegraphics{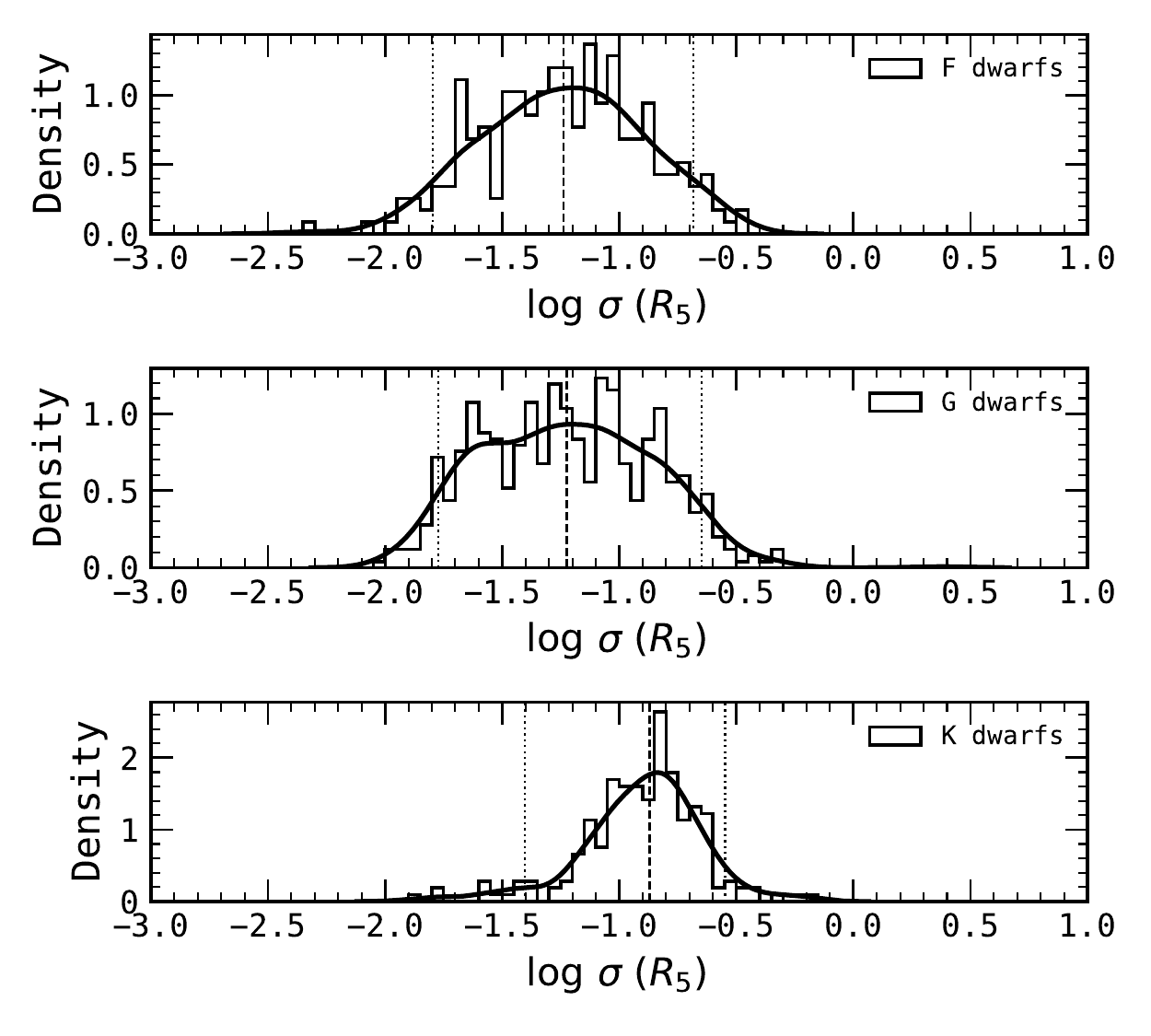}}

        \caption{Distribution of the logarithm of the standard deviation of $R_5$ for MS stars, segregated into the spectral types F (upper panel), G (middle panel), and K (lower panel). The vertical lines are the 50\%  (dashed), 5\%, and 95\% quantiles (dotted).}
        \label{fig:rhk_disp_sptype}
\end{figure}

Figure \ref{fig:rhk_disp_sptype} shows the distribution of the logarithm of the activity dispersion for F dwarfs (upper panel), G dwarfs (middle panel), and K dwarfs (lower panel).
The vertical lines show the 50\% percentile (solid) and the 5\% and 95\% percentiles (dotted).
Both the distributions for F and G dwarfs have similar intervals, but the G dwarf distribution shows a broader peak with a small bump in the lower variability zone near $\log \sigma (R_5) = -1.6$ dex.
The K dwarf distribution has a similar range to those of F and G dwarfs, but is nevertheless very different.
The majority of K dwarfs in this study are in the highest variability zone.
The 50\% percentile of the distribution is clearly higher than those of F and G dwarfs.
There is only a very small tail of low-variability K dwarfs.
This is the reason for the small bump in the distribution of variability for MS stars (Fig. \ref{fig:rhk_disp_lum_class}, lower panel): the contribution of K dwarfs to the distribution is mainly in the more variable region.
The lower and upper end of variability increases from F to G to K dwarfs, suggesting an increase in CE variability with decreasing temperature.
This is interesting because RV variability studies have shown that generally K dwarfs have lower RV variability attributed to stellar activity than G or F dwarfs \citep[e.g.][]{santos2000}.
Furthermore, it is expected that RV variability due to stellar activity decreases with stellar effective temperature \citep[e.g.][]{lovis2011}.
We therefore point out that K dwarfs with  more CE variability in general do not necessarily have higher RV variability.
Further investigation into the source of this variability will be carried out in following papers where we will determine the rotational and activity cycles amplitudes for the different spectral types presented here.

\subsection{Chromospheric emission variability as a function of CE level for main sequence, subgiant, and giant stars} \label{sec:rhk_std_vs_rhk}

\begin{figure}
        \resizebox{\hsize}{!}{\includegraphics{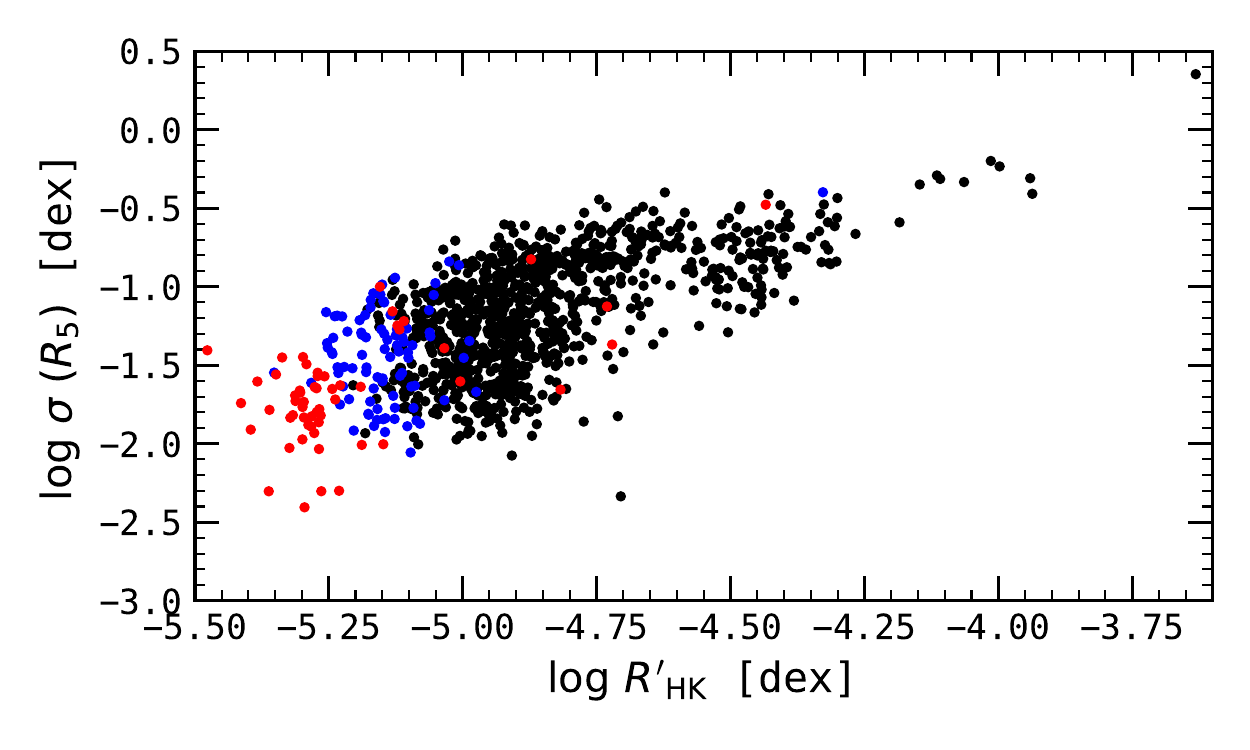}}
        \resizebox{\hsize}{!}{\includegraphics{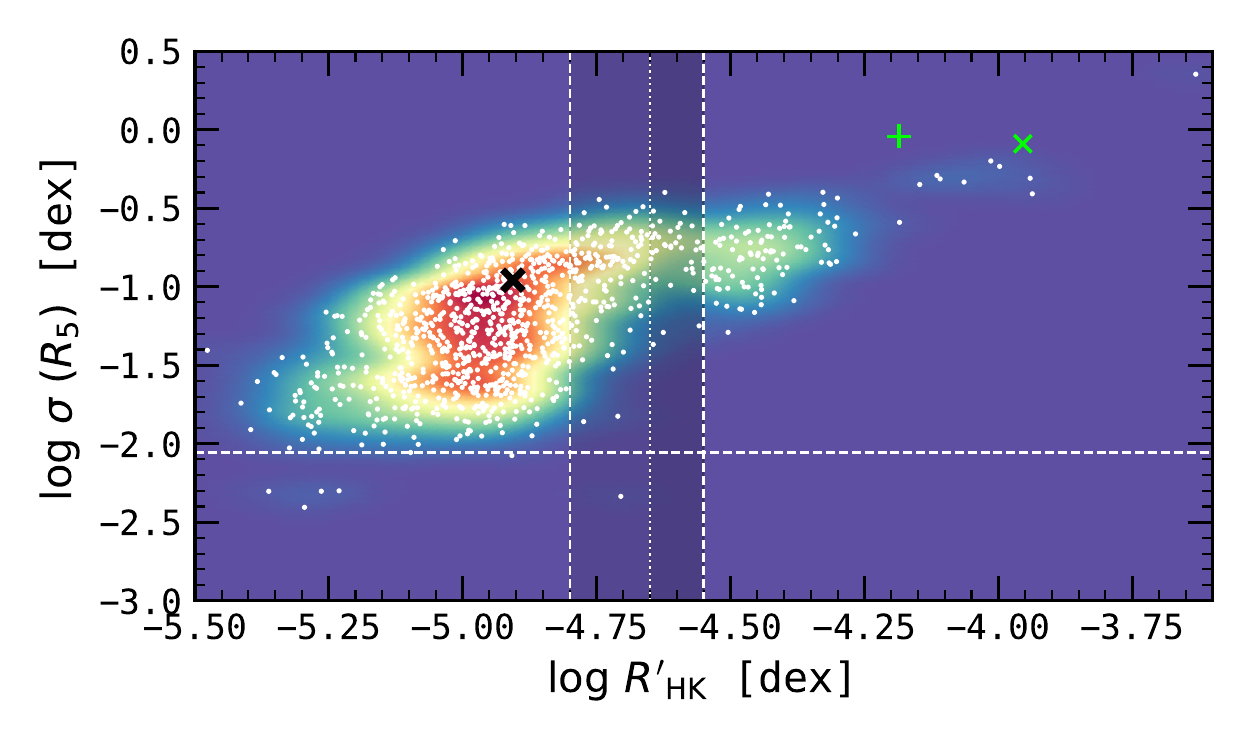}}

        \caption{\textit{Upper panel:} Logarithm of the weighted standard deviation of $R_5$ against median values of $\log R'_\text{HK}$ for stars with more than five nights of observations. Red, blue, and black dots are giants, subgiants, and MS stars. \textit{Lower panel:} Bivariate KDE map of the logarithm of the weighed standard deviation of $R_5$ against median values of $\log R'_\text{HK}$ for stars with more than five nights of observation. Redder areas represent higher density zones. The KDE bandwidth used was 0.07. White dots are stars. The white dashed line marks the dispersion of the star with the least absolute variability, HD\,60532, with $\log \sigma (R_5) = -2.05$ dex. The position of the Sun, based on the values published by \citet{mamajek2008}, is marked with a black `x'. The vertical white lines mark the approximate position of the gap we observe between active and inactive stars. The thinner gap, i.e. the region $-4.55 <$ \rhk $< -4.65$ dex, is located between the dotted line in the middle and the right dashed line, and has a darker shadow. The wider gap, at $-4.8 <$ \rhk $< -4.55$ dex, is located between the two dashed lines and has a lighter shadow. The green cross and `x' mark the positions of LQ\,Hya and EK\,Dra, respectively.}

        \label{fig:rhk_disp_vs_rhk}
\end{figure}

Activity variability is an important parameter in the study of stellar activity.
It can provide information regarding the distribution and filling factors of active regions in the stellar disks, and also the amplitude limits of rotational modulation together with possible activity cycles.
From an exoplanet detection point of view, RV variability produced by activity can inform on the detectability potential for planets around a given star using a specific instrument and/or observation schedule.
In this section, we are interested in deciphering how the activity dispersion varies with activity level, or in other words,  the allowed range in activity variability for a given star with a given
activity level.
We should note that this variability does not include the effects of oscillations and granulation that can affect RV over short temporal scales but are not detected by the $R'_\text{HK}$ chromospheric activity emission ratio.

Figure \ref{fig:rhk_disp_vs_rhk} (upper panel) shows the logarithm of the weighted standard deviation of $R_5$ against the median $\log R'_\text{HK}$.
The higher envelope of activity dispersion decreases with decreasing activity level and evolutional stage: active MS stars have in general higher maximum activity dispersion than inactive MS stars; inactive MS stars have higher dispersion than subgiants, and subgiants have higher maximum dispersion than giants.
The decrease in maximum dispersion is nevertheless more accentuated between giants and inactive MS stars (between around $-5.5$ and $-4.8$ dex) while it appears to reach a plateaux between activity levels of around $-4.75$ and $-4.25$ dex).
The lower envelope of the dispersion appears constant between giants and inactive MS stars with activity levels up to around $-4.8$ dex.
For activity levels higher than around $-4.8$ dex, the lower envelope starts to increase as we enter the VP gap.
This region is occupied mainly by stars with higher activity variability.
The lower envelope stays almost constant in the active level region between $-4.55$ and $-4.25$ dex.
A further increase in the minimum (and maximum) variability can be observed as the activity levels move to the very active region for activity levels $> -4.25$ dex.

The structure of the activity variability--activity level diagram can be better analysed by observing the bivariate KDE map of the two variables (Fig. \ref{fig:rhk_disp_vs_rhk}, lower panel).
The inactive MS region is clearly seen as the most populated region, coloured by an orange-red region and delimited by yellow, between activity levels of around $-5.1$ and $-4.75$ dex.
This region appears to be double (or multi) peaked in density, separating the most variable from the most quiet stars, with a high variability peak near $\log (\sigma (R_5)) = -1.0$ dex and a lower variability peak below $-1.5$ dex (c.f. Fig. \ref{fig:rhk_disp_lum_class}, lower panel, MS KDE in black).
The Sun is located in the higher variability region of the inactive MS stars, with a value of $\log \sigma (R_5)_\odot = -0.958$ dex.

Another feature we can observe is the VP gap that appears to stretch in activity levels from around $-4.55$ to $-4.8$ dex.
This is marked in the figure by three vertical white lines.
The gap appears as a diagonal feature and is thinner at high-activity variability (darker shadow) and wider at lower variability (lighter shadow).
This is why   the original VP-gap in the distribution of activity levels near \rhk$= -4.75$ dex appears blurred (see Fig. \ref{fig:rhk_hist_lum_class}): the thinner region of the gap (higher variability zone) is close to \rhk$= -4.6$ dex, the minimum in the activity distribution, and as the activity level decreases to lower values, the gap spreads as variability decreases, and the density of the stars starts to increase up to values close to \rhk$= -4.8$ dex because of the increasing number of low variability stars.

There is another gap separating active from very active dwarfs at activity levels of around $-4.25$ dex; however we cannot confirm whether or not this gap is devoid of stars because we have a very small number of very active stars in our sample.
These stars have higher upper and lower envelopes than the other activity classes.
As a comparison, we added two well-known very active, fast-rotating stars to Fig. \ref{fig:rhk_disp_vs_rhk} (lower panel), namely EK Dra (G1.5\,V) and LQ Hya (K1\,V).
These stars have $S_{MW}$ and \rhk values in the \citet{borosaikia2018} catalogue, which is a compilation of other \ca-based activity catalogues.
EK\,Dra has four observations while LQ\,Hya has only two.
We note that we limited the stars plotted in Fig. \ref{fig:rhk_disp_vs_rhk} to those with more than five observations.
However, for each of the two stars, the \rhk values have different $B-V$ associated with them, probably because they come from different sources.
Furthermore, no uncertainties or dates are given, and therefore we have no information regarding the time-span of the observations.
To better homogenize these activity values with those from our catalogue, we used the $S_{MW}$ from the Boro Saikia et al., retrieved $B-V$ from Simbad, and recalculated $R'_{HK}$ using the \citet{rutten1984} calibration.
Both EK\,Dra and LQ\,Hya show higher variability than other very active stars in our sample, although the values appear to be in agreement with the very active stars group, in the sense that all these stars have higher variability than the active stars group.
Almost all the very active stars are classified as BY\,Dra-type in Simbad, and as such are expected to have high activity variability.
Contrary to active and inactive stars, this group appears to show only very high variability in activity.

The gaps, or lower density zones, between inactive and active, and between active and very active stars can be explained as regions of fast stellar evolution, where a star stays for a few percent of its lifetime and is thus harder to find in those regions, or as caused by different stellar formation rates in time.

If we assume that, as a star evolves its activity level and variability decreases, stars will move on a diagonal path from top right to bottom left in the diagram shown in Fig. 20.
This means that higher variability active stars cross the VP gap more slowly (thinner gap) than low-variability active stars (wider gap).

There are six stars with very low variability below the lowest long-term variability line (white dashed line in Fig. \ref{fig:rhk_disp_vs_rhk}, lower panel).
Four of them are giants, namely 20\,Mon (K3\,III), $\alpha$\,Ind (K3\,III), $\eta$\,Ser (K2\,III), and HD\,181907 (K3\,III), and two are MS stars, $\alpha$\,CMi (F2\,V) and HD\,91324 (F6\,V).
While all of these stars have less than 16 nights of observations, some of them have a time-span of more than 300 days, and their dispersion might be under the influence of rotational modulation.
All the MS stars in this group (\rhk$> -5$ dex) have their quadratically added individual errors higher than their respective weighted standard deviations and can be considered as having insignificant variability.
On the contrary, all the giant stars in this group (\rhk$< -5$ dex) have their dispersions slightly higher than their errors.
To decipher whether or not those dispersions are significant, we carried out an $F$-test for variability using a procedure following \citet{zechmeister2009}.
The $F$-value used was $F = \sigma_e^2 / \sigma_i^2$ where $\sigma_e$ is the weighted standard deviation and $\sigma_i$ the median internal errors. 
None of the stars resulted in a probability lower than 5\%, meaning that the variability can be explained by the internal errors.
The star with the probability closest to 5\% was HD\,181907 with $P_F = 0.089$ ($F$-value of 3.68).
We therefore consider that even these stars do not show signs of significant variability and that the minimum long-term variability level for MS, subgiant, and giant stars is close to that of HD\,60532, at approximately $\log \sigma (R_5) = -2.05$ dex.

The analysis in this section reveals five important aspects of the distribution of activity variability with activity level:
\begin{enumerate}
    \item There appears to exist at least three different stages of variability, namely the inactive  (\rhk$ < -4.8$ dex), active (between around $-4.55$ and $-4.25$ dex), and very active regions ($> -4.25$ dex). The active and inactive regions are separated by a transition region or extended VP-gap between $-4.8$ and $-4.55$ dex, with a small region of high variability and low density of stars at $-4.6$ dex.
    \item The upper envelope of activity variability of active stars does not change significantly with activity level, and is almost constant between the beginning of the extended VP gap at $-4.8$ dex and the end of the active zone at $-4.25$ dex.
    \item Although the maximum variability shows a large decrease with decreasing activity level for inactive stars, the minimum dispersion is constant, with a minimum close to $\log \sigma (R_5) = -2.0$ dex.
    \item High-variability inactive stars can have variability levels similar to active stars.
    \item The VP gap in the \rhk distribution can be explained in terms of the variability distribution with activity level.
\end{enumerate}

Our most important finding is that inactive stars are not low variability stars by default, and we can find inactive stars showing more variability than active stars.
This is important, for example, for planet search programmes, because limiting stars by their activity level will not necessarily return a low activity variability sample.

\subsection{Chromospheric emission variability as a function of CE level for FGK dwarfs}

\begin{figure}
        \resizebox{\hsize}{!}{\includegraphics{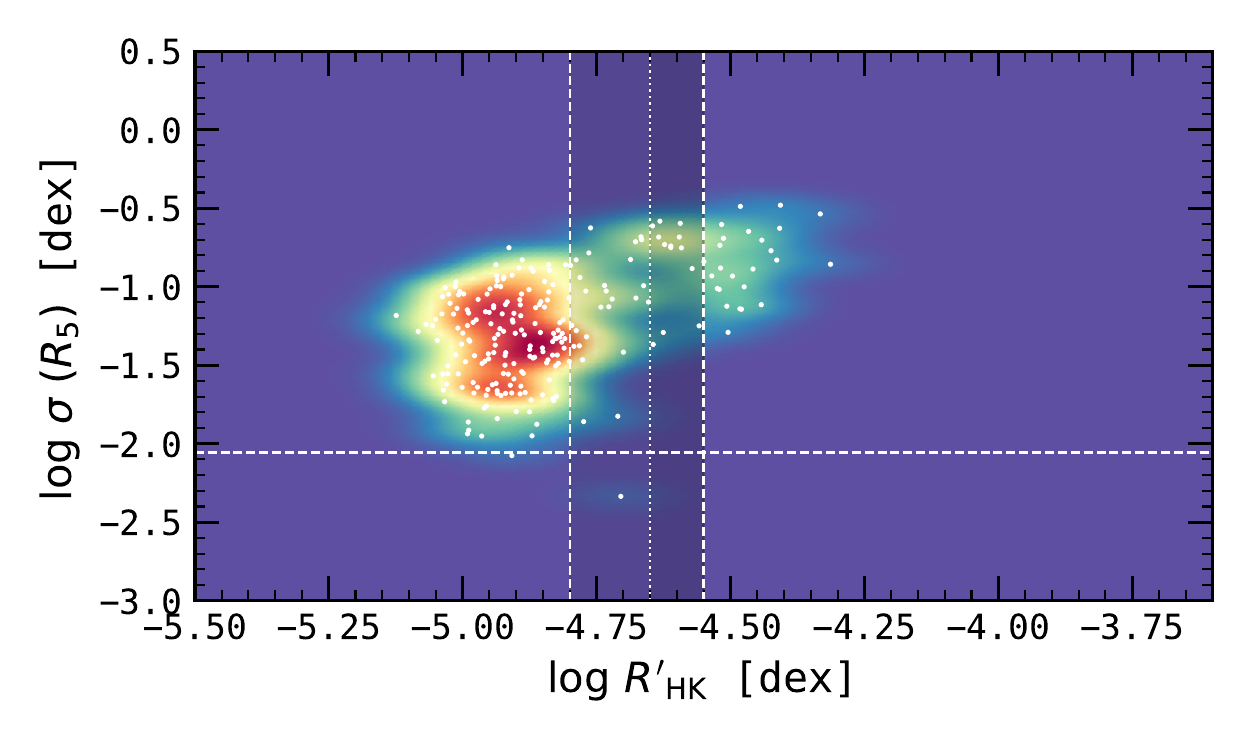}}
        \resizebox{\hsize}{!}{\includegraphics{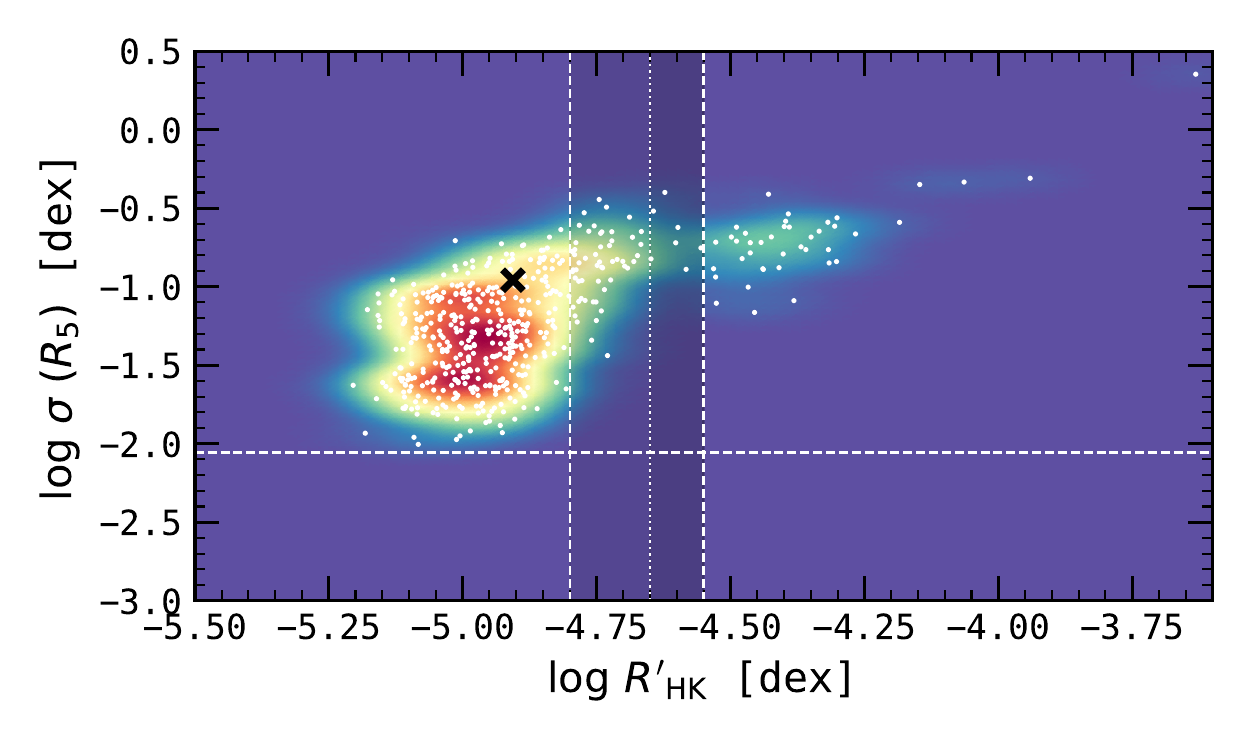}}
        \resizebox{\hsize}{!}{\includegraphics{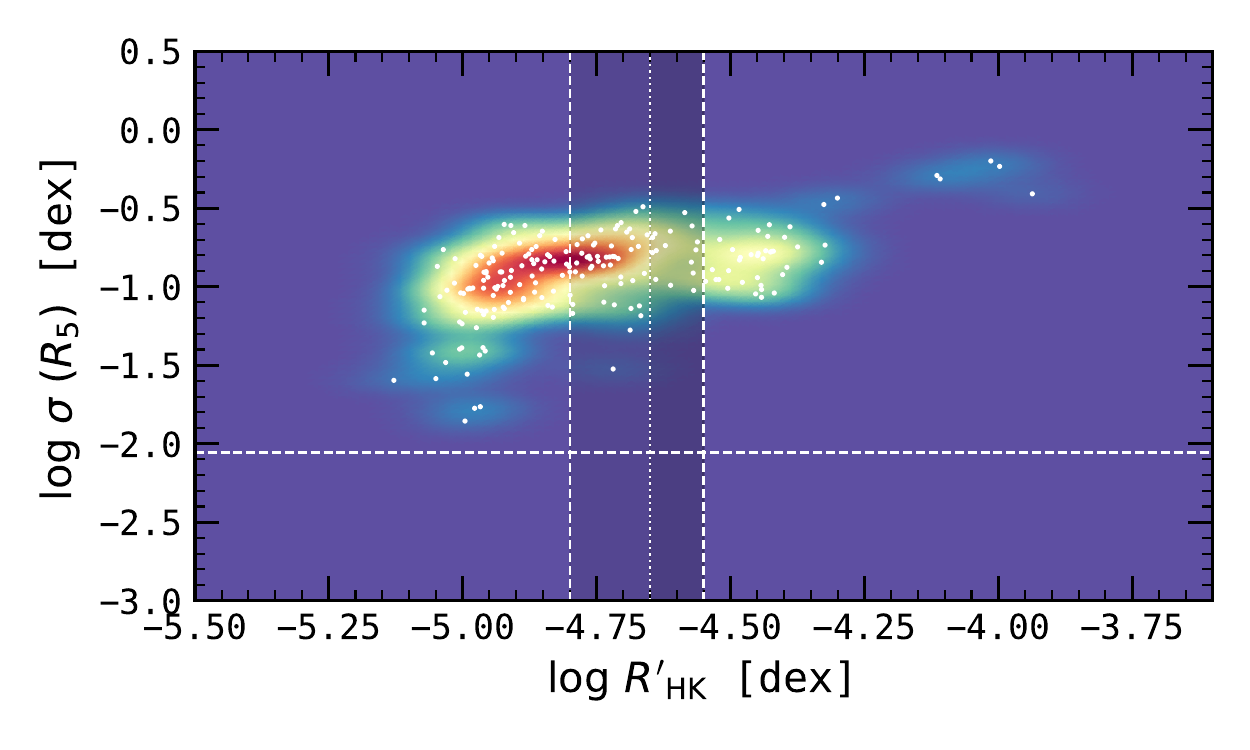}}

    \caption{\textit{Upper panel:} Same as in Fig. \ref{fig:rhk_disp_vs_rhk} for F dwarfs.
        \textit{Middle panel:} Same as in Fig. \ref{fig:rhk_disp_vs_rhk} for G dwarfs.
        \textit{Lower panel:} Same as in Fig. \ref{fig:rhk_disp_vs_rhk} for K dwarfs.}
        \label{fig:rhk_disp_vs_rhk_FGK}
\end{figure}

Bivariate KDE maps for F, G, and K dwarfs are shown in Fig. \ref{fig:rhk_disp_vs_rhk_FGK}.
All three spectral types show two main regions: one highly populated zone of inactive stars separated from another zone of moderately populated active stars by a low-density gap.
In the case of G and K dwarfs there is also a very low-density zone of very active stars around $\log R'_\text{HK} = -4$ dex.

The first noticiable difference between these spectral types is that the vast majority of the inactive K dwarfs have high variability, and there is only a small number of K dwarfs with variability below $\log \sigma (R_5) = -1.2$ dex.
This behaviour, which is also observed in the histograms of the activity variability in Fig. \ref{fig:rhk_disp_sptype}, is not seen in F and G dwarfs.
Furthermore, the maximum variability of active K dwarfs is slightly lower than the maximum variability of the inactive K dwarfs.
In other words, the upper envelope of variability does not decrease significantly with decreasing activity level.
The finding from the previous section that MS stars with low activity are not necessarily low-variability stars is reflected to an even greater degree in K dwarfs, as both active and inactive activity levels show similar variability levels.

The bivariate KDE distribution of inactive F and G dwarfs also seems to be multimodal with several high-density peaks occurring as the variability level decreases.
The nature of this multi-peak distribution of activity variability requires further investigation.
The spread in variability of active F dwarfs is larger than that for both active G and K dwarfs.

The overall distribution of G dwarfs is very similar to that seen in Fig. \ref{fig:rhk_disp_vs_rhk}, meaning that the overall distribution for MS stars is being modelled by the distribution of G dwarfs due to their higher numbers.
The VP gap is similar for the three spectral types; the thinner gap (between the vertical dotted and the right dashed lines in \ref{fig:rhk_disp_vs_rhk_FGK}) is mostly composed of high variability stars, while the wider gap (between dashed vertical lines) contains high and some low variability stars, however less low variability stars in the case of K dwarfs.

Now that we have the bivariate KDE of activity variability and activity level, we can infer the empirical probability distribution of the respective activity variability for a given activity level.
This will enable us to estimate the activity variability for a given activity level.
Another application of the bivariate KDE map is to simulate populations of stars with synthetic activity levels and variability that could be used in population studies.
The methodology on how to use the bivariate KDE to obtain the variability probability density function and synthetic populations is discussed in Appendix \ref{app:2d_kde}.
We also deliver a simple and easy-to-use \verb+python+ code to produce such results using the present  catalogue\footnote{Available at \url{https://github.com/gomesdasilva/rhk_kde}}, although it can be applied to other catalogues.

\section{Conclusions}
\label{sec:conclusions}
The main goal of this project is to characterise the activity of FGK main-sequence and evolved solar-type stars.
In this paper we present a catalogue of homogeneous chromospheric activity levels and stellar parameters for a sample of 1674 FGK main sequence, subgiant, and giant stars using more than 180 000 observations between 2003 and 2019 from the HARPS archive.
This is one of the largest homogeneous catalogues of chromospheric activity both in terms of number of stars and time-span.

One of the most important findings of this work is that, although the upper envelope of the distribution of activity variability with activity level decreases with decreasing activity level, stars with low activity levels do not necessarily have low variability.
This is even more important in the case of K dwarfs, where we find that the vast majority of these types of stars do not show low variability, and inactive K dwarfs have similar levels of activity variability to active K dwarfs.
These results are especially important for planet search programmes where low-variability stars are generally selected based on their activity levels.

The distribution of activity variability appears to have three distinct regimes in the variability-level diagram:  inactive stars between \rhk$=-5.5$ and $-4.8$ dex where the upper envelope of variability is always increasing with increasing activity level but the lower envelope is constant; active stars between \rhk$=-4.55$ and $-4.25$ dex where the upper envelope appears constant while the lower envelope, while at a higher level than for inactive stars, also appears constant; and  very active stars with \rhk$>-4.25$ dex where the variability is always higher than the active regime. The inactive and active regimes are separated by an extended VP gap between $-4.8$ and $-4.55$ dex.

We are also able to explain the shape of the VP gap observed in the \rhk distribution based on the distribution of activity variability as a function of activity level.
The VP gap appears as a diagonal feature in the activity dispersion-level diagram, with different widths depending on the variability of the stars. For stars with high dispersion, the gap is thin, between \rhk$ = -4.65$ and $-4.55$ dex, with a very low number of
stars. As the variability decreases, the gap widens asymmetrically in the direction of the inactive stars down to around \rhk$= -4.8$ dex, marking the transition between active and inactive stars.
This diagonal behaviour explains the distribution of \rhk where the density of stars increases from around $-4.6$ to around $-4.8$ dex.

Our other findings from analysing this catalogue can be summarised as follows:
\begin{itemize}
    \item Our distribution of relative activity dispersion shows that the upper limit is around 22\%, meaning that when measuring activity levels with a small number of observations (not covering rotational modulations and/or activity cycles) one might incur accuracy errors up to 22\% of the median value.
    \item  We confirm the finding \citep[e.g.][]{henry1996, gray2006, jenkins2008} that the distribution of chromospheric activity of MS stars and individual G dwarfs is bimodal, and find a peak near $\log R'_\text{HK} = -4.5$ dex for active stars and another peak near $-4.95$ dex in the inactive star regime.
    These peaks are separated by a low-density zone near \rhk$ =-4.6$ dex.
    \item Subgiant stars occupy a region around $-5.25 \leq \log R'_\text{HK} \leq -5.1$ and giant stars are primarily located around $-5.4 \leq \log R'_\text{HK} \leq -5.25$ dex.
    \item There is a region of very active stars with activity levels above $\log R'_\text{HK} = -4.25$ dex separated from the active region by a gap. This group of stars were previously identified by \citet{henry1996} who suggested they could be young active stars, some of them in close binary systems.
    \item Although planet search programmes tend to bias star selection towards inactive stars, the proportions of CE in our sample are very similar to those found by \citet{henry1996} in their unbiased sample. This reveals that our sample, constituted mainly by planet search programme stars, may be a good representation of stellar activity in the solar neighbourhood.
    Following the activity classification of \citet{henry1996}, our sample contains 1.2\% very active, 28.5\% active, 66.9\% inactive, and 3.5\% very inactive MS stars.
    \item The distribution of activity dispersion appears to be multi-peaked and ranges between around $\log \sigma(R_5) = -2$ and $-0.1$ dex.
    \item The upper envelope of activity dispersion in general always decreases with decreasing activity level.
    However, the lower envelope is constant between the lowest activity levels at around \rhk$-5.5$ and around $-4.8$ dex with a level close to $\log \sigma (R_5) = -2.0$ dex. The lower envelope of dispersion then increases through the VP gap until it stabilises in the active region close to $-4.55$ dex with a dispersion of around $\log \sigma (R_5) = -1.25$ dex. A further increase in the lower envelope of dispersion is observed when activity levels reach the very active zone near \rhk$= -4.25$ dex with log-dispersion levels above $-0.5$ dex.
    \item The activity distribution is triple peaked for K dwarfs, with the additional peak in the VP gap at $\log R'_\text{HK} = -4.75$ dex. This distribution also shows the activity gap near $-4.6$ dex observed in the distribution of MS stars and also in the individual distribution of G dwarfs.
    \item The distribution of activity levels of F dwarfs appears to be double-peaked and similar to that of G dwarfs.
    \item The distribution of activity levels of giant stars presents a tail of higher activity stars that appear to be more massive and younger than the main group. These are intermediate-mass (1.5-4 M$_\odot$) stars that might have evolved quickly into the giant phase without losing angular momentum and thus activity level. Investigation of the rotation rate of these stars should confirm or refute this hypothesis.
    \item We find indications that the population of very inactive MS stars with \rhk$< -5.1$ dex found by \citet{henry1996} are not MS stars in a Maunder Minimum phase but stars showing activity variability that are starting to evolve beyond the MS. We therefore suggest that the $-5.1$ dex activity level might mark a transition stage of evolution from MS to subgiant phase.
    The transition between the subgiant and giant phase appears to be at around \rhk$= -5.25$ dex.
    \item Although having a very low dispersion, $\tau$ Ceti presents a long-term decreasing trend in activity level together with cyclic variability and should not be considered as a `flat' activity standard star.
    Instead, HD\,60532 (F7\,IV) is a long-term chromospherically stable star, with a relative variability in $S_\text{MW}$ of only 0.35\%.
    This star should be used as an activity precision standard rather than $\tau$ Ceti.
    \item The Sun is located in the highest dispersion zone of the MS inactive stars.
    \item Finally, we show a simple way to obtain the probability density function of the activity dispersion from the activity level of a star using the multivariate Gaussian KDE of the two parameters. We also show how to synthesise stellar populations with activity levels and variability based on our multivariate KDE map.
\end{itemize}

We would like to finish with an important note.
These results cannot be extrapolated for the case of RV variability.
For example, oscillations and granulation will contribute to RV variability but are not detected by the \rhk index.
Furthermore, the influence of activity measured in the \rhk index in RV variability is not the same for different spectral types.
While we show that K dwarfs have a high-activity variability, the influence of activity on RV is lower for K dwarfs than for F and G dwarfs \citep[e.g.][]{santos2000, lovis2011}.
We suggest that the activity variability instead of the activity level should be used when analysing the effects of activity in RV observations.

By releasing this catalogue to the community we expect to contribute to further advancing the current knowledge of stellar evolution in general and stellar activity in particular.
In the following papers of this series we will analyse how activity and stellar parameters are connected, including the activity--age relation. We will also study the rotation and activity cycle periods and amplitudes of this sample to understand how they evolve and affect stellar variability for different luminosity classes and spectral types. The activity time-series will be published as and when they are analysed.

\begin{acknowledgements}
        This work was supported by Fundação para a Ciência e a Tecnologia (FCT, Portugal) through national funds and by FEDER - Fundo Europeu de Desenvolvimento Regional through COMPETE2020 - Programa Operacional Competitividade e Internacionalização by these grants: UID/FIS/04434/2019; UIDB/04434/2020; UIDP/04434/2020; PTDC/FIS-AST/32113/2017 \& POCI-01-0145-FEDER-032113; PTDC/FIS-AST/28953/2017 \& POCI-01-0145-FEDER-028953.
        V.A., E.D.M, N.C.S., and S.G.S. also acknowledge the support from FCT through Investigador FCT contracts nr. IF/00650/2015/CP1273/CT0001, IF/00849/2015/CP1273/CT0003, IF/00169/2012/CP0150/CT0002, and IF/00028/2014/CP1215/CT0002, respectively, and POPH/FSE (EC) by FEDER funding through the program ``Programa Operacional de Factores de Competitividade - COMPETE''.
    TLC acknowledges support from the European Union's Horizon 2020 research and innovation programme under the Marie Sk\l{}odowska-Curie grant agreement No.~792848 (PULSATION).
    DB is supported in the form of work contract FCT/MCTES through national funds and by FEDER through COMPETE2020 in connection to these grants: PTDC/FIS-AST/30389/2017 \& POCI-01-0145-FEDER-030389.
    ARB and PdL acknowledge financial support from the ANR 14-CE33-014-01 and from the "Programme National de Physique Stellaire" (PNPS) of CNRS/INSU co-funded by CEA and CNES.

        This work is based on data obtained from the HARPS public data base at the ESO.
        This work has made use of data from the European Space Agency (ESA) mission
        {\it Gaia} (\url{https://www.cosmos.esa.int/gaia}), processed by the {\it Gaia}
        Data Processing and Analysis Consortium (DPAC,
        \url{https://www.cosmos.esa.int/web/gaia/dpac/consortium}). Funding for the DPAC
        has been provided by national institutions, in particular the institutions
        participating in the {\it Gaia} Multilateral Agreement.
    This research has made use of the SIMBAD database, operated at CDS, Strasbourg, France; the VizieR catalogue access tool, CDS, Strasbourg, France; Astropy,\footnote{http://www.astropy.org} a community-developed core Python package for Astronomy \citep{astropy:2013, astropy:2018}; SciPy \citep{scipy2020}; Matplotlib: Visualization with Python \citep{matplotlib2007}; Scikit-learn: Machine Learning in Python \citep{sklearn2011}.
\end{acknowledgements}

\bibliographystyle{aa} 
\bibliography{bibliography.bib}

\begin{thebibliography}{112}
\expandafter\ifx\csname natexlab\endcsname\relax\def\natexlab#1{#1}\fi

\bibitem[{{Akeson} {et~al.}(2013){Akeson}, {Chen}, {Ciardi}, {Crane}, {Good},
  {Harbut}, {Jackson}, {Kane}, {Laity}, {Leifer}, {Lynn}, {McElroy}, {Papin},
  {Plavchan}, {Ram{\'\i}rez}, {Rey}, {von Braun}, {Wittman}, {Abajian}, {Ali},
  {Beichman}, {Beekley}, {Berriman}, {Berukoff}, {Bryden}, {Chan}, {Groom},
  {Lau}, {Payne}, {Regelson}, {Saucedo}, {Schmitz}, {Stauffer}, {Wyatt}, \&
  {Zhang}}]{akeson2013}
{Akeson}, R.~L., {Chen}, X., {Ciardi}, D., {et~al.} 2013, \pasp, 125, 989

\bibitem[{{Alonso} {et~al.}(2009){Alonso}, {Aigrain}, {Pont}, {Mazeh}, \&
  {CoRoT Exoplanet Science Team}}]{alonso2009}
{Alonso}, R., {Aigrain}, S., {Pont}, F., {Mazeh}, T., \& {CoRoT Exoplanet
  Science Team}. 2009, in IAU Symposium, Vol. 253, Transiting Planets, ed.
  F.~{Pont}, D.~{Sasselov}, \& M.~J. {Holman}, 91--96

\bibitem[{{Andreasen} {et~al.}(2017){Andreasen}, {Sousa}, {Tsantaki},
  {Teixeira}, {Mortier}, {Santos}, {Su{\'a}rez-Andr{\'e}s}, {Delgado-Mena}, \&
  {Ferreira}}]{andreasen2017}
{Andreasen}, D.~T., {Sousa}, S.~G., {Tsantaki}, M., {et~al.} 2017, \aap, 600,
  A69

\bibitem[{{Angus} {et~al.}(2015){Angus}, {Aigrain}, {Foreman-Mackey}, \&
  {McQuillan}}]{angus2015}
{Angus}, R., {Aigrain}, S., {Foreman-Mackey}, D., \& {McQuillan}, A. 2015,
  \mnras, 450, 1787

\bibitem[{{Arriagada}(2011)}]{arriagada2011}
{Arriagada}, P. 2011, \apj, 734, 70

\bibitem[{{Astropy Collaboration} {et~al.}(2013){Astropy Collaboration},
  {Robitaille}, {Tollerud}, {Greenfield}, {Droettboom}, {Bray}, {Aldcroft},
  {Davis}, {Ginsburg}, {Price-Whelan}, {Kerzendorf}, {Conley}, {Crighton},
  {Barbary}, {Muna}, {Ferguson}, {Grollier}, {Parikh}, {Nair}, {Unther},
  {Deil}, {Woillez}, {Conseil}, {Kramer}, {Turner}, {Singer}, {Fox}, {Weaver},
  {Zabalza}, {Edwards}, {Azalee Bostroem}, {Burke}, {Casey}, {Crawford},
  {Dencheva}, {Ely}, {Jenness}, {Labrie}, {Lim}, {Pierfederici}, {Pontzen},
  {Ptak}, {Refsdal}, {Servillat}, \& {Streicher}}]{astropy:2013}
{Astropy Collaboration}, {Robitaille}, T.~P., {Tollerud}, E.~J., {et~al.} 2013,
  \aap, 558, A33

\bibitem[{{Baliunas} {et~al.}(1995{\natexlab{a}}){Baliunas}, {Donahue}, {Soon},
  {Gilliland }, \& {Soderblom}}]{baliunas1995b}
{Baliunas}, S.~L., {Donahue}, R.~A., {Soon}, W., {Gilliland }, R., \&
  {Soderblom}, D.~R. 1995{\natexlab{a}}, in American Astronomical Society
  Meeting Abstracts, Vol. 186, American Astronomical Society Meeting Abstracts
  \#186, 21.09

\bibitem[{{Baliunas} {et~al.}(1995{\natexlab{b}}){Baliunas}, {Donahue}, {Soon},
  {Horne}, {Frazer}, {Woodard-Eklund}, {Bradford}, {Rao}, {Wilson}, {Zhang},
  {Bennett}, {Briggs}, {Carroll}, {Duncan}, {Figueroa}, {Lanning}, {Misch},
  {Mueller}, {Noyes}, {Poppe}, {Porter}, {Robinson}, {Russell}, {Shelton},
  {Soyumer}, {Vaughan}, \& {Whitney}}]{baliunas1995a}
{Baliunas}, S.~L., {Donahue}, R.~A., {Soon}, W.~H., {et~al.}
  1995{\natexlab{b}}, \apj, 438, 269

\bibitem[{{Barnes}(2003)}]{barnes2003}
{Barnes}, S.~A. 2003, \apj, 586, 464

\bibitem[{{Barnes}(2010)}]{barnes2010}
{Barnes}, S.~A. 2010, \apj, 722, 222

\bibitem[{{Bennett} {et~al.}(2012){Bennett}, {Sumi}, {Bond}, {Kamiya}, {Abe},
  {Botzler}, {Fukui}, {Furusawa}, {Itow}, {Korpela}, {Kilmartin}, {Ling},
  {Masuda}, {Matsubara}, {Miyake}, {Muraki}, {Ohnishi}, {Rattenbury}, {Saito},
  {Sullivan}, {Suzuki}, {Sweatman}, {Tristram}, {Wada}, {Yock}, \& {MOA
  Collaboration}}]{bennett2012}
{Bennett}, D.~P., {Sumi}, T., {Bond}, I.~A., {et~al.} 2012, \apj, 757, 119

\bibitem[{{Boro Saikia} {et~al.}(2018){Boro Saikia}, {Marvin}, {Jeffers},
  {Reiners}, {Cameron}, {Marsden}, {Petit}, {Warnecke}, \&
  {Yadav}}]{borosaikia2018}
{Boro Saikia}, S., {Marvin}, C.~J., {Jeffers}, S.~V., {et~al.} 2018, \aap, 616,
  A108

\bibitem[{{Bovy} {et~al.}(2016){Bovy}, {Rix}, {Green}, {Schlafly}, \&
  {Finkbeiner}}]{bovy2016}
{Bovy}, J., {Rix}, H.-W., {Green}, G.~M., {Schlafly}, E.~F., \& {Finkbeiner},
  D.~P. 2016, \apj, 818, 130

\bibitem[{{Bressan} {et~al.}(2012){Bressan}, {Marigo}, {Girardi}, {Salasnich},
  {Dal Cero}, {Rubele}, \& {Nanni}}]{bressan2012}
{Bressan}, A., {Marigo}, P., {Girardi}, L., {et~al.} 2012, \mnras, 427, 127

\bibitem[{{Brun} \& {Browning}(2017)}]{brun2017}
{Brun}, A.~S. \& {Browning}, M.~K. 2017, Living Reviews in Solar Physics, 14, 4

\bibitem[{{Choi} {et~al.}(2016){Choi}, {Dotter}, {Conroy}, {Cantiello},
  {Paxton}, \& {Johnson}}]{choi2016}
{Choi}, J., {Dotter}, A., {Conroy}, C., {et~al.} 2016, \apj, 823, 102

\bibitem[{{Ciardi} {et~al.}(2011){Ciardi}, {von Braun}, {Bryden}, {van Eyken},
  {Howell}, {Kane}, {Plavchan}, {Ram{\'\i}rez}, \& {Stauffer}}]{ciardi2011}
{Ciardi}, D.~R., {von Braun}, K., {Bryden}, G., {et~al.} 2011, \aj, 141, 108

\bibitem[{{Collier Cameron} {et~al.}(2019){Collier Cameron}, {Mortier},
  {Phillips}, {Dumusque}, {Haywood}, {Langellier}, {Watson}, {Cegla}, {Costes},
  {Charbonneau}, {Coffinet}, {Latham}, {Lopez-Morales}, {Malavolta},
  {Maldonado}, {Micela}, {Milbourne}, {Molinari}, {Saar}, {Thompson},
  {Buchschacher}, {Cecconi}, {Cosentino}, {Ghedina}, {Glenday}, {Gonzalez},
  {Li}, {Lodi}, {Lovis}, {Pepe}, {Poretti}, {Rice}, {Sasselov}, {Sozzetti},
  {Szentgyorgyi}, {Udry}, \& {Walsworth}}]{colliercameron2019}
{Collier Cameron}, A., {Mortier}, A., {Phillips}, D., {et~al.} 2019, \mnras,
  487, 1082

\bibitem[{{da Silva} {et~al.}(2006){da Silva}, {Girardi}, {Pasquini},
  {Setiawan}, {von der L{\"u}he}, {de Medeiros}, {Hatzes}, {D{\"o}llinger}, \&
  {Weiss}}]{dasilva2006}
{da Silva}, L., {Girardi}, L., {Pasquini}, L., {et~al.} 2006, \aap, 458, 609

\bibitem[{{de Laverny} {et~al.}(2013){de Laverny}, {Recio-Blanco}, {Worley},
  {De Pascale}, {Hill}, \& {Bijaoui}}]{delaverny2013}
{de Laverny}, P., {Recio-Blanco}, A., {Worley}, C.~C., {et~al.} 2013, The
  Messenger, 153, 18

\bibitem[{{De Pascale} {et~al.}(2014){De Pascale}, {Worley}, {de Laverny},
  {Recio-Blanco}, {Hill}, \& {Bijaoui}}]{depascale2014}
{De Pascale}, M., {Worley}, C.~C., {de Laverny}, P., {et~al.} 2014, \aap, 570,
  A68

\bibitem[{{Delgado Mena} {et~al.}(2017){Delgado Mena}, {Tsantaki}, {Adibekyan},
  {Sousa}, {Santos}, {Gonz{\'a}lez Hern{\'a}ndez}, \&
  {Israelian}}]{delgadomena2017}
{Delgado Mena}, E., {Tsantaki}, M., {Adibekyan}, V.~Z., {et~al.} 2017, \aap,
  606, A94

\bibitem[{{Dotter}(2016)}]{dotter2016}
{Dotter}, A. 2016, \apjs, 222, 8

\bibitem[{{Drimmel} {et~al.}(2003){Drimmel}, {Cabrera-Lavers}, \&
  {L{\'o}pez-Corredoira}}]{drimmel2003}
{Drimmel}, R., {Cabrera-Lavers}, A., \& {L{\'o}pez-Corredoira}, M. 2003, \aap,
  409, 205

\bibitem[{{Duncan} {et~al.}(1991){Duncan}, {Vaughan}, {Wilson}, {Preston},
  {Frazer}, {Lanning}, {Misch}, {Mueller}, {Soyumer}, {Woodard}, {Baliunas},
  {Noyes}, {Hartmann}, {Porter}, {Zwaan}, {Middelkoop}, {Rutten}, \&
  {Mihalas}}]{duncan1991}
{Duncan}, D.~K., {Vaughan}, A.~H., {Wilson}, O.~C., {et~al.} 1991, \apjs, 76,
  383

\bibitem[{{Faria} {et~al.}(2020){Faria}, {Adibekyan}, {Amazo-G{\'o}mez},
  {Barros}, {Camacho}, {Demangeon}, {Figueira}, {Mortier}, {Oshagh}, {Pepe},
  {Santos}, {Gomes da Silva}, {Costa Silva}, {Sousa}, {Ulmer-Moll}, \&
  {Viana}}]{faria2020}
{Faria}, J.~P., {Adibekyan}, V., {Amazo-G{\'o}mez}, E.~M., {et~al.} 2020, \aap,
  635, A13

\bibitem[{{Figueira} {et~al.}(2010){Figueira}, {Marmier}, {Bonfils}, {di
  Folco}, {Udry}, {Santos}, {Lovis}, {M{\'e}gevand}, {Melo}, {Pepe}, {Queloz},
  {S{\'e}gransan}, {Triaud}, \& {Viana Almeida}}]{figueira2010}
{Figueira}, P., {Marmier}, M., {Bonfils}, X., {et~al.} 2010, \aap, 513, L8

\bibitem[{{Gaia Collaboration} {et~al.}(2018){Gaia Collaboration}, {Brown},
  {Vallenari}, {Prusti}, {de Bruijne}, {Babusiaux}, {Bailer-Jones}, {Biermann},
  {Evans}, {Eyer}, {Jansen}, {Jordi}, {Klioner}, {Lammers}, {Lindegren},
  {Luri}, {Mignard}, {Panem}, {Pourbaix}, {Randich}, {Sartoretti}, {Siddiqui},
  {Soubiran}, {van Leeuwen}, {Walton}, {Arenou}, {Bastian}, {Cropper},
  {Drimmel}, {Katz}, {Lattanzi}, {Bakker}, {Cacciari}, {Casta{\~n}eda},
  {Chaoul}, {Cheek}, {De Angeli}, {Fabricius}, {Guerra}, {Holl}, {Masana},
  {Messineo}, {Mowlavi}, {Nienartowicz}, {Panuzzo}, {Portell}, {Riello},
  {Seabroke}, {Tanga}, {Th{\'e}venin}, {Gracia-Abril}, {Comoretto},
  {Garcia-Reinaldos}, {Teyssier}, {Altmann}, {Andrae}, {Audard},
  {Bellas-Velidis}, {Benson}, {Berthier}, {Blomme}, {Burgess}, {Busso},
  {Carry}, {Cellino}, {Clementini}, {Clotet}, {Creevey}, {Davidson}, {De
  Ridder}, {Delchambre}, {Dell'Oro}, {Ducourant},
  {Fern{\'a}ndez-Hern{\'a}ndez}, {Fouesneau}, {Fr{\'e}mat}, {Galluccio},
  {Garc{\'\i}a-Torres}, {Gonz{\'a}lez-N{\'u}{\~n}ez}, {Gonz{\'a}lez-Vidal},
  {Gosset}, {Guy}, {Halbwachs}, {Hambly}, {Harrison}, {Hern{\'a}ndez},
  {Hestroffer}, {Hodgkin}, {Hutton}, {Jasniewicz}, {Jean-Antoine-Piccolo},
  {Jordan}, {Korn}, {Krone-Martins}, {Lanzafame}, {Lebzelter}, {L{\"o}ffler},
  {Manteiga}, {Marrese}, {Mart{\'\i}n-Fleitas}, {Moitinho}, {Mora}, {Muinonen},
  {Osinde}, {Pancino}, {Pauwels}, {Petit}, {Recio-Blanco}, {Richards},
  {Rimoldini}, {Robin}, {Sarro}, {Siopis}, {Smith}, {Sozzetti}, {S{\"u}veges},
  {Torra}, {van Reeven}, {Abbas}, {Abreu Aramburu}, {Accart}, {Aerts},
  {Altavilla}, {{\'A}lvarez}, {Alvarez}, {Alves}, {Anderson}, {Andrei},
  {Anglada Varela}, {Antiche}, {Antoja}, {Arcay}, {Astraatmadja}, {Bach},
  {Baker}, {Balaguer-N{\'u}{\~n}ez}, {Balm}, {Barache}, {Barata}, {Barbato},
  {Barblan}, {Barklem}, {Barrado}, {Barros}, {Barstow}, {Bartholom{\'e}
  Mu{\~n}oz}, {Bassilana}, {Becciani}, {Bellazzini}, {Berihuete}, {Bertone},
  {Bianchi}, {Bienaym{\'e}}, {Blanco-Cuaresma}, {Boch}, {Boeche}, {Bombrun},
  {Borrachero}, {Bossini}, {Bouquillon}, {Bourda}, {Bragaglia}, {Bramante},
  {Breddels}, {Bressan}, {Brouillet}, {Br{\"u}semeister}, {Brugaletta},
  {Bucciarelli}, {Burlacu}, {Busonero}, {Butkevich}, {Buzzi}, {Caffau},
  {Cancelliere}, {Cannizzaro}, {Cantat-Gaudin}, {Carballo}, {Carlucci},
  {Carrasco}, {Casamiquela}, {Castellani}, {Castro-Ginard}, {Charlot},
  {Chemin}, {Chiavassa}, {Cocozza}, {Costigan}, {Cowell}, {Crifo}, {Crosta},
  {Crowley}, {Cuypers}, {Dafonte}, {Damerdji}, {Dapergolas}, {David}, {David},
  {de Laverny}, {De Luise}, {De March}, {de Martino}, {de Souza}, {de Torres},
  {Debosscher}, {del Pozo}, {Delbo}, {Delgado}, {Delgado}, {Di Matteo},
  {Diakite}, {Diener}, {Distefano}, {Dolding}, {Drazinos}, {Dur{\'a}n},
  {Edvardsson}, {Enke}, {Eriksson}, {Esquej}, {Eynard Bontemps}, {Fabre},
  {Fabrizio}, {Faigler}, {Falc{\~a}o}, {Farr{\`a}s Casas}, {Federici},
  {Fedorets}, {Fernique}, {Figueras}, {Filippi}, {Findeisen}, {Fonti},
  {Fraile}, {Fraser}, {Fr{\'e}zouls}, {Gai}, {Galleti}, {Garabato},
  {Garc{\'\i}a-Sedano}, {Garofalo}, {Garralda}, {Gavel}, {Gavras}, {Gerssen},
  {Geyer}, {Giacobbe}, {Gilmore}, {Girona}, {Giuffrida}, {Glass}, {Gomes},
  {Granvik}, {Gueguen}, {Guerrier}, {Guiraud}, {Guti{\'e}rrez-S{\'a}nchez},
  {Haigron}, {Hatzidimitriou}, {Hauser}, {Haywood}, {Heiter}, {Helmi}, {Heu},
  {Hilger}, {Hobbs}, {Hofmann}, {Holland}, {Huckle}, {Hypki}, {Icardi},
  {Jan{\ss}en}, {Jevardat de Fombelle}, {Jonker}, {Juh{\'a}sz}, {Julbe},
  {Karampelas}, {Kewley}, {Klar}, {Kochoska}, {Kohley}, {Kolenberg},
  {Kontizas}, {Kontizas}, {Koposov}, {Kordopatis}, {Kostrzewa-Rutkowska},
  {Koubsky}, {Lambert}, {Lanza}, {Lasne}, {Lavigne}, {Le Fustec}, {Le
  Poncin-Lafitte}, {Lebreton}, {Leccia}, {Leclerc}, {Lecoeur-Taibi},
  {Lenhardt}, {Leroux}, {Liao}, {Licata}, {Lindstr{\o}m}, {Lister}, {Livanou},
  {Lobel}, {L{\'o}pez}, {Managau}, {Mann}, {Mantelet}, {Marchal}, {Marchant},
  {Marconi}, {Marinoni}, {Marschalk{\'o}}, {Marshall}, {Martino}, {Marton},
  {Mary}, {Massari}, {Matijevi{\v{c}}}, {Mazeh}, {McMillan}, {Messina},
  {Michalik}, {Millar}, {Molina}, {Molinaro}, {Moln{\'a}r}, {Montegriffo},
  {Mor}, {Morbidelli}, {Morel}, {Morris}, {Mulone}, {Muraveva}, {Musella},
  {Nelemans}, {Nicastro}, {Noval}, {O'Mullane}, {Ord{\'e}novic},
  {Ord{\'o}{\~n}ez-Blanco}, {Osborne}, {Pagani}, {Pagano}, {Pailler},
  {Palacin}, {Palaversa}, {Panahi}, {Pawlak}, {Piersimoni}, {Pineau}, {Plachy},
  {Plum}, {Poggio}, {Poujoulet}, {Pr{\v{s}}a}, {Pulone}, {Racero}, {Ragaini},
  {Rambaux}, {Ramos-Lerate}, {Regibo}, {Reyl{\'e}}, {Riclet}, {Ripepi}, {Riva},
  {Rivard}, {Rixon}, {Roegiers}, {Roelens}, {Romero-G{\'o}mez}, {Rowell},
  {Royer}, {Ruiz-Dern}, {Sadowski}, {Sagrist{\`a} Sell{\'e}s}, {Sahlmann},
  {Salgado}, {Salguero}, {Sanna}, {Santana-Ros}, {Sarasso}, {Savietto},
  {Schultheis}, {Sciacca}, {Segol}, {Segovia}, {S{\'e}gransan}, {Shih},
  {Siltala}, {Silva}, {Smart}, {Smith}, {Solano}, {Solitro}, {Sordo}, {Soria
  Nieto}, {Souchay}, {Spagna}, {Spoto}, {Stampa}, {Steele},
  {Steidelm{\"u}ller}, {Stephenson}, {Stoev}, {Suess}, {Surdej}, {Szabados},
  {Szegedi-Elek}, {Tapiador}, {Taris}, {Tauran}, {Taylor}, {Teixeira},
  {Terrett}, {Teyssand ier}, {Thuillot}, {Titarenko}, {Torra Clotet}, {Turon},
  {Ulla}, {Utrilla}, {Uzzi}, {Vaillant}, {Valentini}, {Valette}, {van Elteren},
  {Van Hemelryck}, {van Leeuwen}, {Vaschetto}, {Vecchiato}, {Veljanoski},
  {Viala}, {Vicente}, {Vogt}, {von Essen}, {Voss}, {Votruba}, {Voutsinas},
  {Walmsley}, {Weiler}, {Wertz}, {Wevers}, {Wyrzykowski}, {Yoldas},
  {{\v{Z}}erjal}, {Ziaeepour}, {Zorec}, {Zschocke}, {Zucker}, {Zurbach}, \&
  {Zwitter}}]{GaiaDR22018}
{Gaia Collaboration}, {Brown}, A.~G.~A., {Vallenari}, A., {et~al.} 2018, \aap,
  616, A1

\bibitem[{{Gaia Collaboration} {et~al.}(2016){Gaia Collaboration}, {Brown},
  {Vallenari}, {Prusti}, {de Bruijne}, {Mignard}, {Drimmel}, {Babusiaux},
  {Bailer-Jones}, {Bastian}, {Biermann}, {Evans}, {Eyer}, {Jansen}, {Jordi},
  {Katz}, {Klioner}, {Lammers}, {Lindegren}, {Luri}, {O'Mullane}, {Panem},
  {Pourbaix}, {Randich}, {Sartoretti}, {Siddiqui}, {Soubiran}, {Valette}, {van
  Leeuwen}, {Walton}, {Aerts}, {Arenou}, {Cropper}, {H{\o}g}, {Lattanzi},
  {Grebel}, {Holland}, {Huc}, {Passot}, {Perryman}, {Bramante}, {Cacciari},
  {Casta{\~n}eda}, {Chaoul}, {Cheek}, {De Angeli}, {Fabricius}, {Guerra},
  {Hern{\'a}ndez}, {Jean-Antoine-Piccolo}, {Masana}, {Messineo}, {Mowlavi},
  {Nienartowicz}, {Ord{\'o}{\~n}ez-Blanco}, {Panuzzo}, {Portell}, {Richards},
  {Riello}, {Seabroke}, {Tanga}, {Th{\'e}venin}, {Torra}, {Els},
  {Gracia-Abril}, {Comoretto}, {Garcia-Reinaldos}, {Lock}, {Mercier},
  {Altmann}, {Andrae}, {Astraatmadja}, {Bellas-Velidis}, {Benson}, {Berthier},
  {Blomme}, {Busso}, {Carry}, {Cellino}, {Clementini}, {Cowell}, {Creevey},
  {Cuypers}, {Davidson}, {De Ridder}, {de Torres}, {Delchambre}, {Dell'Oro},
  {Ducourant}, {Fr{\'e}mat}, {Garc{\'\i}a-Torres}, {Gosset}, {Halbwachs},
  {Hambly}, {Harrison}, {Hauser}, {Hestroffer}, {Hodgkin}, {Huckle}, {Hutton},
  {Jasniewicz}, {Jordan}, {Kontizas}, {Korn}, {Lanzafame}, {Manteiga},
  {Moitinho}, {Muinonen}, {Osinde}, {Pancino}, {Pauwels}, {Petit},
  {Recio-Blanco}, {Robin}, {Sarro}, {Siopis}, {Smith}, {Smith}, {Sozzetti},
  {Thuillot}, {van Reeven}, {Viala}, {Abbas}, {Abreu Aramburu}, {Accart},
  {Aguado}, {Allan}, {Allasia}, {Altavilla}, {{\'A}lvarez}, {Alves},
  {Anderson}, {Andrei}, {Anglada Varela}, {Antiche}, {Antoja}, {Ant{\'o}n},
  {Arcay}, {Bach}, {Baker}, {Balaguer-N{\'u}{\~n}ez}, {Barache}, {Barata},
  {Barbier}, {Barblan}, {Barrado y Navascu{\'e}s}, {Barros}, {Barstow},
  {Becciani}, {Bellazzini}, {Bello Garc{\'\i}a}, {Belokurov}, {Bendjoya},
  {Berihuete}, {Bianchi}, {Bienaym{\'e}}, {Billebaud}, {Blagorodnova},
  {Blanco-Cuaresma}, {Boch}, {Bombrun}, {Borrachero}, {Bouquillon}, {Bourda},
  {Bouy}, {Bragaglia}, {Breddels}, {Brouillet}, {Br{\"u}semeister},
  {Bucciarelli}, {Burgess}, {Burgon}, {Burlacu}, {Busonero}, {Buzzi}, {Caffau},
  {Cambras}, {Campbell}, {Cancelliere}, {Cantat-Gaudin}, {Carlucci},
  {Carrasco}, {Castellani}, {Charlot}, {Charnas}, {Chiavassa}, {Clotet},
  {Cocozza}, {Collins}, {Costigan}, {Crifo}, {Cross}, {Crosta}, {Crowley},
  {Dafonte}, {Damerdji}, {Dapergolas}, {David}, {David}, {De Cat}, {de Felice},
  {de Laverny}, {De Luise}, {De March}, {de Martino}, {de Souza}, {Debosscher},
  {del Pozo}, {Delbo}, {Delgado}, {Delgado}, {Di Matteo}, {Diakite},
  {Distefano}, {Dolding}, {Dos Anjos}, {Drazinos}, {Duran}, {Dzigan},
  {Edvardsson}, {Enke}, {Evans}, {Eynard Bontemps}, {Fabre}, {Fabrizio},
  {Faigler}, {Falc{\~a}o}, {Farr{\`a}s Casas}, {Federici}, {Fedorets},
  {Fern{\'a}ndez-Hern{\'a}ndez}, {Fernique}, {Fienga}, {Figueras}, {Filippi},
  {Findeisen}, {Fonti}, {Fouesneau}, {Fraile}, {Fraser}, {Fuchs}, {Gai},
  {Galleti}, {Galluccio}, {Garabato}, {Garc{\'\i}a-Sedano}, {Garofalo},
  {Garralda}, {Gavras}, {Gerssen}, {Geyer}, {Gilmore}, {Girona}, {Giuffrida},
  {Gomes}, {Gonz{\'a}lez-Marcos}, {Gonz{\'a}lez-N{\'u}{\~n}ez},
  {Gonz{\'a}lez-Vidal}, {Granvik}, {Guerrier}, {Guillout}, {Guiraud},
  {G{\'u}rpide}, {Guti{\'e}rrez-S{\'a}nchez}, {Guy}, {Haigron},
  {Hatzidimitriou}, {Haywood}, {Heiter}, {Helmi}, {Hobbs}, {Hofmann}, {Holl},
  {Holland }, {Hunt}, {Hypki}, {Icardi}, {Irwin}, {Jevardat de Fombelle},
  {Jofr{\'e}}, {Jonker}, {Jorissen}, {Julbe}, {Karampelas}, {Kochoska},
  {Kohley}, {Kolenberg}, {Kontizas}, {Koposov}, {Kordopatis}, {Koubsky},
  {Krone-Martins}, {Kudryashova}, {Kull}, {Bachchan}, {Lacoste-Seris}, {Lanza},
  {Lavigne}, {Le Poncin-Lafitte}, {Lebreton}, {Lebzelter}, {Leccia}, {Leclerc},
  {Lecoeur-Taibi}, {Lemaitre}, {Lenhardt}, {Leroux}, {Liao}, {Licata},
  {Lindstr{\o}m}, {Lister}, {Livanou}, {Lobel}, {L{\"o}ffler}, {L{\'o}pez},
  {Lorenz}, {MacDonald}, {Magalh{\~a}es Fernandes}, {Managau}, {Mann},
  {Mantelet}, {Marchal}, {Marchant}, {Marconi}, {Marinoni}, {Marrese},
  {Marschalk{\'o}}, {Marshall}, {Mart{\'\i}n-Fleitas}, {Martino}, {Mary},
  {Matijevi{\v{c}}}, {Mazeh}, {McMillan}, {Messina}, {Michalik}, {Millar},
  {Mirand a}, {Molina}, {Molinaro}, {Molinaro}, {Moln{\'a}r}, {Moniez},
  {Montegriffo}, {Mor}, {Mora}, {Morbidelli}, {Morel}, {Morgenthaler},
  {Morris}, {Mulone}, {Muraveva}, {Musella}, {Narbonne}, {Nelemans},
  {Nicastro}, {Noval}, {Ord{\'e}novic}, {Ordieres-Mer{\'e}}, {Osborne},
  {Pagani}, {Pagano}, {Pailler}, {Palacin}, {Palaversa}, {Parsons}, {Pecoraro},
  {Pedrosa}, {Pentik{\"a}inen}, {Pichon}, {Piersimoni}, {Pineau}, {Plachy},
  {Plum}, {Poujoulet}, {Pr{\v{s}}a}, {Pulone}, {Ragaini}, {Rago}, {Rambaux},
  {Ramos-Lerate}, {Ranalli}, {Rauw}, {Read}, {Regibo}, {Reyl{\'e}}, {Ribeiro},
  {Rimoldini}, {Ripepi}, {Riva}, {Rixon}, {Roelens}, {Romero-G{\'o}mez},
  {Rowell}, {Royer}, {Ruiz-Dern}, {Sadowski}, {Sagrist{\`a} Sell{\'e}s},
  {Sahlmann}, {Salgado}, {Salguero}, {Sarasso}, {Savietto}, {Schultheis},
  {Sciacca}, {Segol}, {Segovia}, {Segransan}, {Shih}, {Smareglia}, {Smart},
  {Solano}, {Solitro}, {Sordo}, {Soria Nieto}, {Souchay}, {Spagna}, {Spoto},
  {Stampa}, {Steele}, {Steidelm{\"u}ller}, {Stephenson}, {Stoev}, {Suess},
  {S{\"u}veges}, {Surdej}, {Szabados}, {Szegedi-Elek}, {Tapiador}, {Taris},
  {Tauran}, {Taylor}, {Teixeira}, {Terrett}, {Tingley}, {Trager}, {Turon},
  {Ulla}, {Utrilla}, {Valentini}, {van Elteren}, {Van Hemelryck}, {van
  Leeuwen}, {Varadi}, {Vecchiato}, {Veljanoski}, {Via}, {Vicente}, {Vogt},
  {Voss}, {Votruba}, {Voutsinas}, {Walmsley}, {Weiler}, {Weingrill}, {Wevers},
  {Wyrzykowski}, {Yoldas}, {{\v{Z}}erjal}, {Zucker}, {Zurbach}, {Zwitter},
  {Alecu}, {Allen}, {Allende Prieto}, {Amorim}, {Anglada-Escud{\'e}},
  {Arsenijevic}, {Azaz}, {Balm}, {Beck}, {Bernstein}, {Bigot}, {Bijaoui},
  {Blasco}, {Bonfigli}, {Bono}, {Boudreault}, {Bressan}, {Brown}, {Brunet},
  {Bunclark}, {Buonanno}, {Butkevich}, {Carret}, {Carrion}, {Chemin},
  {Ch{\'e}reau}, {Corcione}, {Darmigny}, {de Boer}, {de Teodoro}, {de Zeeuw},
  {Delle Luche}, {Domingues}, {Dubath}, {Fodor}, {Fr{\'e}zouls}, {Fries},
  {Fustes}, {Fyfe}, {Gallardo}, {Gallegos}, {Gardiol}, {Gebran}, {Gomboc},
  {G{\'o}mez}, {Grux}, {Gueguen}, {Heyrovsky}, {Hoar}, {Iannicola}, {Isasi
  Parache}, {Janotto}, {Joliet}, {Jonckheere}, {Keil}, {Kim}, {Klagyivik},
  {Klar}, {Knude}, {Kochukhov}, {Kolka}, {Kos}, {Kutka}, {Lainey}, {LeBouquin},
  {Liu}, {Loreggia}, {Makarov}, {Marseille}, {Martayan}, {Martinez-Rubi},
  {Massart}, {Meynadier}, {Mignot}, {Munari}, {Nguyen}, {Nordlander}, {Ocvirk},
  {O'Flaherty}, {Olias Sanz}, {Ortiz}, {Osorio}, {Oszkiewicz}, {Ouzounis},
  {Palmer}, {Park}, {Pasquato}, {Peltzer}, {Peralta}, {P{\'e}turaud},
  {Pieniluoma}, {Pigozzi}, {Poels}, {Prat}, {Prod'homme}, {Raison}, {Rebordao},
  {Risquez}, {Rocca-Volmerange}, {Rosen}, {Ruiz-Fuertes}, {Russo}, {Sembay},
  {Serraller Vizcaino}, {Short}, {Siebert}, {Silva}, {Sinachopoulos}, {Slezak},
  {Soffel}, {Sosnowska}, {Strai{\v{z}}ys}, {ter Linden}, {Terrell}, {Theil},
  {Tiede}, {Troisi}, {Tsalmantza}, {Tur}, {Vaccari}, {Vachier}, {Valles}, {Van
  Hamme}, {Veltz}, {Virtanen}, {Wallut}, {Wichmann}, {Wilkinson}, {Ziaeepour},
  \& {Zschocke}}]{GaiaDR12016}
{Gaia Collaboration}, {Brown}, A.~G.~A., {Vallenari}, A., {et~al.} 2016, \aap,
  595, A2

\bibitem[{{Gomes da Silva} {et~al.}(2018){Gomes da Silva}, {Figueira},
  {Santos}, \& {Faria}}]{gomesdasilva2018}
{Gomes da Silva}, J., {Figueira}, P., {Santos}, N., \& {Faria}, J. 2018, The
  Journal of Open Source Software, 3, 667

\bibitem[{{Gomes da Silva} {et~al.}(2014){Gomes da Silva}, {Santos}, {Boisse},
  {Dumusque}, \& {Lovis}}]{gomesdasilva2014}
{Gomes da Silva}, J., {Santos}, N.~C., {Boisse}, I., {Dumusque}, X., \&
  {Lovis}, C. 2014, \aap, 566, A66

\bibitem[{{Gomes da Silva} {et~al.}(2011){Gomes da Silva}, {Santos}, {Bonfils},
  {Delfosse}, {Forveille}, \& {Udry}}]{gomesdasilva2011}
{Gomes da Silva}, J., {Santos}, N.~C., {Bonfils}, X., {et~al.} 2011, \aap, 534,
  A30+

\bibitem[{{Gomes da Silva} {et~al.}(2012){Gomes da Silva}, {Santos}, {Bonfils},
  {Delfosse}, {Forveille}, {Udry}, {Dumusque}, \& {Lovis}}]{gomesdasilva2012}
{Gomes da Silva}, J., {Santos}, N.~C., {Bonfils}, X., {et~al.} 2012, \aap, 541,
  A9

\bibitem[{{Gray} {et~al.}(2006){Gray}, {Corbally}, {Garrison}, {McFadden},
  {Bubar}, {McGahee}, {O'Donoghue}, \& {Knox}}]{gray2006}
{Gray}, R.~O., {Corbally}, C.~J., {Garrison}, R.~F., {et~al.} 2006, \aj, 132,
  161

\bibitem[{{Gray} {et~al.}(2003){Gray}, {Corbally}, {Garrison}, {McFadden}, \&
  {Robinson}}]{gray2003}
{Gray}, R.~O., {Corbally}, C.~J., {Garrison}, R.~F., {McFadden}, M.~T., \&
  {Robinson}, P.~E. 2003, \aj, 126, 2048

\bibitem[{{Green} {et~al.}(2015){Green}, {Schlafly}, {Finkbeiner}, {Rix},
  {Martin}, {Burgett}, {Draper}, {Flewelling}, {Hodapp}, {Kaiser}, {Kudritzki},
  {Magnier}, {Metcalfe}, {Price}, {Tonry}, \& {Wainscoat}}]{green2015}
{Green}, G.~M., {Schlafly}, E.~F., {Finkbeiner}, D.~P., {et~al.} 2015, \apj,
  810, 25

\bibitem[{{Hall} {et~al.}(2007){Hall}, {Lockwood}, \& {Skiff}}]{hall2007}
{Hall}, J.~C., {Lockwood}, G.~W., \& {Skiff}, B.~A. 2007, \aj, 133, 862

\bibitem[{{Hartmann} {et~al.}(1984){Hartmann}, {Soderblom}, {Noyes}, {Burnham},
  \& {Vaughan}}]{hartmann1984}
{Hartmann}, L., {Soderblom}, D.~R., {Noyes}, R.~W., {Burnham}, N., \&
  {Vaughan}, A.~H. 1984, \apj, 276, 254

\bibitem[{{Haywood} {et~al.}(2016){Haywood}, {Collier Cameron}, {Unruh},
  {Lovis}, {Lanza}, {Llama}, {Deleuil}, {Fares}, {Gillon}, {Moutou}, {Pepe},
  {Pollacco}, {Queloz}, \& {S{\'e}gransan}}]{haywood2016}
{Haywood}, R.~D., {Collier Cameron}, A., {Unruh}, Y.~C., {et~al.} 2016, \mnras,
  457, 3637

\bibitem[{{Henry} {et~al.}(1996){Henry}, {Soderblom}, {Donahue}, \&
  {Baliunas}}]{henry1996}
{Henry}, T.~J., {Soderblom}, D.~R., {Donahue}, R.~A., \& {Baliunas}, S.~L.
  1996, \aj, 111, 439

\bibitem[{Hunter(2007)}]{matplotlib2007}
Hunter, J.~D. 2007, Computing in Science \& Engineering, 9, 90

\bibitem[{{Isaacson} \& {Fischer}(2010)}]{isaacson2010}
{Isaacson}, H. \& {Fischer}, D. 2010, \apj, 725, 875

\bibitem[{{Jenkins} {et~al.}(2008){Jenkins}, {Jones}, {Pavlenko}, {Pinfield},
  {Barnes}, \& {Lyubchik}}]{jenkins2008}
{Jenkins}, J.~S., {Jones}, H.~R.~A., {Pavlenko}, Y., {et~al.} 2008, \aap, 485,
  571

\bibitem[{{Jenkins} {et~al.}(2006){Jenkins}, {Jones}, {Tinney}, {Butler},
  {McCarthy}, {Marcy}, {Pinfield}, {Carter}, \& {Penny}}]{jenkins2006}
{Jenkins}, J.~S., {Jones}, H.~R.~A., {Tinney}, C.~G., {et~al.} 2006, \mnras,
  372, 163

\bibitem[{{Jenkins} {et~al.}(2011){Jenkins}, {Murgas}, {Rojo}, {Jones},
  {Day-Jones}, {Jones}, {Clarke}, {Ruiz}, \& {Pinfield}}]{jenkins2011}
{Jenkins}, J.~S., {Murgas}, F., {Rojo}, P., {et~al.} 2011, \aap, 531, A8

\bibitem[{{Jurgenson} {et~al.}(2016){Jurgenson}, {Fischer}, {McCracken},
  {Sawyer}, {Szymkowiak}, {Davis}, {Muller}, \& {Santoro}}]{jurgenson2016}
{Jurgenson}, C., {Fischer}, D., {McCracken}, T., {et~al.} 2016, in Society of
  Photo-Optical Instrumentation Engineers (SPIE) Conference Series, Vol. 9908,
  \procspie, 99086T

\bibitem[{Kass \& Raftery(1995)}]{kass1995}
Kass, R.~E. \& Raftery, A.~E. 1995, Journal of the American Statistical
  Association, 90, 773

\bibitem[{{Kippenhahn} {et~al.}(2012){Kippenhahn}, {Weigert}, \&
  {Weiss}}]{kippenhahn2012}
{Kippenhahn}, R., {Weigert}, A., \& {Weiss}, A. 2012, {Stellar Structure and
  Evolution}

\bibitem[{{Kurucz}(1993)}]{kurucz1993}
{Kurucz}, R.~L. 1993, Physica Scripta Volume T, 47, 110

\bibitem[{{Leighton}(1959)}]{leighton1959}
{Leighton}, R.~B. 1959, \apj, 130, 366

\bibitem[{{Lorenzo-Oliveira} {et~al.}(2018){Lorenzo-Oliveira}, {Freitas},
  {Mel{\'e}ndez}, {Bedell}, {Ram{\'\i}rez}, {Bean}, {Asplund}, {Spina},
  {Dreizler}, {Alves-Brito}, \& {Casagrande}}]{lorenzooliveira2018}
{Lorenzo-Oliveira}, D., {Freitas}, F.~C., {Mel{\'e}ndez}, J., {et~al.} 2018,
  \aap, 619, A73

\bibitem[{{Lorenzo-Oliveira} {et~al.}(2016){Lorenzo-Oliveira}, {Porto de
  Mello}, \& {Schiavon}}]{lorenzooliveira2016}
{Lorenzo-Oliveira}, D., {Porto de Mello}, G.~F., \& {Schiavon}, R.~P. 2016,
  \aap, 594, L3

\bibitem[{{Lovis} {et~al.}(2011){Lovis}, {Dumusque}, {Santos}, {Bouchy},
  {Mayor}, {Pepe}, {Queloz}, {S{\'e}gransan}, \& {Udry}}]{lovis2011}
{Lovis}, C., {Dumusque}, X., {Santos}, N.~C., {et~al.} 2011, ArXiv e-prints
  [\eprint[arXiv]{1107.5325}]

\bibitem[{{Luhn} {et~al.}(2020){Luhn}, {Wright}, {Howard}, \&
  {Isaacson}}]{luhn2020}
{Luhn}, J.~K., {Wright}, J.~T., {Howard}, A.~W., \& {Isaacson}, H. 2020, \aj,
  159, 235

\bibitem[{{Mamajek} \& {Hillenbrand}(2008)}]{mamajek2008}
{Mamajek}, E.~E. \& {Hillenbrand}, L.~A. 2008, \apj, 687, 1264

\bibitem[{{Marshall} {et~al.}(2006){Marshall}, {Robin}, {Reyl{\'e}},
  {Schultheis}, \& {Picaud}}]{marshall2006}
{Marshall}, D.~J., {Robin}, A.~C., {Reyl{\'e}}, C., {Schultheis}, M., \&
  {Picaud}, S. 2006, \aap, 453, 635

\bibitem[{{Mayor} {et~al.}(2011){Mayor}, {Marmier}, {Lovis}, {Udry},
  {S{\'e}gransan}, {Pepe}, {Benz}, {Bertaux}, {Bouchy}, {Dumusque}, {Lo Curto},
  {Mordasini}, {Queloz}, \& {Santos}}]{mayor2011}
{Mayor}, M., {Marmier}, M., {Lovis}, C., {et~al.} 2011, ArXiv e-prints
  [\eprint[arXiv]{1109.2497}]

\bibitem[{{Mayor} {et~al.}(2003){Mayor}, {Pepe}, {Queloz}, {Bouchy},
  {Rupprecht}, {Lo Curto}, {Avila}, {Benz}, {Bertaux}, {Bonfils}, {Dall},
  {Dekker}, {Delabre}, {Eckert}, {Fleury}, {Gilliotte}, {Gojak}, {Guzman},
  {Kohler}, {Lizon}, {Longinotti}, {Lovis}, {Megevand}, {Pasquini}, {Reyes},
  {Sivan}, {Sosnowska}, {Soto}, {Udry}, {van Kesteren}, {Weber}, \&
  {Weilenmann}}]{mayor2003}
{Mayor}, M., {Pepe}, F., {Queloz}, D., {et~al.} 2003, The Messenger, 114, 20

\bibitem[{{Meunier} {et~al.}(2010){Meunier}, {Desort}, \&
  {Lagrange}}]{meunier2010}
{Meunier}, N., {Desort}, M., \& {Lagrange}, A.-M. 2010, \aap, 512, A39+

\bibitem[{{Meunier} {et~al.}(2017){Meunier}, {Lagrange}, {Mbemba Kabuiku},
  {Alex}, {Mignon}, \& {Borgniet}}]{meunier2017a}
{Meunier}, N., {Lagrange}, A.~M., {Mbemba Kabuiku}, L., {et~al.} 2017, \aap,
  597, A52

\bibitem[{{Middelkoop}(1982)}]{middelkoop1982}
{Middelkoop}, F. 1982, \aap, 107, 31

\bibitem[{{Middelkoop} \& {Zwaan}(1981)}]{middelkoop1981}
{Middelkoop}, F. \& {Zwaan}, C. 1981, \aap, 101, 26

\bibitem[{{Milbourne} {et~al.}(2019){Milbourne}, {Haywood}, {Phillips}, {Saar},
  {Cegla}, {Cameron}, {Costes}, {Dumusque}, {Langellier}, {Latham},
  {Maldonado}, {Malavolta}, {Mortier}, {Palumbo}, {Thompson}, {Watson},
  {Bouchy}, {Buchschacher}, {Cecconi}, {Charbonneau}, {Cosentino}, {Ghedina},
  {Glenday}, {Gonzalez}, {Li}, {Lodi}, {L{\'o}pez-Morales}, {Lovis}, {Mayor},
  {Micela}, {Molinari}, {Pepe}, {Piotto}, {Rice}, {Sasselov}, {S{\'e}gransan},
  {Sozzetti}, {Szentgyorgyi}, {Udry}, \& {Walsworth}}]{milbourne2019}
{Milbourne}, T.~W., {Haywood}, R.~D., {Phillips}, D.~F., {et~al.} 2019, \apj,
  874, 107

\bibitem[{{Mortier} {et~al.}(2014){Mortier}, {Sousa}, {Adibekyan}, {Brand
  {\~a}o}, \& {Santos}}]{mortier2014}
{Mortier}, A., {Sousa}, S.~G., {Adibekyan}, V.~Z., {Brand {\~a}o}, I.~M., \&
  {Santos}, N.~C. 2014, \aap, 572, A95

\bibitem[{{Noyes} {et~al.}(1984){Noyes}, {Hartmann}, {Baliunas}, {Duncan}, \&
  {Vaughan}}]{noyes1984}
{Noyes}, R.~W., {Hartmann}, L.~W., {Baliunas}, S.~L., {Duncan}, D.~K., \&
  {Vaughan}, A.~H. 1984, \apj, 279, 763

\bibitem[{{Oshagh} {et~al.}(2012){Oshagh}, {Bou{\'e}}, {Haghighipour},
  {Montalto}, {Figueira}, \& {Santos}}]{oshagh2012}
{Oshagh}, M., {Bou{\'e}}, G., {Haghighipour}, N., {et~al.} 2012, \aap, 540, A62

\bibitem[{{Oshagh} {et~al.}(2013){Oshagh}, {Santos}, {Boisse}, {Bou{\'e}},
  {Montalto}, {Dumusque}, \& {Haghighipour}}]{oshagh2013}
{Oshagh}, M., {Santos}, N.~C., {Boisse}, I., {et~al.} 2013, \aap, 556, A19

\bibitem[{{Paxton} {et~al.}(2011){Paxton}, {Bildsten}, {Dotter}, {Herwig},
  {Lesaffre}, \& {Timmes}}]{Paxton2011}
{Paxton}, B., {Bildsten}, L., {Dotter}, A., {et~al.} 2011, \apjs, 192, 3

\bibitem[{{Paxton} {et~al.}(2013){Paxton}, {Cantiello}, {Arras}, {Bildsten},
  {Brown}, {Dotter}, {Mankovich}, {Montgomery}, {Stello}, {Timmes}, \&
  {Townsend}}]{Paxton2013}
{Paxton}, B., {Cantiello}, M., {Arras}, P., {et~al.} 2013, \apjs, 208, 4

\bibitem[{{Paxton} {et~al.}(2015){Paxton}, {Marchant}, {Schwab}, {Bauer},
  {Bildsten}, {Cantiello}, {Dessart}, {Farmer}, {Hu}, {Langer}, {Townsend},
  {Townsley}, \& {Timmes}}]{Paxton2015}
{Paxton}, B., {Marchant}, P., {Schwab}, J., {et~al.} 2015, \apjs, 220, 15

\bibitem[{{Pecaut} \& {Mamajek}(2013)}]{pecaut2013}
{Pecaut}, M.~J. \& {Mamajek}, E.~E. 2013, \apjs, 208, 9

\bibitem[{Pedregosa {et~al.}(2011)Pedregosa, Varoquaux, Gramfort, Michel,
  Thirion, Grisel, Blondel, Prettenhofer, Weiss, Dubourg, Vanderplas, Passos,
  Cournapeau, Brucher, Perrot, \& Duchesnay}]{sklearn2011}
Pedregosa, F., Varoquaux, G., Gramfort, A., {et~al.} 2011, Journal of Machine
  Learning Research, 12, 2825

\bibitem[{{Pepe} {et~al.}(2020){Pepe}, {Cristiani}, {Rebolo}, {Santos},
  {Dekker}, {Cabral}, {Di Marcantonio}, {Figueira}, {Lo Curto}, {Lovis},
  {Mayor}, {M{\'e}gevand}, {Molaro}, {Riva}, {Zapatero Osorio}, {Amate},
  {Manescau}, {Pasquini}, {Zerbi}, {Adibekyan}, {Abreu}, {Affolter}, {Alibert},
  {Aliverti}, {Allart}, {Allende Prieto}, {{\'A}lvarez}, {Alves}, {Avila},
  {Baldini}, {Bandy}, {Barros}, {Benz}, {Bianco}, {Borsa}, {Bourrier},
  {Bouchy}, {Broeg}, {Calderone}, {Cirami}, {Coelho}, {Conconi}, {Coretti},
  {Cumani}, {Cupani}, {D'Odorico}, {Damasso}, {Deiries}, {Delabre},
  {Demangeon}, {Dumusque}, {Ehrenreich}, {Faria}, {Fragoso}, {Genolet},
  {Genoni}, {G{\'e}nova Santos}, {Gonz{\'a}lez Hern{\'a}ndez}, {Hughes},
  {Iwert}, {Kerber}, {Knudstrup}, {Landoni}, {Lavie}, {Lillo-Box}, {Lizon},
  {Maire}, {Martins}, {Mehner}, {Micela}, {Modigliani}, {Monteiro}, {Monteiro},
  {Moschetti}, {Murphy}, {Nunes}, {Oggioni}, {Oliveira}, {Oshagh}, {Pall{\'e}},
  {Pariani}, {Poretti}, {Rasilla}, {Rebord{\~a}o}, {Redaelli}, {Santana
  Tschudi}, {Santin}, {Santos}, {S{\'e}gransan}, {Schmidt}, {Segovia},
  {Sosnowska}, {Sozzetti}, {Sousa}, {Span{\`o}}, {Su{\'a}rez Mascare{\~n}o},
  {Tabernero}, {Tenegi}, {Udry}, \& {Zanutta}}]{pepe2020}
{Pepe}, F., {Cristiani}, S., {Rebolo}, R., {et~al.} 2020, arXiv e-prints,
  arXiv:2010.00316

\bibitem[{{Pepe} {et~al.}(2002){Pepe}, {Mayor}, {Rupprecht}, {Avila},
  {Ballester}, {Beckers}, {Benz}, {Bertaux}, {Bouchy}, {Buzzoni}, {Cavadore},
  {Deiries}, {Dekker}, {Delabre}, {D'Odorico}, {Eckert}, {Fischer}, {Fleury},
  {George}, {Gilliotte}, {Gojak}, {Guzman}, {Koch}, {Kohler}, {Kotzlowski},
  {Lacroix}, {Le Merrer}, {Lizon}, {Lo Curto}, {Longinotti}, {Megevand},
  {Pasquini}, {Petitpas}, {Pichard}, {Queloz}, {Reyes}, {Richaud}, {Sivan},
  {Sosnowska}, {Soto}, {Udry}, {Ureta}, {van Kesteren}, {Weber}, {Weilenmann},
  {Wicenec}, {Wieland}, {Christensen-Dalsgaard}, {Dravins}, {Hatzes},
  {K{\"u}rster}, {Paresce}, \& {Penny}}]{pepe2002}
{Pepe}, F., {Mayor}, M., {Rupprecht}, G., {et~al.} 2002, The Messenger, 110, 9

\bibitem[{{Perryman} {et~al.}(1997){Perryman}, {Lindegren}, {Kovalevsky},
  {Hog}, {Bastian}, {Bernacca}, {Creze}, {Donati}, {Grenon}, {Grewing}, {van
  Leeuwen}, {van der Marel}, {Mignard}, {Murray}, {Le Poole}, {Schrijver},
  {Turon}, {Arenou}, {Froeschle}, \& {Petersen}}]{hipparcos1997}
{Perryman}, M.~A.~C., {Lindegren}, L., {Kovalevsky}, J., {et~al.} 1997, \aap,
  500, 501

\bibitem[{{Pont} {et~al.}(2007){Pont}, {Gilliland}, {Moutou}, {Charbonneau},
  {Bouchy}, {Brown}, {Mayor}, {Queloz}, {Santos}, \& {Udry}}]{pont2007}
{Pont}, F., {Gilliland}, R.~L., {Moutou}, C., {et~al.} 2007, \aap, 476, 1347

\bibitem[{{Pont} {et~al.}(2008){Pont}, {Knutson}, {Gilliland}, {Moutou}, \&
  {Charbonneau}}]{pont2008}
{Pont}, F., {Knutson}, H., {Gilliland}, R.~L., {Moutou}, C., \& {Charbonneau},
  D. 2008, \mnras, 385, 109

\bibitem[{{Price-Whelan} {et~al.}(2018){Price-Whelan}, {Sip{\H{o}}cz},
  {G{\"u}nther}, {Lim}, {Crawford}, {Conseil}, {Shupe}, {Craig}, {Dencheva},
  {Ginsburg}, {VanderPlas}, {Bradley}, {P{\'e}rez-Su{\'a}rez}, {de Val-Borro},
  {Paper Contributors}, {Aldcroft}, {Cruz}, {Robitaille}, {Tollerud},
  {Coordination Committee}, {Ardelean}, {Babej}, {Bach}, {Bachetti}, {Bakanov},
  {Bamford}, {Barentsen}, {Barmby}, {Baumbach}, {Berry}, {Biscani}, {Boquien},
  {Bostroem}, {Bouma}, {Brammer}, {Bray}, {Breytenbach}, {Buddelmeijer},
  {Burke}, {Calderone}, {Cano Rodr{\'\i}guez}, {Cara}, {Cardoso}, {Cheedella},
  {Copin}, {Corrales}, {Crichton}, {D{\textquoteright}Avella}, {Deil},
  {Depagne}, {Dietrich}, {Donath}, {Droettboom}, {Earl}, {Erben}, {Fabbro},
  {Ferreira}, {Finethy}, {Fox}, {Garrison}, {Gibbons}, {Goldstein}, {Gommers},
  {Greco}, {Greenfield}, {Groener}, {Grollier}, {Hagen}, {Hirst}, {Homeier},
  {Horton}, {Hosseinzadeh}, {Hu}, {Hunkeler}, {Ivezi{\'c}}, {Jain}, {Jenness},
  {Kanarek}, {Kendrew}, {Kern}, {Kerzendorf}, {Khvalko}, {King}, {Kirkby},
  {Kulkarni}, {Kumar}, {Lee}, {Lenz}, {Littlefair}, {Ma}, {Macleod},
  {Mastropietro}, {McCully}, {Montagnac}, {Morris}, {Mueller}, {Mumford},
  {Muna}, {Murphy}, {Nelson}, {Nguyen}, {Ninan}, {N{\"o}the}, {Ogaz}, {Oh},
  {Parejko}, {Parley}, {Pascual}, {Patil}, {Patil}, {Plunkett}, {Prochaska},
  {Rastogi}, {Reddy Janga}, {Sabater}, {Sakurikar}, {Seifert}, {Sherbert},
  {Sherwood-Taylor}, {Shih}, {Sick}, {Silbiger}, {Singanamalla}, {Singer},
  {Sladen}, {Sooley}, {Sornarajah}, {Streicher}, {Teuben}, {Thomas},
  {Tremblay}, {Turner}, {Terr{\'o}n}, {van Kerkwijk}, {de la Vega}, {Watkins},
  {Weaver}, {Whitmore}, {Woillez}, {Zabalza}, \& {Contributors}}]{astropy:2018}
{Price-Whelan}, A.~M., {Sip{\H{o}}cz}, B.~M., {G{\"u}nther}, H.~M., {et~al.}
  2018, \aj, 156, 123

\bibitem[{{Queloz} {et~al.}(2001){Queloz}, {Henry}, {Sivan}, {Baliunas},
  {Beuzit}, {Donahue}, {Mayor}, {Naef}, {Perrier}, \& {Udry}}]{queloz2001}
{Queloz}, D., {Henry}, G.~W., {Sivan}, J.~P., {et~al.} 2001, \aap, 379, 279

\bibitem[{{Rutten}(1984)}]{rutten1984}
{Rutten}, R.~G.~M. 1984, \aap, 130, 353

\bibitem[{{Saar} \& {Donahue}(1997)}]{saar1997}
{Saar}, S.~H. \& {Donahue}, R.~A. 1997, \apj, 485, 319

\bibitem[{{Santos} {et~al.}(2000){Santos}, {Mayor}, {Naef}, {Pepe}, {Queloz},
  {Udry}, \& {Blecha}}]{santos2000}
{Santos}, N.~C., {Mayor}, M., {Naef}, D., {et~al.} 2000, \aap, 361, 265

\bibitem[{{Santos} {et~al.}(2014){Santos}, {Mortier}, {Faria}, {Dumusque},
  {Adibekyan}, {Delgado-Mena}, {Figueira}, {Benamati}, {Boisse}, {Cunha},
  {Gomes da Silva}, {Lo Curto}, {Lovis}, {Martins}, {Mayor}, {Melo}, {Oshagh},
  {Pepe}, {Queloz}, {Santerne}, {S{\'e}gransan}, {Sozzetti}, {Sousa}, \&
  {Udry}}]{santos2014}
{Santos}, N.~C., {Mortier}, A., {Faria}, J.~P., {et~al.} 2014, \aap, 566, A35

\bibitem[{{Santos} {et~al.}(2013){Santos}, {Sousa}, {Mortier}, {Neves},
  {Adibekyan}, {Tsantaki}, {Delgado Mena}, {Bonfils}, {Israelian}, {Mayor}, \&
  {Udry}}]{santos2013}
{Santos}, N.~C., {Sousa}, S.~G., {Mortier}, A., {et~al.} 2013, \aap, 556, A150

\bibitem[{{Schatzman}(1962)}]{schatzman1962}
{Schatzman}, E. 1962, Annales d'Astrophysique, 25, 18

\bibitem[{{Schwarz}(1978)}]{schwarz1978}
{Schwarz}, G. 1978, Annals of Statistics, 6, 461

\bibitem[{{Skumanich}(1972)}]{skumanich1972}
{Skumanich}, A. 1972, \apj, 171, 565

\bibitem[{{Skumanich} {et~al.}(1975){Skumanich}, {Smythe}, \&
  {Frazier}}]{skumanich1975}
{Skumanich}, A., {Smythe}, C., \& {Frazier}, E.~N. 1975, \apj, 200, 747

\bibitem[{{Sneden}(1973)}]{sneden1973}
{Sneden}, C.~A. 1973, PhD thesis, THE UNIVERSITY OF TEXAS AT AUSTIN.

\bibitem[{{Sousa} {et~al.}(2018){Sousa}, {Adibekyan}, {Delgado-Mena}, {Santos},
  {Andreasen}, {Ferreira}, {Tsantaki}, {Barros}, {Demangeon}, {Israelian},
  {Faria}, {Figueira}, {Mortier}, {Brand{\~a}o}, {Montalto}, {Rojas-Ayala}, \&
  {Santerne}}]{sousa2018}
{Sousa}, S.~G., {Adibekyan}, V., {Delgado-Mena}, E., {et~al.} 2018, \aap, 620,
  A58

\bibitem[{{Sousa} {et~al.}(2015){Sousa}, {Santos}, {Adibekyan}, {Delgado-Mena},
  \& {Israelian}}]{sousa2015}
{Sousa}, S.~G., {Santos}, N.~C., {Adibekyan}, V., {Delgado-Mena}, E., \&
  {Israelian}, G. 2015, \aap, 577, A67

\bibitem[{{Sousa} {et~al.}(2014){Sousa}, {Santos}, {Adibekyan}, {Delgado-Mena},
  {Tabernero}, {Gonz{\'a}lez Hern{\'a}ndez}, {Montes}, {Smiljanic}, {Korn},
  {Bergemann}, {Soubiran}, \& {Mikolaitis}}]{sousa2014}
{Sousa}, S.~G., {Santos}, N.~C., {Adibekyan}, V., {et~al.} 2014, \aap, 561, A21

\bibitem[{{Sousa} {et~al.}(2007){Sousa}, {Santos}, {Israelian}, {Mayor}, \&
  {Monteiro}}]{sousa2007}
{Sousa}, S.~G., {Santos}, N.~C., {Israelian}, G., {Mayor}, M., \& {Monteiro},
  M.~J.~P.~F.~G. 2007, \aap, 469, 783

\bibitem[{{Sousa} {et~al.}(2011){Sousa}, {Santos}, {Israelian}, {Mayor}, \&
  {Udry}}]{sousa2011}
{Sousa}, S.~G., {Santos}, N.~C., {Israelian}, G., {Mayor}, M., \& {Udry}, S.
  2011, \aap, 533, A141

\bibitem[{{Sousa} {et~al.}(2008){Sousa}, {Santos}, {Mayor}, {Udry},
  {Casagrande}, {Israelian}, {Pepe}, {Queloz}, \& {Monteiro}}]{sousa2008}
{Sousa}, S.~G., {Santos}, N.~C., {Mayor}, M., {et~al.} 2008, \aap, 487, 373

\bibitem[{{Strassmeier} {et~al.}(2000){Strassmeier}, {Washuettl}, {Granzer},
  {Scheck}, \& {Weber}}]{strassmeier2000}
{Strassmeier}, K., {Washuettl}, A., {Granzer}, T., {Scheck}, M., \& {Weber}, M.
  2000, \aaps, 142, 275

\bibitem[{{Su{\'a}rez Mascare{\~n}o} {et~al.}(2015){Su{\'a}rez Mascare{\~n}o},
  {Rebolo}, {Gonz{\'a}lez Hern{\'a}ndez}, \& {Esposito}}]{suarezmascareno2015}
{Su{\'a}rez Mascare{\~n}o}, A., {Rebolo}, R., {Gonz{\'a}lez Hern{\'a}ndez},
  J.~I., \& {Esposito}, M. 2015, \mnras, 452, 2745

\bibitem[{{Su{\'a}rez Mascare{\~n}o} {et~al.}(2016){Su{\'a}rez Mascare{\~n}o},
  {Rebolo}, {Gonz{\'a}lez Hern{\'a}ndez}, \& {Esposito}}]{suarezmascareno2016}
{Su{\'a}rez Mascare{\~n}o}, A., {Rebolo}, R., {Gonz{\'a}lez Hern{\'a}ndez},
  J.~I., \& {Esposito}, M. 2016, \mnras, 457, 2604

\bibitem[{{Tsantaki} {et~al.}(2013){Tsantaki}, {Sousa}, {Adibekyan}, {Santos},
  {Mortier}, \& {Israelian}}]{tsantaki2013}
{Tsantaki}, M., {Sousa}, S.~G., {Adibekyan}, V.~Z., {et~al.} 2013, \aap, 555,
  A150

\bibitem[{{van Leeuwen}(2007)}]{hipparcos2007}
{van Leeuwen}, F. 2007, \aap, 474, 653

\bibitem[{{van Saders} {et~al.}(2016){van Saders}, {Ceillier}, {Metcalfe},
  {Silva Aguirre}, {Pinsonneault}, {Garc{\'\i}a}, {Mathur}, \&
  {Davies}}]{vanSanders2016}
{van Saders}, J.~L., {Ceillier}, T., {Metcalfe}, T.~S., {et~al.} 2016, \nat,
  529, 181

\bibitem[{{Vaughan} \& {Preston}(1980)}]{vaughan1980}
{Vaughan}, A.~H. \& {Preston}, G.~W. 1980, \pasp, 92, 385

\bibitem[{{Vaughan} {et~al.}(1978){Vaughan}, {Preston}, \&
  {Wilson}}]{vaughan1978}
{Vaughan}, A.~H., {Preston}, G.~W., \& {Wilson}, O.~C. 1978, \pasp, 90, 267

\bibitem[{Virtanen {et~al.}(2020)Virtanen, Gommers, Oliphant, Haberland, Reddy,
  Cournapeau, Burovski, Peterson, Weckesser, Bright, {van der Walt}, Brett,
  Wilson, Millman, Mayorov, Nelson, Jones, Kern, Larson, Carey, Polat, Feng,
  Moore, {VanderPlas}, Laxalde, Perktold, Cimrman, Henriksen, Quintero, Harris,
  Archibald, Ribeiro, Pedregosa, {van Mulbregt}, \& {SciPy 1.0
  Contributors}}]{scipy2020}
Virtanen, P., Gommers, R., Oliphant, T.~E., {et~al.} 2020, Nature Methods, 17,
  261

\bibitem[{{Weber} \& {Davis}(1967)}]{weber1967}
{Weber}, E.~J. \& {Davis}, Leverett, J. 1967, \apj, 148, 217

\bibitem[{{White} \& {Livingston}(1981)}]{white1981}
{White}, O.~R. \& {Livingston}, W.~C. 1981, \apj, 249, 798

\bibitem[{{Wilson}(1968)}]{wilson1968}
{Wilson}, O.~C. 1968, \apj, 153, 221

\bibitem[{{Wright}(2004)}]{wright2004b}
{Wright}, J.~T. 2004, \aj, 128, 1273

\bibitem[{{Wright} {et~al.}(2004){Wright}, {Marcy}, {Butler}, \&
  {Vogt}}]{wright2004}
{Wright}, J.~T., {Marcy}, G.~W., {Butler}, R.~P., \& {Vogt}, S.~S. 2004, \apjs,
  152, 261

\bibitem[{{Zechmeister} \& {K{\"u}rster}(2009)}]{zechmeister2009_gls}
{Zechmeister}, M. \& {K{\"u}rster}, M. 2009, \aap, 496, 577

\bibitem[{{Zechmeister} {et~al.}(2009){Zechmeister}, {K{\"u}rster}, \&
  {Endl}}]{zechmeister2009}
{Zechmeister}, M., {K{\"u}rster}, M., \& {Endl}, M. 2009, \aap, 505, 859

\bibitem[{{Zhao} {et~al.}(2013){Zhao}, {Oswalt}, {Zhao}, {Lu}, {Luo}, \&
  {Zhang}}]{zhao2013}
{Zhao}, J.~K., {Oswalt}, T.~D., {Zhao}, G., {et~al.} 2013, \aj, 145, 140

\end{thebibliography}

\begin{appendix}
\section{Measuring chromospheric activity with ACTIN}
\label{app:actin}
We computed activity indices based on the \ion{Ca}{ii} H\&K lines using the open-source package \verb+ACTIN+\footnote{\url{https://github.com/gomesdasilva/ACTIN}} \citep{gomesdasilva2018}.
In the following we briefly describe how fluxes and activity indices are calculated by the code.

\subsection{Flux calculation}
The wavelength space is a grid where each step (pixel) has a finite size.
For a given bandpass, the number of pixels that will fall inside it will be an integer number, and the associated wavelength range between the reddest and the bluest pixel inside the bandpass will be, generally, smaller than the bandpass range. Consequently, the value of the flux inside a bandpass tends to be underestimated unless the flux on the fraction of pixels on the redder and bluer part of the bandpass is considered.

In \verb+ACTIN+ this is taken into account by linear interpolation using the \verb+scipy+ \verb+interp1d+ module.
The step used in the interpolation was $1\cdot10^{-5}$~$\AA$ ($100 \times$ smaller than the wavelength resolution of 0.01 $\AA$ for HARPS s1d spectra).
The flux per wavelength step, $F_\lambda$, is obtained by multiplying the interpolated flux, $F_{\lambda_\text{inter}}$, by the the ratio of the interpolation step to the average original wavelength resolution, $R$, at the bandpass region:
\begin{eqnarray}
    F_\lambda = F_{\lambda_\text{inter}} \cdot R.
\end{eqnarray}
The total integrated flux inside the effective bandwidth $\Delta\lambda = \lambda_\text{max} - \lambda_\text{min}$ is then
\begin{eqnarray}
        F_{\Delta\lambda} = \frac{1}{\Delta\lambda} \sum^{\lambda_\text{max}}_{\lambda=\lambda_\text{min}} B_\lambda \cdot F_\lambda
,\end{eqnarray}
where $B_\lambda$ is the bandpass function and $F_\lambda$ the flux in each interpolated step.

The errors on flux are derived via error propagation and results in
\begin{eqnarray}
    \sigma_{F_{\Delta\lambda}} = \frac{1}{\Delta\lambda} \sqrt{ \sum^{\lambda_\text{max}}_{\lambda=\lambda_\text{min}} B_\lambda^2 \cdot \sigma_{F_\lambda}^2},
\end{eqnarray}
where $\sigma_{F_\lambda} = \sqrt{F_\lambda}$ is the photon noise uncertainty on the flux.

Square and triangular bandpass filters come pre-installed and can be used to calculate the fluxes.
The code flags unphysical values of negative flux in the bandpasses that may occur, mainly in the bluer region of the \ion{Ca}{ii} H\&K lines, when using very low-signal-to-noise-ratio spectra.

\subsection{Indices calculation}
\verb+ACTIN+ calculates indices by dividing the average flux in the activity sensitive lines by the average flux in reference regions,
\begin{eqnarray}
        I = \frac{\sum^N_{i=1} F_i}{\sum^M_{j=1} R_j},
\end{eqnarray}
where $F_i$ is the flux in the activity sensitive line $i$, $R_j$ the flux in reference region $j$, $N$ the number of activity lines used, and $M$ the number of reference regions.

The flux errors only take into account photon noise.
The index (photon) error is, by error propagation,
\begin{eqnarray}
        \sigma_I = \frac{1}{\sum^M_{j=1} R_j} \sqrt{\sum^N_{i=1} \sigma_{F_i}^2 + I^2 \sum^M_{j=1} \sigma_{R_j}^2}.
\end{eqnarray}
The code is able to calculate indices with any number of lines and reference bands as long as they respect the wavelength range of the spectra used.
It comes with four widely used activity indices pre-installed, $S_\text{\ion{Ca}{ii}}$, $I_{\text{H}\alpha}$, $I_{\ion{Na}{i}}$ and $I_{\ion{He}{i}}$, using line parameters and bandpasses described in \citet{gomesdasilva2011} (except for $S_\text{CaII}$ which uses a triangular bandpass filter instead of the square bandpass used by the authors), and it is very easy to modify line parameters and add more lines and indices to the pre-installed list.

\section{Using the bivariate KDE map to obtain activity variability and simulate activity levels and variability for stellar populations}
\label{app:2d_kde}

\subsection{Obtaining the activity variability probability distribution for a given activity level}
If the activity level of a star is known, it is possible to estimate the probability density function of the activity variability for that activity level from the bivariate KDE distribution of $\log \sigma(R_5)$ and $\log R'_\text{HK}$ derived in \S \ref{sec:rhk_std_vs_rhk}.

We use $\alpha$ Cen B as an example.
This star, a K1 dwarf, has a well-characterised activity level of $\log R'_\text{HK} = -4.96$ dex and variability of $\log \sigma (R_5) = -0.907$ dex.
We can therefore look up on the KDE map the closest bin of activity to this value and obtain the KDE of the variability (the probability density function) in that bin.
We used a bin resolution of $\sim$0.01 dex and selected only main sequence K dwarfs.

\begin{figure}
        \resizebox{\hsize}{!}{\includegraphics{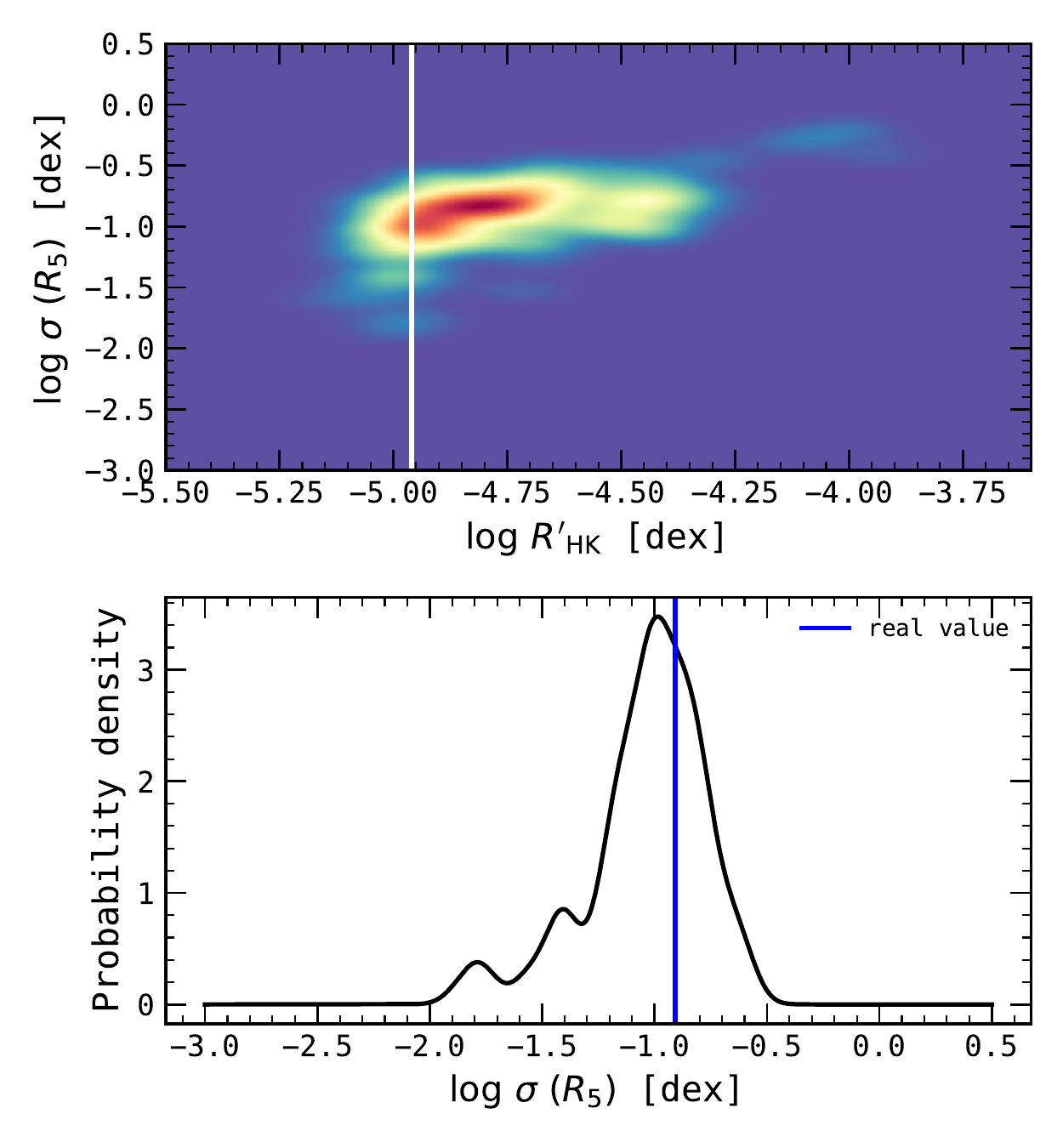}}

        \caption{\textit{Upper panel:} Bivariate KDE of the logarithm of the weighted standard deviation of $R_5$ against median values of $\log R'_\text{HK}$ for stars with more than five nights of observation. Redder colours indicate higher density. The white vertical line indicates the activity level of $\alpha$ Cen B. \textit{Lower panel:} Activity variability probability function for the $\alpha$ Cen B activity level. The vertical blue line marks the value of the weighted $R_5$ standard deviation of $\alpha$ Cen B obtained from the $R_5$ time series.}
        \label{fig:rhk_std_est}
\end{figure}

Figure. \ref{fig:rhk_std_est} shows the results of this method.
The upper panel is the bivariate KDE of activity variability and level for all stars with a white line marking the $\alpha$ Cen B activity level and region of variability associated with it.
The variability KDE is shown in the lower panel (black) where the variability obtained from the time-series is marked by a blue vertical line.
These limits represent the most probable range of variability for stars with activity levels similar to that of $\alpha$ Cen B.
This shows that $\alpha$ Cen B has less activity variability than other stars with the same activity level.
This method could be used to estimate the activity variability KDE and most probable values for any FGK star with activity levels between $-5.5$ and $-3.6$ dex.
We should note that our sample might not be representative of the overall population of FGK stars: for example, giant stars and fast rotating stars are under-represented due to the dominating presence of stars from planet search programs in our sample.
However, our main sequence FGK sample appears to be a good representative of the solar neighbourhood population (see \S \ref{sec:rhk_dist}).

\subsection{Simulating stellar populations with activity level and variability values}
The bivariate KDE distribution we calculated in \S \ref{sec:rhk_std_vs_rhk} can also be used to generate populations of stars with $R'_\text{HK}$ activity levels and variability.
The method is very simple:
\begin{enumerate}
        \item Calculate the $\log R'_\text{HK}$ KDE from the activity distribution and sample values from it;
        \item For every sampled activity level value, obtain the $\log \sigma(R_5)$ KDE from the bivariate KDE map;
        \item Sample one value from each $\log \sigma(R_5)$ KDE obtained in 2.
\end{enumerate}
This will result in a population of stars with synthetic $\log R'_\text{HK}$ and $\log \sigma(R_5)$ values.
Depending on the KDE map used, different synthetic populations of stars can be obtained (assuming enough datapoints can be generated to represent the desired population).

In this example we simulate 1 500 main sequence K stars using the bivariate KDE of 212 K dwarfs.
We construct the KDE map using only K dwarfs, with more than five nights of observation, and using a KDE bandwidth of 0.05 to better constrain the density structure.

\begin{figure}
        \resizebox{\hsize}{!}{\includegraphics{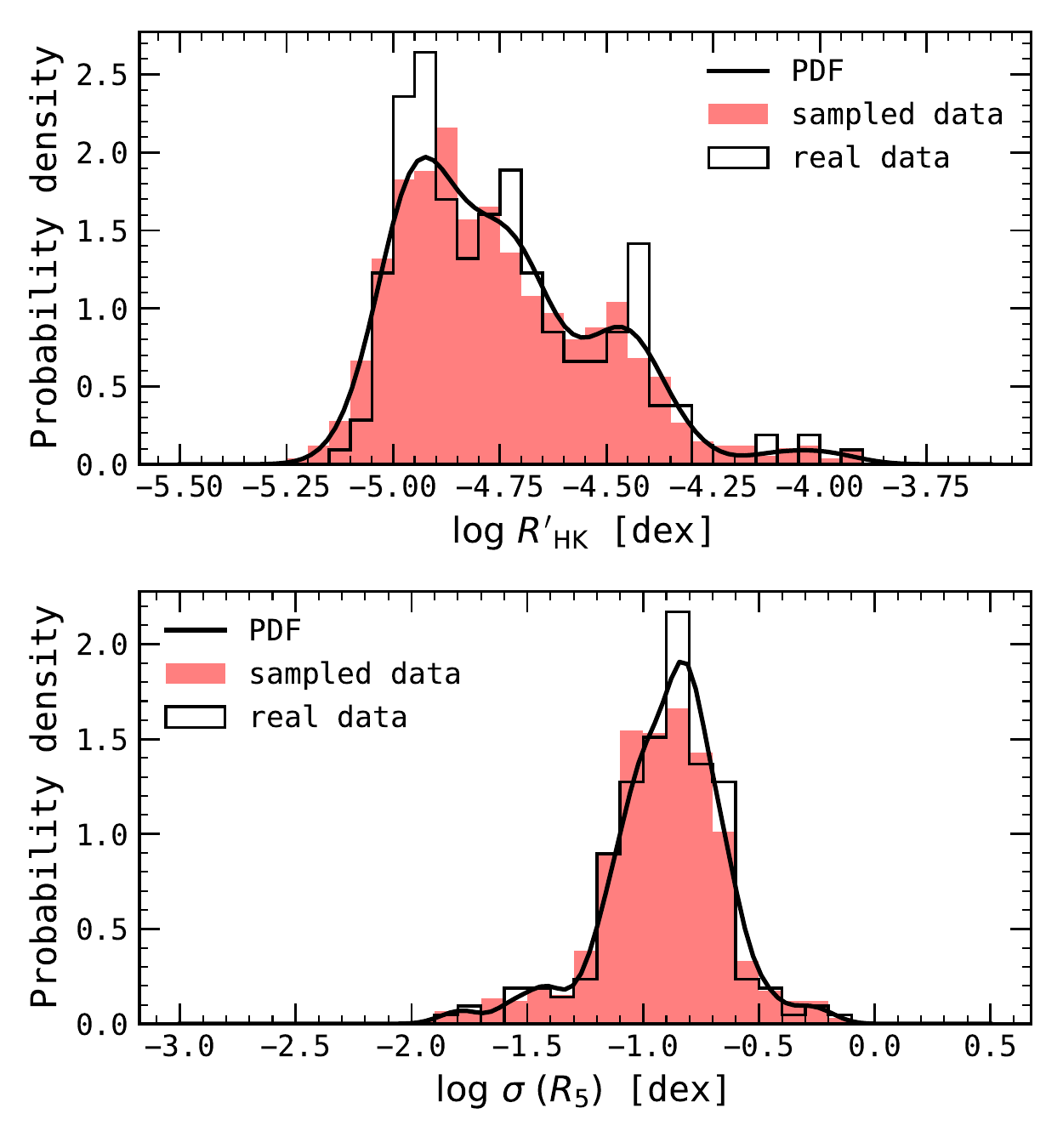}}

        \caption{\textit{Upper panel:} Distribution of median $\log R'_\text{HK}$ values for K dwarfs with more than five nights of observation (black histogram).
        The solid black line is the KDE of the distribution using a bandwidth of 0.05. The red histogram shows the sampled values from the distribution.
        \textit{Lower panel:} Same as upper panel but for the weighted $\log \sigma(R_5)$ values.}
        \label{fig:rhk_sim_hists}
\end{figure}

\begin{figure}
        \resizebox{\hsize}{!}{\includegraphics{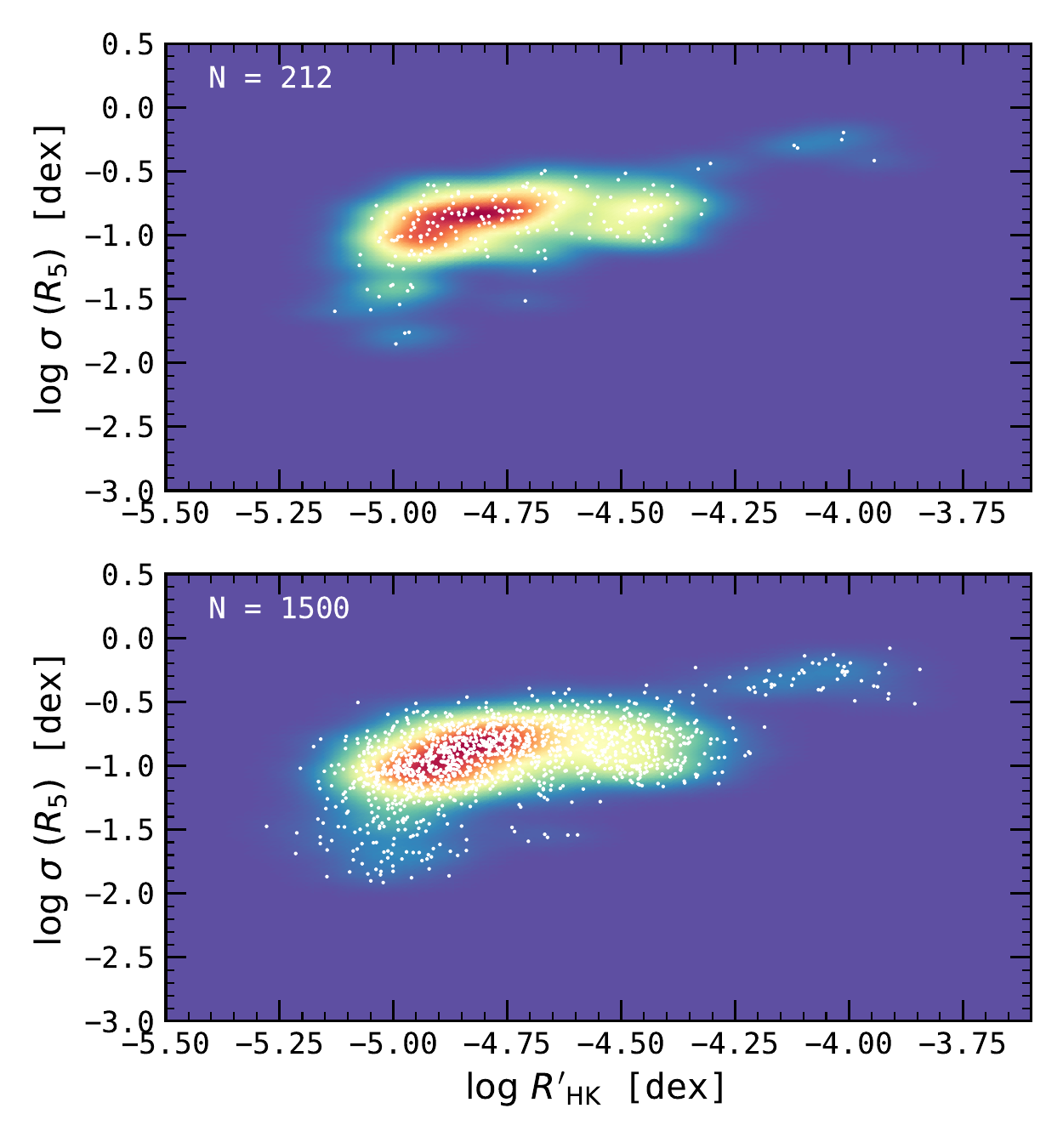}}

        \caption{\textit{Upper panel:} Bivariate KDE distribution of $\log \sigma(R_5)$ against $\log R'_\text{HK}$ for our sample of K dwarfs with more than five nights of observation. \textit{Lower panel:} Synthetic bivariate KDE of 1 500 K dwarfs. White dots are stars. Redder regions have higher density.}
        \label{fig:rhk_sim_maps}
\end{figure}

The results are shown in Figs. \ref{fig:rhk_sim_hists} and \ref{fig:rhk_sim_maps}.
Figure \ref{fig:rhk_sim_hists} shows the activity level (upper panel) and variability (lower panel) distributions (black) for K dwarfs and respective sampled values (red).
Figure \ref{fig:rhk_sim_maps} presents the real K dwarf bivariate KDE (upper panel) and synthetic bivariate KDE  (lower panel).
The synthetic KDE map follows the empirical map very closely and all the major structures are present in the map.

\end{appendix}

\end{document}